\documentclass[a4paper,12pt,twoside]{report}
\usepackage[left=2cm,right=2cm,top=2cm,bottom=3cm]{geometry}

\usepackage{graphicx}
\usepackage{verbatim}
\usepackage{latexsym}
\usepackage{mathchars}
\usepackage{setspace}
\usepackage{cite} 
\usepackage{comment}

\usepackage[]{amsmath}
\usepackage{amssymb}
\usepackage{mathrsfs}
\usepackage[]{mathtools}
\usepackage[]{bm}
\usepackage{braket}
\usepackage{esint}
\usepackage{cases}

\usepackage[]{graphicx}
\usepackage[colorlinks=true]{hyperref}
\usepackage[caption=false]{subfig}

\usepackage{titling}
\usepackage{fancyhdr}

\input{blocked.sty}
\input{uhead.sty}
\input{boxit.sty}
\input{icthesis.sty}

\newcommand{\NN}{{\sf I\kern-0.14emN}}   
\newcommand{\ZZ}{{\sf Z\kern-0.45emZ}}   
\newcommand{\QQQ}{{\sf C\kern-0.48emQ}}   
\newcommand{\RR}{{\sf I\kern-0.14emR}}   

\newcommand{\normallinespacing}{\renewcommand{\baselinestretch}{1.5} \normalsize}

\newcommand{\narrowlinespacing}{\renewcommand{\baselinestretch}{1.0} \normalsize}

\newcommand{\syncc}{~\stackrel{\textstyle \rhd\kern-0.57em\lhd}{\scriptstyle L}~}

\usepackage{fontawesome}

\newcommand{\figref}[1]{\figurename~\ref{#1}}
\newcommand{\tabref}[1]{\tablename~\ref{#1}}

\newcommand{\gl}{{\raisebox{.15em}[0em][0em]{$\scriptscriptstyle{>}$}\hspace{-.52em}\raisebox{-.15em}[0em][0em]{$\scriptscriptstyle{<}$}}}
\newcommand{\ssl}{{\scriptscriptstyle{<}}}
\newcommand{\ssg}{{\scriptscriptstyle{>}}}
\newcommand{\ssm}{{\scriptscriptstyle{-}}}
\newcommand{\ssp}{{\scriptscriptstyle{+}}}
\newcommand{\Inc}{\ssm\ssg}
\newcommand{\Refl}{\ssp\ssg}
\newcommand{\Tra}{\ssm\ssl}
\newcommand{\cc}{\raisebox{.3em}{$\scriptstyle{*}$}}

\newcommand{\inv}{\raisebox{.3em}{$\scriptscriptstyle{-1}$}}

\newcommand{\sgn}{\operatorname{sgn}}

\DeclareMathOperator*{\SumInt}{%
  \mathchoice%
  {\ooalign{$\displaystyle\sum$\cr\hidewidth$\displaystyle\int$\hidewidth\cr}}
  {\ooalign{\raisebox{.14\height}{\scalebox{.7}{$\textstyle\sum$}}\cr\hidewidth$\textstyle\int$\hidewidth\cr}}
  {\ooalign{\raisebox{.2\height}{\scalebox{.6}{$\scriptstyle\sum$}}\cr$\scriptstyle\int$\cr}}
  {\ooalign{\raisebox{.2\height}{\scalebox{.6}{$\scriptstyle\sum$}}\cr$\scriptstyle\int$\cr}}
}

\newcommand{\crT}{{\rotatebox[origin=c]{-90}{$\scriptscriptstyle{\top}$}}}
\newcommand{\rT}{{\hspace{.1em}\rotatebox[origin=c]{90}{$\scriptscriptstyle{\top}$}}}

\usepackage{tikz}
\usetikzlibrary{arrows,arrows.meta}
\newcommand\uscirc[1][1]{%
  \begin{tikzpicture}[scale=0.16]
    \coordinate (center) at (0,0.1);
    \draw[black,-] (0, 0.0) + (center) arc (0:+180:0.8);
  \end{tikzpicture}
  {\hspace{.1em}}
}

\newcommand{\num}[2]{\#_{\hspace{-.1em}#1}^{\mathrm{#2}}}

\newcommand{\kpara}{k_\parallel}
\newcommand{\K}[3]{K_{\mathbf{#1}_{#2}}^{#3}}

\newcommand{\epsv}{\epsilon^{\scriptscriptstyle \mathrm{v}}}
\newcommand{\epsm}{\epsilon^{\scriptscriptstyle \mathrm{m}}}
\newcommand{\epsi}{\epsilon^{\scriptscriptstyle \mathrm{i}}}
\newcommand{\epstau}{\epsilon^{\scriptscriptstyle \tau}}

\begin{document}

\title{\LARGE {\bf 
    Dynamical metasurfaces:
    electromagnetic properties and instabilities
  }\\
  \vspace*{6mm}
}

\author{Daigo Oue}
\submitdate{\today}

\normallinespacing
\maketitle

\preface
\cleardoublepage
\section*{Declaration of authorship}
\subsection*{Statement of Originality}
I hereby declare that everything in this thesis is my own work, except where mentioned.

\subsection*{Copyright Declaration}
Under this licence, you may copy and redistribute the material in any medium or format. You may also create and distribute modified versions of the work. This is on the condition that; you credit the author, do not use it for commercial purposes and share any derivative works under the same licence.

When reusing or sharing this work, ensure you make the licence terms clear to others by naming the licence and linking to the licence text. Where a work has been adapted, you should indicate that the work has been changed and describe those changes.

Please seek permission from the copyright holder for uses of this work that are not included in this licence or permitted under UK Copyright Law.

\newpage
\addcontentsline{toc}{chapter}{Abstract}

\begin{abstract}
  In this thesis,
  I analyse the electromagnetic properties of dynamical metasurfaces and find two critical phenomena.
  The first is the Casimir-induced instability of a deformable metallic film.
  In general,
  two charge-neutral interfaces attract with or repel each other due to the contribution from the zero-point fluctuation of the electromagnetic field between them, namely, the Casimir effect.
  The effects of perturbative interface corrugation on the Casimir energy in the film system is studied by the proximity force approximation with dispersion correction. If the corrugation period exceeds a critical value,
  the Casimir effect dominates the surface tension and brings about structural instability.
  The second is \v{C}erenkov radiation in the vacuum from a time-varying, corrugated surface.
  Travelling faster than light brings about electromagnetic shock waves, \v{C}erenkov radiation.
  Since light is the fastest object in a vacuum,
  it has been considered that \v{C}erenkov radiation is emitted only in the presence of some refractive index.
  Here,
  I propose mimicking a series of particles travelling faster than light in a vacuum by dynamical surface corrugation to find \v{C}erenkov radiation in a vacuum from the surface.
  The dynamical corrugation induces an effective current source on the surface with an external electrostatic field applied.
  When the corrugation profile is of travelling wave type,
  the source can be regarded as a series of dipoles virtually travelling along the surface.
  If the phase velocity of the travelling wave profile exceeds the speed of light, and so do the dipoles,
  they emit \v{C}erenkov radiation in a vacuum.
\end{abstract}

\cleardoublepage

\addcontentsline{toc}{chapter}{Acknowledgements}

\begin{acknowledgements}
  First of all, 
  I would like to thank my supervisor, Prof.~Sir John Brian Pendry,
  for kindly accepting me as a PhD student,
  although I had never conducted theoretical studies and had little knowledge before joining his group.
  He kindly taught me theoretical physics from the beginning.
  I could not fill a lot of gaps in my knowledge without his support.
  His physical insights and mathematical ability always inspired me to deepen my understanding and broaden my horizons.

  A big thank goes to Dr.~Kun Ding, my co-supervisor.
  He continuously supported me in many ways.
  When I had a hard problem, he helped me divide it into small problems that I can address.
  When I was mentally tired, he concentrated on listening to me.
  Every time I had a question and sent him a message, he gave me swift, constructive advice.
  Without his kind support, I could not reach this point.

  I would like to express my gratitude to current and former members of our group:
  Dr.~Paloma Arroyo Huidobro,
  Dr.~Yao-Ting Wang,
  Dr.~Fan Yang,
  Dr.~Emanuele Galiffi,
  and
  Lizhen Lu.
  They helped me to initiate my PhD project.
  I had fruitful discussions with them that move the project forward.

  I would also thank my office mates.
  Samuel Palmer kindly helped me to improve my academic writing in English.
  I could not generate any manuscript from my sole project without his support.
  Yiming Lai often spare his time for discussion with me.
  I always enjoyed the discussion on physics.
  Tom Hodson supported us as our representative.
  Although I have seldom appeared in the college,
  I got along with him.

  I am thankful to Prof.~Satoshi Kawata, my previous supervisor, who recommended that I study at Imperial.
  He kindly allowed me to use his office during my stay back in Japan due to COVID-19,
  and I could continue my PhD project even in a difficult situation.
  I am also thankful to Ai Shimode and Nobune Toba, secretaries to Prof.~Kawata.
  They warmly welcomed me there.

  Special thanks to my collaborators:
  Prof.~Mamoru Matsuo,
  Dr.~Yuya Ominato,
  Dr.~Junji Fujimoto,
  and 
  Dr.~Hideaki Nishikawa.
  I enjoyed the collaboration projects very much.
  Besides,
  they gave me helpful advice when I am worried about my career.

  I thank Dr.~Masaru Hongo for his informal lecture on quantum field theory which has directly motivated me to write Sec.~4.1,
  where the PFA method is described from the field theoretical point of view. 
  His lecture made me recognise that the path integral formulation is a sophisticated way to analyse fields in the presence of boundaries,
  as in the Casimir type problems.

  I am appreciative of friends of mine in Japan,
  who were/are in their PhD courses:
  Prof.~Suguru Shimomura,
  Dr.~Toshiki Kubo,
  Dr.~Ryo Kato,
  Taiki Matsushita,
  Dr.~Toshihiro Ota, 
  and
  Teppei Suzuki.
  Their activities stimulated me to continue studying abroad.

  Last but not least, 
  I want to thank my wife, Yuri.
  She cheered me up when I had a hard time.
  She blessed me when I had good news.
  Without her devoted support during this COVID-19 situation,
  I could not continue to remotely study abroad or write this thesis.
\end{acknowledgements}

\cleardoublepage

\begin{dedication}
  This thesis is dedicated to my wife Yuri.
  I am truly happy to have you in my life.
\end{dedication}

\clearpage

\narrowlinespacing

\vspace*{4mm}

`Nature creates curved lines while humans create straight lines.'\\
\\
\emph{Hideki Yukawa%
}

\normallinespacing

\body

\chapter{Introduction}
More than 100 years ago,
Nichols and Hull experimentally confirmed the radiation force exerted by thermal illumination
\cite{nichols1903pressure, nichols1903pressure2nd}.
At that time,
the intensity of the light source was not high, and the measured force was not so large that there had not been any specific application for a while.
In 1960,
laser, a high-intensity light source, was invented in Bell laboratories 
\footnote{
  See, for example, \cite{nelson2010bell} for the controversy over the priority in the invention of lasers.
}
\cite{collins1960coherence,schawlow1958infrared}.
That is what Ashkin utilised to exert large radiation forces and manipulate microparticles in the same laboratories
\cite{ashkin1970acceleration,ashkin1986observation}.
His studies pioneered optical manipulation.

When I was a bachelor student before studying wave physics,
I got interested in the radiation force because I could not believe that a particle with zero mass has momentum and exerts a force.
Reproducing Ashkin's experiment,
I got more interested in the radiation force and started to learn wave physics, focusing on the radiation force.

The study of wave physics made me fascinated by the quantum aspects of light.
Besides reproducing Ashkin's experiment,
I conducted a verification experiment of Bell's inequality to confirm the quantumness of light.
Then,
I had questions in my mind:
is there any force exerted by virtual photons (vacuum fluctuation)?
That is nothing but the Casimir force.
I wanted to freely dig the physics around the force mediated by vacuum fluctuation.
This was when I decided to switch my career from the experimental side to the theoretical side.
I started working on Casimir physics with Prof. Sir John Pendry at Imperial College London.

This thesis presents two critical phenomena at dynamical surfaces,
 starting from the Casimir effect on a deformable surface.

In the first part of this thesis,
we study the effects of the Casimir force on the structural stability of a deformable metallic film at the nanoscale.
Since the Casimir force pulls together two neutral interfaces,
small corrugation introduced on the film surface could grow.
On the other hand,
corrugation increases the surface area and hence the surface tension that keeps the planar surface energetically favourable.
However,
it is not the case at the nanoscale.
Surface plasmon modes at two interfaces of the film are strongly hybridised when the film is very thin,
of the order of 10 [nm].
This hybridisation causes the asymmetric level repulsion,
leading to net lowering of the Casimir energy to overcome the surface tension and structural instability of the film.

In the second part of the thesis,
we study the properties of a deformable surface with a dynamical modulation.
It has been studied that the vacuum fluctuation brings about not only the attractive force between two bodies but also a frictional force between two relatively moving bodies,
the Casimir friction.
What happens if one temporally modulates the surface itself instead of moving the bodies? Keeping this question in mind,
we investigate the electromagnetic properties of the dynamically modulated surface.
We will find the dynamical surface emits \v{C}erenkov radiation,
which triggers the Casimir friction.

\section{List of Publications}
This thesis is drawn from the following publications:
\begin{enumerate}
  \item 
    Kun Ding, \underline{Daigo Oue}, C. T. Chan and J. B. Pendry
    ``Casimir-Induced Instabilities at Metallic Surfaces and Interfaces,''
    \textit{Physical Review Letters} \textbf{126},
    046802 (2021).
  \item
    \underline{Daigo Oue}, Kun Ding and J. B. Pendry
    ``Calculating spatiotemporally modulated surfaces: A dynamical differential formalism,''
    \textit{Physical Review A} \textbf{104},
    013509 (2021).
  \item 
    \underline{Daigo Oue}, Kun Ding and J. B. Pendry
    ``\v{C}erenkov radiation in vacuum from a superluminal grating,''
    \textit{Physical Review Research} \textbf{4},
    013064 (2022).
\end{enumerate}

The first publication is on the Casimir effect on a metallic film.
The first half of this thesis is based on the first publication.

The second and third publications are on asymmetric diffraction and \v{C}erenkov radiation from a dynamically modulated dielectric surface.
The latter half of this thesis is based on these publications.

The following manuscripts were generated from my sole project during my PhD course but not included in this thesis:
\begin{itemize}
  \item \underline{Daigo Oue} 
    ``Complex-angle analysis of electromagnetic waves on interfaces,''
    \textit{Journal of Physics: Conference Series} \textbf{1220},
    012058 (2019).
  \item \underline{Daigo Oue}
    ``Dielectric loss induced excess momentum and anomalous spin of light''
    \textit{Complex Light and Optical Forces XIII} \textbf{10935},
    109350D (2019).
  \item \underline{Daigo Oue}
    ``Dissipation effect on optical force and torque near interfaces,''
    \textit{Journal of Optics} \textbf{21},
    065601 (2019).
\end{itemize}

\section{Notation}
\label{sec:notation}
The notation adopted in this thesis is listed in Table \ref{tab:notation}.
\begin{table}[htbp]
  \centering
  \caption{
    Important symbols and formulas used throughout this thesis.
  }
  \label{tab:notation}
  \begin{tabular}{cc}
    \hline\hline
    Symbol 
    & 
    Meaning
    \\ 
    \hline 
    \noalign{\vskip .5em}
    $\vec{\mathcal{E}}_{\mathbf{x},z}^\Lambda$,
    $\vec{\mathcal{H}}_{\mathbf{x},z}^\Lambda$
    &
    electric and magnetic fields evaluated in the real space and time domain
    \\
    $\Lambda$
    &
    label specifying a mode in the real space
    \\
    $\vec{E}_{\mathbf{k},z}^{\sigma\tau}$,
    $\vec{H}_{\mathbf{k},z}^{\sigma\tau}$
    &
    electric and magnetic fields evaluated in the reciprocal space
    \\
    $\sigma$
    &
    label specifying to which direction the mode propagates
    \\
    $\tau$
    &
    label specifying in which medium the mode lives
    \\
    $\lambda$
    &
    polarisation index
    \\
    $\vec{e}_{\lambda,\mathbf{k}}^{\sigma\tau}$,
    $\vec{h}_{\lambda,\mathbf{k}}^{\sigma\tau}$
    &
    linear polarisation basis vectors
    \\
    $\vec{k}_\parallel$
    &
    parallel component of a wave vector
    \\
    $K_\mathbf{k}^\tau$
    &
    wavenumber in the $z$ direction in a medium labeled by $\tau$
    \\
    $k_0$
    &
    radiation wavenumber
    \\
    $\epsilon^\tau, \mu^\tau$
    &
    permittivity and permeability of a medium labeled by $\tau$
    \\
    $r_{\lambda,\mathbf{k}}$,
    $t_{\lambda,\mathbf{k}}$
    &
    Fresnel reflection and transmission coefficients for the $\lambda$ polarisation
    \\
    $a_\mathbf{x}$
    &
    boundary profile
    \\
    $\vec{n}$,
    $\vec{t}_{1,2}$
    &
    normal and tangential vectors
    \\
    $g, \Omega$
    & 
    spatial and temporal frequencies of surface modulation
    \\ 
    \noalign{\vskip .5em}
    \hline\hline
  \end{tabular}
\end{table}

\section{Abbreviations}
\label{sec:abbreviation}
The abbreviations adopted in this thesis are listed in Table \ref{tab:abbreviation}.
\begin{table}[htbp]
  \centering
  \caption{Abbreviation used in this thesis.}
  \label{tab:abbreviation}
  \begin{tabular}{cc}
    \hline\hline
    abbreviation & meaning
    \\
    \hline
    \noalign{\vskip .5em}
    TE & transverse electric
    \\
    TM & transverse magnetic
    \\
    DC & direct current
    \\
    PFA & proximity force approximation
    \\
    \noalign{\vskip .5em}
    \hline\hline
  \end{tabular}
\end{table}

\part{Casimir-induced instabilities of metallic thin films}
\chapter{Surface plasmon polaritons in a metallic film}
\label{ch:background_casimir}
\section{Plasmonics}
\label{sec:plasmonics}
\begin{figure}
  \centering
  \includegraphics[width=.8\linewidth]
  {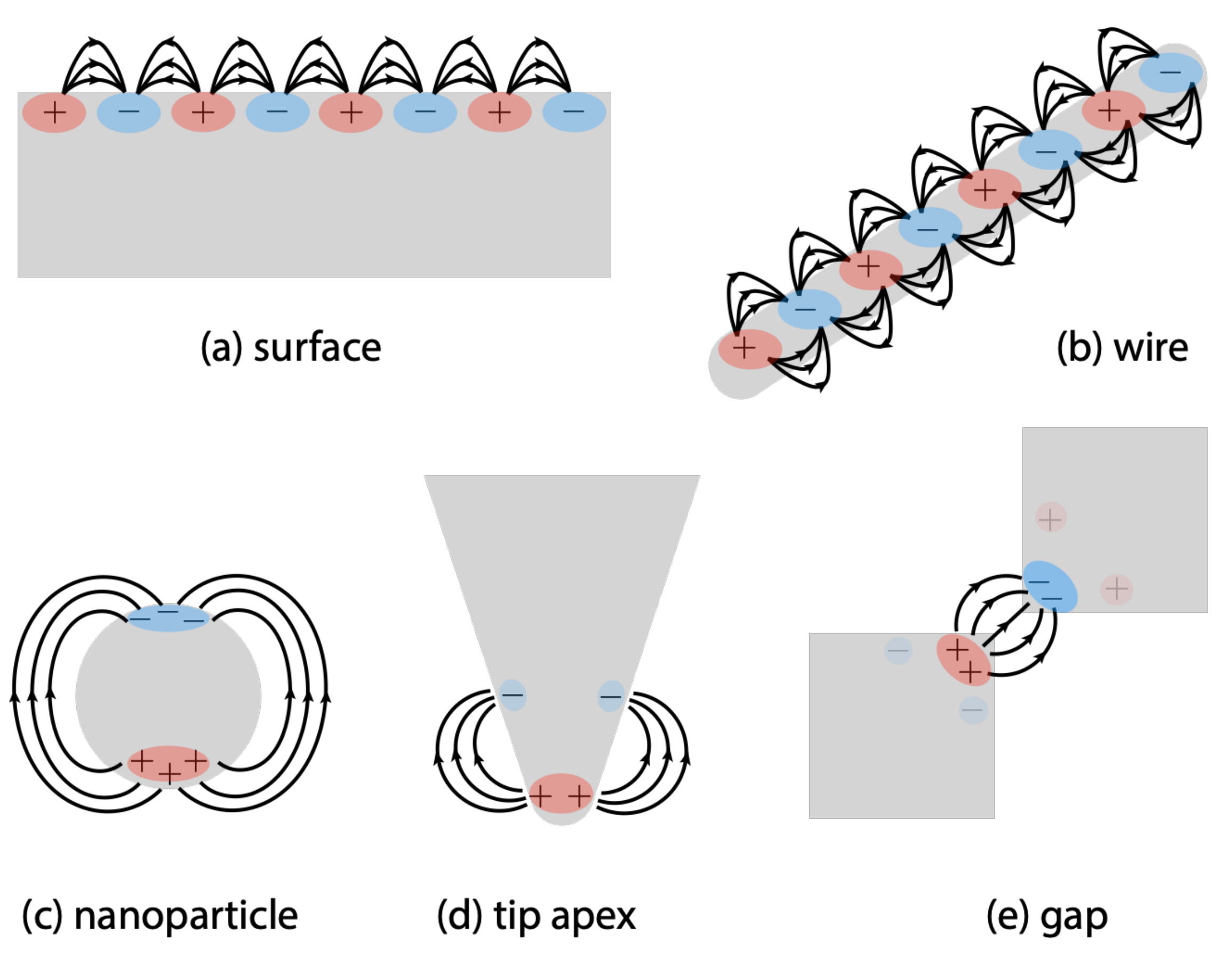}
  \caption{
    Electron oscillation interacting with electromagnetic fields in various configurations.
    Propagating electron oscillation interacting with electromagnetic fields at (a) a metallic surface and (b) wire.
    Localised electron oscillation interacting with the fields (c) at a metallic nanoparticle, (d) near the apex of a metallic tip and (e) at a gap between metallic media.
  }
  \label{fig:plasmonics}
\end{figure}

Plasmonics is a research field where the interaction between metallic media and electromagnetic fields is studied, focusing on collective electronic  oscillations in the metallic media.
In \figref{fig:plasmonics},
various configurations used in plasmonics are shown.
The black arrows are field lines,
and blue (red) circles represent electron (hole) clouds.
Propagating oscillations of electrons can be interacting with electromagnetic fields at metallic surfaces and wires,
which are applied to energy and information transfer at the nanoscale
\cite{%
  takahara1997guiding,%
  falk2010near%
}.
Localised oscillations of electrons coupled with the fields at metallic nanoparticles, tip apices and gaps are utilised in microscopy, spectroscopy and sensing,
\cite{%
  stiles2008surface,%
  kawata2017nano,%
  verma2017tip%
}.
Recently, 
it has been demonstrated that those configurations work in the quantum regime
\cite{%
  tame2013quantum,%
  fitzgerald2016quantum%
}.

Plasmonics has become a role model that provides a minimal description of the medium-field interaction and is a good starting point for studying electromagnetism in media. Indeed, it has inspired the studies on the field interaction with lattice vibrations (i.e.~phonons) 
\cite{%
  caldwell2015low%
}
and often becomes a starting point to investigate dispersion effects on Casimir type problems
\cite{%
intravaia2005surface,%
intravaia2007role,%
guerout2014derivation,%
iizuka2019casimir%
}.

The field dynamics is described by Maxwell-Heaviside equations,
\begin{align}
  \nabla \cdot \vec{\mathcal{E}} 
  &= -\nabla \cdot \vec{\mathcal{P}}/\epsilon_0,
  \label{eq:gauss_E}
  \\
  \nabla \cdot \vec{\mathcal{H}} 
  &= -\nabla \cdot \vec{\mathcal{M}},
  \label{eq:gauss_H}
  \\
  \nabla \times \vec{\mathcal{E}} 
  + \mu_0 \frac{\partial}{\partial t} \vec{\mathcal{H}}
  &= -\mu_0 \frac{\partial}{\partial t} \vec{\mathcal{M}},
  \label{eq:faraday}
  \\
  \nabla \times \vec{\mathcal{H}} 
  - \epsilon_0 \frac{\partial}{\partial t} \vec{\mathcal{E}}
  &= +\frac{\partial}{\partial t} \vec{\mathcal{P}},
  \label{eq:ampere}
\end{align}
where $\mathcal{P}$ is electric polarisation and $\mathcal{M}$ is magnetisation (magnetic polarisation).
When the field propagates in a medium,
nonequilibrium dynamics is triggered in the medium and brings about electric and/or magnetic polarisation, which gives back actions to the field 
[i.e.~$\mathcal{P}\neq 0$ and/or $\mathcal{M}\neq 0$ in Eqs.~(\ref{eq:gauss_E}--\ref{eq:ampere})].
There are many ways to calculate the electromagnetic response of a medium,
$\mathcal{P}$ and $\mathcal{M}$,
including hydrodynamic theory 
\cite{%
  ding2018optical,%
  baghramyan2021laplacian,%
  mortensen2021mesoscopic%
}
and the density functional theory
\cite{%
  marini2009yambo,%
  deslippe2012berkeleygw,%
  leng2016gw,%
  prandini2019simple,%
  sato2014numerical,%
  yabana2012time%
}.

One of the simple but powerful models,
which is often used in plasmonics and successfully describes the electromagnetic responses of dielectric and metallic media,
is Drude-Lorentz model
\cite{%
  maier2007plasmonics%
}.
In the dielectric and metallic media,
the electric response dominates over the magnetic one.
It stems from the dynamics of electrons,
which can be described by the equation of motion of each electron,
\begin{align}
  \ddot{\vec{r}}_t
  +{\omega_0}^2 \vec{r}_t
  +\gamma \dot{\vec{r}}_t
  = 
  -\frac{q_e}{m_e}
  \vec{\mathcal{E}}_{t},
  \label{eq:drude-lorentz}
\end{align}
where the right-hand side corresponds to the electric part of the Lorentz force that dominates the dielectric response.
We have defined the position of electron $\vec{r}_t$ in the time domain, the electron mass $m_e$ and the electron charge $-q_e$.
The eigenfrequency of an electron $\omega_0$ and the damping constant $\gamma$ are determined by microscopic origins such as interband transitions and electron-phonon scattering \cite{maier2007plasmonics}.
We have introduced the dot and double dot symbols to represent the first and second-order derivatives with respect to time.
Note that $\vec{\mathcal{E}}_{t}$ is the electric field acting on the electron whose spatial variation is neglected because the typical wavelength of the electromagnetic field is much larger than that of electrons.

One can directly solve the field equations (\ref{eq:gauss_E}--\ref{eq:ampere}) and the electron equation of motion \eqref{eq:drude-lorentz} simultaneously by numerical methods in the time or frequency domain
\cite{%
  raman2010photonic,%
  raman2011perturbation,%
  shin2012instantaneous,%
  raman2013upper%
}.
Alternatively, 
in the frequency domain,
we can embed the electron dynamics in a dielectric response function $\epsilon(\omega)$ to proceed with the analytical approach.
Applying the temporal Fourier transform to the equation of motion produces the equation of motion in the frequency domain,
\begin{align}
  -\omega^2 \tilde{\vec{r}}_\omega
  +{\omega_0}^2 \tilde{\vec{r}}_\omega
  -i \omega \gamma \tilde{\vec{r}}_\omega
  = 
  -\frac{q_e}{m_e}
  \vec{E}_{\omega},
  \\
  \tilde{\vec{r}}_\omega
  =
  \frac{-q_e/m_e}{{\omega_0}^2 - i \omega \gamma - \omega^2}
  \tilde{\vec{E}}_{\omega}.
\end{align}
The resulting electric polarisation induced by the field at a position $\vec{r}$ is written as
\begin{align}
  \vec{P}_\omega
  &= -n q_e \tilde{\vec{r}}_\omega 
  = 
  \frac{{\omega_\mathrm{p}}^2}{{\omega_0}^2 - i \omega \gamma - \omega^2}
  \epsilon_0
  \vec{E}_{\omega},
\end{align}
where we have defined the electron density $n$ and the permittivity of the free space $\epsilon_0$
and the plasma frequency 
$\omega_\mathrm{p} \equiv \sqrt{n{q_e}^2/(m_e\epsilon_0)}$.
We can find the dielectric function $\epsilon(\omega)$ as following:
\begin{align}
  \vec{D}_\omega 
  &= 
  \epsilon_0 \vec{E}_\omega 
  +
  \vec{P}_\omega,
  = 
  \epsilon_0
  \left(
    1 + 
    \frac{{\omega_\mathrm{p}}^2}{{\omega_0}^2 - i \omega \gamma - \omega^2}
  \right)
  \vec{E}_\omega,
  \\
  \epsilon(\omega)
  &=
  1 + 
  \frac{{\omega_\mathrm{p}}^2}{{\omega_0}^2 - i \omega \gamma - \omega^2}.
  \label{eq:epsilon_DL}
\end{align}
Note that we have used the constitutive relation between the electric flux density and the field strength
[i.e.~$
\vec{D}_\omega = \epsilon(\omega) \epsilon_0 \vec{E}_\omega
$].

In order to obtain an effective dielectric function of a metallic medium,
we can take the free-electron limit in the Drude-Lorentz dielectric function \eqref{eq:epsilon_DL},
\begin{align}
  \omega_0, \gamma \longrightarrow 0^+,
\end{align}
where $0^+$ is a positive infinitesimal,
and have
\begin{align}
  \epsilon^\mathrm{m}
  = 1 - \frac{\omega_\mathrm{p}^2}{\omega^2 + i0^+}
  \label{eq:eps^m}
\end{align}
This plasma permittivity \eqref{eq:eps^m} is one of the simplest dielectric functions often used in plasmonics.
From the expression,
we can find that the permittivity is negative below the plasma frequency $\omega_\mathrm{p}$,
where electromagnetic waves cannot enter the metallic medium,
while it is positive above the plasma frequency, and thus the metal behaves as a dielectric medium.

In the following section,
we study the electromagnetic response of a thin metallic film,
taking the plasma permittivity \eqref{eq:eps^m} and the corresponding macroscopic Maxwell-Heaviside equations,
\begin{align}
  \nabla \cdot \vec{\mathcal{E}} 
  &= 0,
  \label{eq:gauss_E_macro}
  \\
  \nabla \cdot \vec{\mathcal{H}} 
  &= 0,
  \label{eq:gauss_H_macro}
  \\
  \nabla \times \vec{\mathcal{E}} 
  + \mu_0 \frac{\partial}{\partial t} \vec{\mathcal{H}}
  &= 0,
  \label{eq:faraday_macro}
  \\
  \nabla \times \vec{\mathcal{H}} 
  - \epsilon_0 \frac{\partial}{\partial t} 
  \varepsilon \circledast \vec{\mathcal{E}}
  &= 0,
  \label{eq:ampere_macro}
\end{align}
where we have introduced a convolution
\begin{align}
  \varepsilon \circledast \vec{\mathcal{E}}
  :=
  \int_{-\infty}^{t}
  \varepsilon_{t-t'} \vec{\mathcal{E}}_{t'}
  \hspace{.1em}\mathrm{d}t'.
\end{align}
Remind that the electromagnetic response of the media is dominated by electric polarisation.
The electric response is embedded in the dielectric function in the time domain,
\begin{align}
  \varepsilon_{t-t'} 
  =
  \int
  \epsilon(\omega)
  e^{-i\omega (t-t')}
  \mathrm{d}\omega/2\pi.
\end{align}
If we neglect the frequency dependence of the dielectric function (dispersion), the convolution becomes just a scalar multiplication,
\begin{align}
  \frac{\mathrm{d}\epsilon(\omega)}{\mathrm{d}\omega} = 0
  \quad
  &\Rightarrow
  \quad
  \varepsilon_{t-t'}
  =
  \epsilon \delta_{t-t'}
  \\
  &\Rightarrow
  \quad
  \varepsilon \circledast \vec{\mathcal{E}}
  =
  \int_{-\infty}^{t}
  \varepsilon_{t-t'} \vec{\mathcal{E}}_{t'}
  \hspace{.1em}\mathrm{d}t'
  =
  \epsilon
  \vec{\mathcal{E}}
\end{align}

\section{Scattering and transfer matrix method for layered systems}
\label{sec:sca_trans_mat}
One systematic way to calculate wave scattering phenomena at an interface between two media is scattering formalism.
In this formalism,
we firstly find the eigenmodes and expand the wave in a series of eigenmodes in each medium to identify incoming and outgoing components with respect to the interface.
We can write the outgoing amplitudes in terms of the incoming wave amplitudes by writing all boundary conditions at the interface in a matrix-vector form.

A popular example based on this type of approach is the calculation of Mie scattering
\cite{%
  mie1908beitrage,%
  bohren2008absorption%
},
In the Mie problem,
the eigenmode expansion is performed in the spherical coordinate system, and then the boundary conditions are imposed to calculate light scattering by small particles. 

\begin{figure}[htbp]
  \centering
  \includegraphics[width=\linewidth]
  {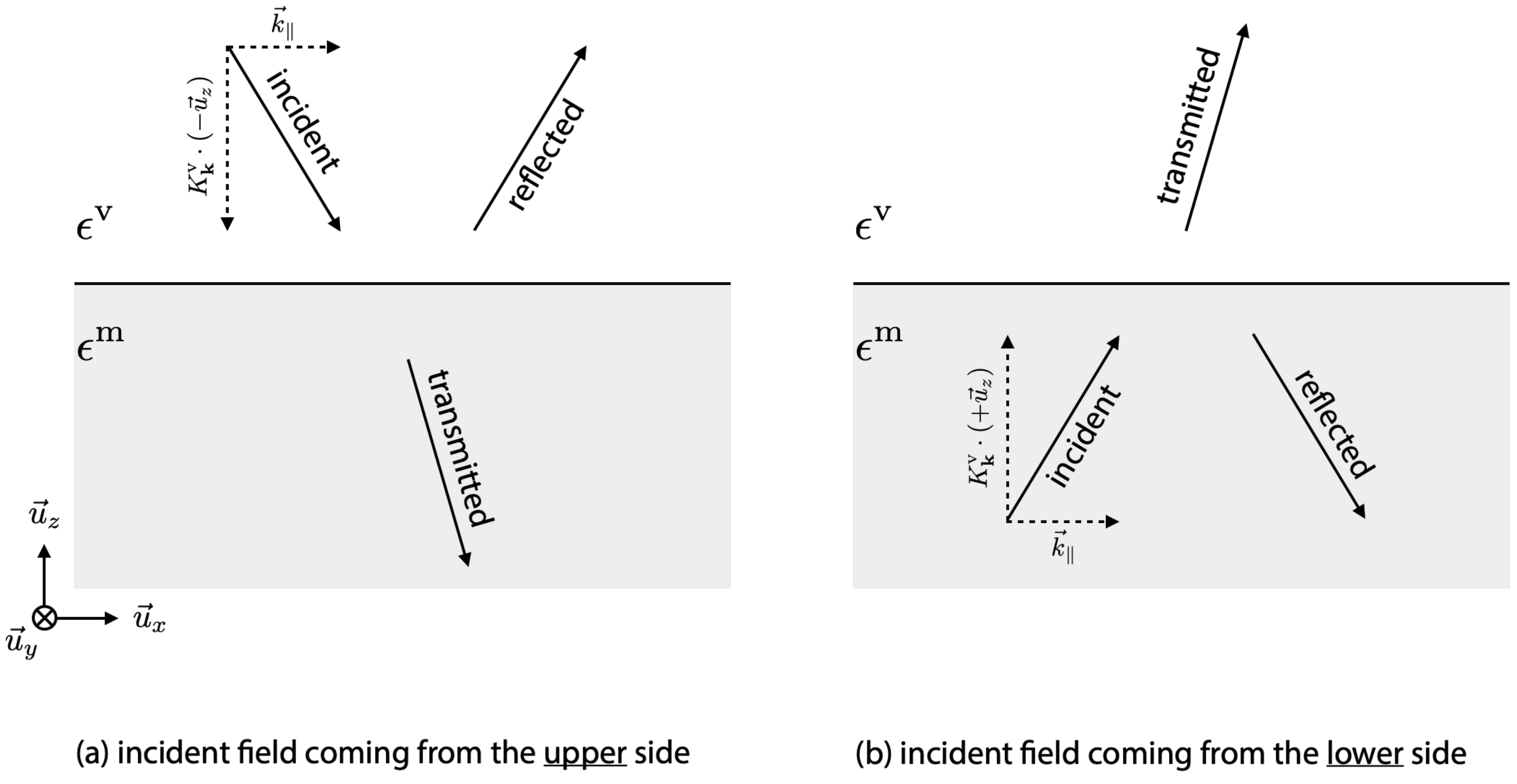}
  \caption{
    Reflection and transmission of the electromagnetic field at a flat interface between two media.
    The incident field can come from either (a) upper or (b) lower side.
  }
  \label{fig:flat}
\end{figure}

In the following, we will see another example,
the calculation of reflection and transmission at a flat surface,
leading to Fresnel coefficients (\figref{fig:flat}).
We assume each medium is homogeneous and characterised by a dielectric function.
The incident field can come from either the upper or lower side.
In the flat surface problem,
we can work on the Cartesian coordinate system.

\subsection{Eigenmode expansion}
Here,
we consider the eigenmode expansion of the fields in the Cartesian coordinate system in order to investigate the reflection and transmission at the flat interface between two media (\figref{fig:flat}).
Since each medium is translational invariant in the $x$ and $y$ directions as well as time invariant,
we can write the electric and magnetic fields in the Fourier integral forms with respect to $x$, $y$ and $t$,
\begin{align}
    \vec{\mathcal{E}}{}_{\mathbf{x},z} 
    &= \int
    \vec{E}{}_{\mathbf{k},z}
    e^{i\mathbf{k}\cdot \mathbf{x}}
    \mathrm{d}\mathbf{k},
    \quad
    \vec{\mathcal{H}}{}_{\mathbf{x},z} 
    = \int
    \vec{H}{}_{\mathbf{k},z},
    e^{i\mathbf{k}\cdot \mathbf{x}}
    \mathrm{d}\mathbf{k}
  \label{eq:FT[E],FT[H]}
\end{align}
where we have collected the frequency and the parallel component of the wave vector in a single vector,
\begin{align}
  \mathbf{k} 
  = (k_x, k_y, ik_0),
  \quad
  k_0 \equiv \omega/c,
\end{align}
and introduced a shorthand notation,
$\mathrm{d}\mathbf{k} := \mathrm{d}\omega \mathrm{d}k_x \mathrm{d}k_y/(2\pi)^3$
and
$\mathbf{x} = (x,y,ict)$.
Note that we keep the $z$ coordinate on the right-hand side because we have an interface in the direction.
The electric and magnetic fields are observable, real quantities 
(e.g.~$
\vec{\mathcal{E}}{}_{\mathbf{x},z}^{\hspace{.1em} \cc}
= 
\vec{\mathcal{E}}{}_{\mathbf{x},z}
$)
and thus we have the following constraints:
\begin{align}
  \vec{E}{}_{\mathbf{k},z}
  &= 
  \vec{E}{}_{-\mathbf{k},z}^{\hspace{.1em}\cc},
  \quad
  \vec{H}{}_{\mathbf{k},z}
  = 
  \vec{H}{}_{-\mathbf{k},z}^{\hspace{.1em}\cc}.
  \label{eq:constraint_vec_E}
\end{align}

By substituting the integral expressions \eqref{eq:FT[E],FT[H]} into the macroscopic Maxwell--Heaviside equations (\ref{eq:gauss_E_macro}--\ref{eq:ampere_macro}),
we can derive a vectorial wave equation for the electric field,
\begin{align}
  \left[
    -\frac{\partial^2}{\partial z^2}
    -
    \left(
      \epsilon^\tau {k_0}^2 - {k_\parallel}^2
    \right)
  \right]
  \vec{E}_{\mathbf{k},z}^\tau
  &= 0,
  \label{eq:ODE_E_z}
\end{align}
where 
$\tau \in \{\mathrm{v}, \mathrm{m}\}$
is a label specifying the medium and hence the dielectric function,
and we have defined the parallel wave vector 
$\vec{k}_\parallel = k_x \vec{u}_x + k_y \vec{u}_y$ 
and the corresponding parallel wavenumber 
$k_\parallel = \sqrt{{k_x}^2 + {k_y}^2}$.
We can find the solution to \eqref{eq:ODE_E_z} in each medium,
\begin{align}
  \vec{E}{}_{\mathbf{k},z}^{\tau}
  = 
  \sum_{\sigma=\pm}
  \vec{E}{}_{\mathbf{k},0}^{\sigma \tau}
  e^{i\sigma K_{\mathbf{k}}^\tau z}
  \label{eq:E_kz}
\end{align}
where $\sigma \in \{+,-\}$ specifies the propagation direction for a given $\mathbf{k}$ and allows us to identify whether the field is incoming or outgoing with respect to the interface.
\tabref{tab:incoming_outgoing} shows which pair of labels $(\sigma,\tau)$ corresponds to the incoming/outgoing field.
For example,
a field propagating downward ($\sigma = -$) in the upper medium ($\tau = \mathrm{v}$) is incoming.
\begin{table}[tbp]
  \centering
  \caption{Two labels $(\sigma, \tau)$ and the corresponding field}
  \label{tab:incoming_outgoing}
  \begin{tabular}{c|c}
    $(\sigma, \tau)$ & incoming or outgoing
    \\ \hline \hline
    $(+, \mathrm{v})$ & outgoing
    \\
    $(-, \mathrm{v})$ & incoming
    \\
    $(+, \mathrm{m})$ & outgoing
    \\
    $(-, \mathrm{m})$ & incoming
  \end{tabular}
\end{table}
The wavenumber in the $z$ direction is defined by
\begin{align}
  K_{\mathbf{k}}^\tau
  &=
  \operatorname{sgn}(\omega)
  \operatorname{Re}
  \sqrt{
    \frac{\omega^2}{c^2}\epsilon^\tau - {k_\parallel}^2
  }
  +
  i\operatorname{Im}
  \sqrt{
    \frac{\omega^2}{c^2}\epsilon^\tau - {k_\parallel}^2
  }.
  \label{eq:K}
\end{align}
The prefactor at the real part reflects that the negative frequency mode acquires the opposite phase to the positive counterpart \cite{pendry2008time}.
Note that the group velocity is invariant under the frequency inversion $\omega \mapsto -\omega$ while the wavenumber is flipped with the inversion.
Let us consider a simple case $(\vec{k}_\parallel, \epsilon^\tau) = (0,1)$.
We can confirm that the group velocity is independent of the signature of the frequency $\operatorname{sgn}(\omega)$,
\begin{align}
  K_\mathbf{k}^\tau = \operatorname{sgn}(\omega) \sqrt{\frac{\omega^2}{c^2}}
  = \operatorname{sgn}(\omega) \Big|\ \frac{\omega}{c}\ \Big|
  = \frac{\omega}{c}
  \quad
  \Rightarrow
  \quad
  v_\mathrm{g}
  = \frac{\mathrm{d}\omega}{\mathrm{d}K_\mathbf{k}^\tau} = c.
\end{align}
We can find the phase velocity from the exponent that is dependent on the signature of the frequency,
\begin{align}
  \vec{E}{}_{\mathbf{k},z}^{\tau}
  = 
  \sum_{\sigma=\pm}
  \vec{E}{}_{\mathbf{k},0}^{\sigma \tau}
  \exp
  \bigg[i\sigma K_\mathbf{k}^\tau \vec{u}_z \cdot 
    \bigg(
      \vec{r} -
      v_\mathrm{ph}
      \frac{\sigma K_\mathbf{k}^\tau \vec{u}_z}{|\sigma K_\mathbf{k}^\tau|}t
    \bigg)
  \bigg],
  \qquad
  v_\mathrm{ph} = \frac{\omega}{|\sigma K_\mathbf{k}^\tau|} 
  \propto \operatorname{sgn}(\omega).
\end{align}

Since the wavenumber in the $z$ direction satisfies
$
K_{-\mathbf{k}}^{\tau\cc}
= -K_{\mathbf{k}}^\tau
$,
the vectorial coefficients are subject to the following constraint:
\begin{align}
  \vec{E}{}_{\mathbf{k},0}^{\sigma \tau \cc}
  = 
  \vec{E}{}_{-\mathbf{k},0}^{\sigma \tau}.
  \label{eq:constraint_vec_E0}
\end{align}
We have another constraint for the vectorial coefficients due to the transversality condition,
\begin{align}
  i(
  k_x \vec{u}_x 
  + k_y \vec{u}_y 
  + \sigma K_{\mathbf{k}}^\tau \vec{u}_z
  )
  \cdot 
  \vec{E}{}_{\mathbf{k},0}^{\sigma\tau} = 0.
  \label{eq:transversality}
\end{align}
This equation means the vectorial coefficients are orthogonal to the wave vector defined by
\begin{align}
  \vec{k}_{\mathbf{k}}^{\sigma \tau} 
  &=
  k_x \vec{u}_x + k_y \vec{u}_y + \sigma K_{\mathbf{k}}^\tau \vec{u}_z
  \label{eq:wave_vector}
\end{align}
for each mode labeled by a set $\{\mathbf{k}, \sigma, \tau\}$.
We can find two orthonormal polarisation vectors given in terms of the wave vector,
\begin{align}
  \vec{e}_{\lambda,\mathbf{k}}^{\sigma\tau}
  &=
  \begin{cases}{}
    \displaystyle{
      \frac{\operatorname{sgn}(\omega) \vec{k}_\mathbf{k}^{\sigma\tau} \times \vec{u}_z}
      {|\vec{k}_\mathbf{k}^{\sigma\tau} \times \vec{u}_z|}
    }
    &
    (\lambda = s),
    \vspace{.5em}
    \\
    \displaystyle{
      \frac{\operatorname{sgn}(\omega) \vec{k}_\mathbf{k}^{\sigma\tau} \times \vec{e}_{s,\mathbf{k}}^{\sigma\tau}}
      {|\vec{k}_\mathbf{k}^{\sigma\tau} \times \vec{e}_{s,\mathbf{k}}^{\sigma\tau}|}
    }
    &
    (\lambda = p).
  \end{cases}
  \label{eq:e_lambda}
\end{align}
which satisfy the transversality condition \eqref{eq:transversality}.
These vectors, $\vec{e}_s$ and $\vec{e}_p$, correspond to conventional transverse electric (TE) and transverse magnetic (TM) modes and satisfy
\begin{align}
\vec{e}_{\lambda,-\mathbf{k}}^{\sigma\tau\cc}
=
\vec{e}_{\lambda,\mathbf{k}}^{\sigma\tau}
\qquad
(\lambda = s,p).
\label{eq:constraint_e_lambda}
\end{align}
We can write the vectorial coefficient as a linear combination of the polarisation vectors,
\begin{align}
  \vec{E}{}_{\mathbf{k},0}^{\sigma\tau}
  &= 
  \sum_{\lambda=s,p} 
  E_{\lambda,\mathbf{k}}^{\sigma\tau}
  \vec{e}_{\lambda,\mathbf{k}}^{\hspace{.2em}\sigma\tau},
  \label{eq:E_k0}
\end{align}
where the scalar coefficient
$
E_{\lambda,\mathbf{k}}^{\sigma\tau}
$
correspond to the electric modal amplitude in a mode labeled by a set $\{\lambda,\mathbf{k},\sigma,\tau\}$.
From Eqs.~(\ref{eq:constraint_vec_E0}, \ref{eq:constraint_e_lambda}),
we can find that the modal amplitude satisfies the same type of constraints,
\begin{align}
  E_{\lambda,\mathbf{k}}^{\sigma\tau}
  &= 
  E_{\lambda,-\mathbf{k}}^{\sigma\tau\cc}
  \qquad
  (\lambda = s,p).
  \label{eq:constraint_vec_E_lambda}
\end{align}

Finally,
we can write
\begin{align}
  \begin{cases}{}
  \vec{\mathcal{E}}{}_{\mathbf{x},z} 
  = \displaystyle{\int}
  \vec{E}{}_{\mathbf{k},z},
  e^{i\mathbf{k}\cdot \mathbf{x}}
  \mathrm{d}\mathbf{k},
  \\
  \vec{E}{}_{\mathbf{k},z}
  =
  \displaystyle{\sum_{\sigma=\pm}}
  \left[
    \vec{E}{}_{\mathbf{k},0}^{\sigma \ssg}
    \Theta(+z)
    e^{i\sigma K_{\mathbf{k}}^\ssg z}
    +
    \vec{E}{}_{\mathbf{k},0}^{\sigma \ssl}
    \Theta(-z)
    e^{i\sigma K_{\mathbf{k}}^\ssl z}
  \right],
  \\
  \vec{E}{}_{\mathbf{k},0}^{\sigma \tau}
  =
  \displaystyle{\sum_{\lambda=s,p}}
  E_{\lambda,\mathbf{k}}^{\sigma\tau}
  \vec{e}_{\lambda,\mathbf{k}}^{\hspace{.2em}\sigma\tau},
  \end{cases}
  \label{eq:E_expanded}
\end{align}
where $\Theta(z)$ is the Heaviside unit step function.
This completes the eigenmode expansion of the electric field $\vec{\mathcal{E}}_{\mathbf{x},z}$ in each medium.
The similar calculation can be performed for the magnetic field $\vec{\mathcal{H}}_{\mathbf{x},z}$,
and we can write
\begin{align}
  \begin{cases}{}
  \vec{\mathcal{H}}{}_{\mathbf{x},z} 
  = \displaystyle{\int}
  \vec{H}{}_{\mathbf{k},z},
  e^{i\mathbf{k}\cdot \mathbf{x}}
  \mathrm{d}\mathbf{k},
  \\
  \vec{H}{}_{\mathbf{k},z}
  =
  \displaystyle{\sum_{\sigma=\pm}}
  \left[
    \vec{H}{}_{\mathbf{k},0}^{\sigma \ssg}
    \Theta(+z)
    e^{i\sigma K_{\mathbf{k}}^\ssg z}
    +
    \vec{H}{}_{\mathbf{k},0}^{\sigma \ssl}
    \Theta(-z)
    e^{i\sigma K_{\mathbf{k}}^\ssl z}
  \right],
  \\
  \vec{H}{}_{\mathbf{k},0}^{\sigma \tau}
  =
  \displaystyle{\sum_{\lambda=s,p}}
  H_{\lambda,\mathbf{k}}^{\sigma\tau}
  \vec{h}_{\lambda,\mathbf{k}}^{\sigma\tau},
  \end{cases}
  \label{eq:H_expanded}
\end{align}
where we have introduced another basis,
\begin{align}
  \begin{pmatrix}
    \vec{h}_{s,\mathbf{k}}^{\sigma\tau}\\
    \vec{h}_{p,\mathbf{k}}^{\sigma\tau}
  \end{pmatrix}
  &=
  \begin{pmatrix}
    0 & 1\\
    -1 & 0
  \end{pmatrix}
  \begin{pmatrix}
    \vec{e}_{s,\mathbf{k}}^{\hspace{.2em}\sigma\tau}\\
    \vec{e}_{p,\mathbf{k}}^{\hspace{.2em}\sigma\tau}
  \end{pmatrix}.
  \label{eq:e_h}
\end{align}
Note that we have the magnetic counterpart of the constraint \eqref{eq:constraint_vec_E_lambda},
\begin{align}
  H_{\lambda,\mathbf{k}}^{\sigma\tau}
  &= 
  H_{\lambda,-\mathbf{k}}^{\sigma\tau\cc}
  \qquad
  (\lambda = s,p).
  \label{eq:constraint_vec_H_lambda}
\end{align}
Note also that the magnetic field amplitude can be associated with the electric one via the characteristic impedance,
\begin{align}
  E_{\lambda,\mathbf{k}}^{\sigma \tau}
  =
  Z_{\lambda,\mathbf{k}}^\tau
  H_{\lambda,\mathbf{k}}^{\sigma \tau},
  \quad
  Z_{\lambda, \mathbf{k}}^\tau
  &=
  Z_0
  \times
  \begin{cases}{}
    \displaystyle{
      \frac{1}{\kappa^\tau}
      \sqrt{\frac{{k_0}^2}{|K_{\mathbf{k}}^\tau|^2 + {k_\parallel}^2}}
    }
    &
    (\lambda = s),
    \vspace{.5em}
    \\
    \displaystyle{
      \frac{1}{\epsilon^\tau}
      \sqrt{\frac{|K_{\mathbf{k}}^\tau|^2 + {k_\parallel}^2}{{k_0}^2}}
    }
    &
    (\lambda = p),
  \end{cases}
\end{align}
where we should be careful that the impedance depends on the polarisation state as well as the parallel wavenumber and the frequency.
Note that we substitute $\kappa^\tau = (\mu^\tau)^{-1} = 1$ in the current case because we are focusing on the dielectric response.

\subsection{Reflection and transmission at a single interface}
\label{subsec:single}
Here, we derive the boundary conditions that give the outgoing modal amplitudes in terms of the incoming ones.
\begin{figure}[htbp]
  \centering
  \includegraphics[width=.5\linewidth]
  {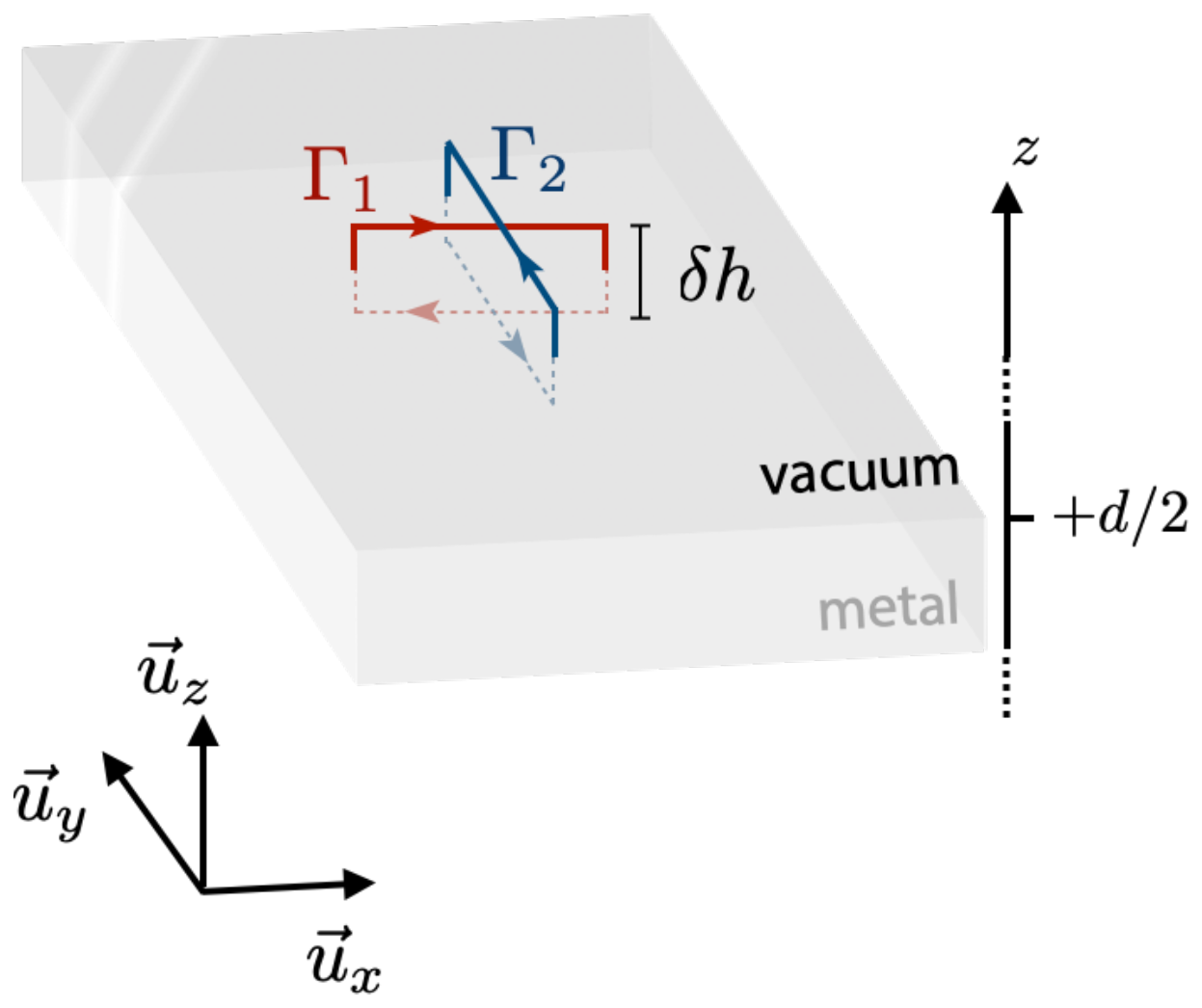}
  \caption{
    Two integration paths $\Gamma_{1,2}$ enclosing the interface,
    which provide Maxwell's boundary conditions.
  }
  \label{fig:metal_single}
\end{figure}
Our configuration is composed of two media (\figref{fig:flat}),
Each of them fills the space above and below the boundary at $z=+d/2$ and is characterised by
\begin{align}
  \epsilon 
  =
  \begin{cases}{}
    \epsv & (z > +d/2),
    \\
    \epsm & (z < +d/2).
  \end{cases}
\end{align}
Note that we set the position of the interface at $z=+d/2$ for later convenience.
By integrating the Maxwell's equations on paths $\Gamma_{1,2}$ enclosing the metal interface at $z=+d/2$ (\figref{fig:metal_single}),
\begin{align}
  \begin{cases}{}
    \displaystyle{
      \lim_{\delta h \rightarrow 0}
      \oint_{\Gamma_{1,2}}
    }
    \vec{\mathcal{E}}_{\mathbf{x},z}
    \cdot 
    \mathrm{d}\vec{r}
    & =
    -\displaystyle{
      \lim_{\delta h \rightarrow 0}
      \oiint_{\Gamma_{1,2}}
    }
      \frac{\partial}{\partial t}
      \mu_0 
      \vec{\mathcal{H}}_{\mathbf{x},z}
    \cdot 
    \mathrm{d}\vec{S},
    \vspace{.5em}
    \\
    \displaystyle{
      \lim_{\delta h \rightarrow 0}
      \oint_{\Gamma_{1,2}}
    }
    Z_0 \vec{\mathcal{H}}_{\mathbf{x},z}
    \cdot 
    \mathrm{d}\vec{r}
    &=
    +\displaystyle{
      \lim_{\delta h \rightarrow 0}
      \oiint_{\Gamma_{1,2}}
    }
    \frac{1}{c}
    \frac{\partial}{\partial t}
    \varepsilon
    \circledast
    \vec{\mathcal{E}}_{\mathbf{x},z}
    \cdot 
    \mathrm{d}\vec{S},
  \end{cases}
  \label{eq:boundary_condition_flat_int}
\end{align}
where the second equations is multiplied by the characteristic impedance of free space so that the two equations are written in the same unit.
We can find the tangential components of electric and magnetic fields are continuous at the interface, 
\begin{align}
  \begin{cases}{}
    \vec{u}_x \cdot
    \left(
      \vec{\mathcal{E}}_{\mathbf{x},+\frac{d}{2}+0} 
      - \vec{\mathcal{E}}_{\mathbf{x},+\frac{d}{2}-0}
    \right)
    = 0,
    &
    \vec{u}_y \cdot
    \left(
      \vec{\mathcal{E}}_{\mathbf{x},+\frac{d}{2}+0} 
      - \vec{\mathcal{E}}_{\mathbf{x},+\frac{d}{2}-0}
    \right)
    = 0,
    \\
    \vec{u}_x \cdot
    Z_0
    \left(
      \vec{\mathcal{H}}_{\mathbf{x},+\frac{d}{2}+0} 
      - \vec{\mathcal{H}}_{\mathbf{x},+\frac{d}{2}-0}
    \right)
    = 0,
    &
    \vec{u}_y \cdot
    Z_0
    \left(
      \vec{\mathcal{H}}_{\mathbf{x},+\frac{d}{2}+0} 
      - \vec{\mathcal{H}}_{\mathbf{x},+\frac{d}{2}-0}
    \right)
    = 0.
  \end{cases}
  \label{eq:boundary_condition_flat}
\end{align}
Here, we have denoted the electric and magnetic fields just above and below the interface as 
\begin{align}
  \vec{\mathcal{E}}_{\mathbf{x},+\frac{d}{2}\pm 0}
  &\equiv
  \displaystyle{\lim_{\delta h \rightarrow 0}} 
  \vec{\mathcal{E}}_{\mathbf{x},+\frac{d}{2}\pm \delta h},
  \\
  \vec{\mathcal{H}}_{\mathbf{x},+\frac{d}{2}\pm 0} 
  &\equiv
  \displaystyle{\lim_{\delta h \rightarrow 0}} 
  \vec{\mathcal{H}}_{\mathbf{x},+\frac{d}{2}\pm \delta h}. 
\end{align}
Note that the right-hand side of Eq.~\eqref{eq:boundary_condition_flat_int} goes to zero when we take the limit if we consider static media and the permittivity does not suddenly change
(i.e.~$\dot{\varepsilon} < \infty$),
\begin{align}
  \frac{\partial}{\partial t}
  \varepsilon \circledast \vec{\mathcal{E}}
  = \int_{-\infty}^t
  \dot{\varepsilon}_{t-t'}
  \vec{\mathcal{E}}_{t'}
  \mathrm{d}t'
\end{align}
As we will see in Chapter \ref{ch:df}, we should update Eq.~\eqref{eq:boundary_condition_flat} in time-dependent systems where the time derivative returns the Dirac delta function.

We substitute the Fourier representation \eqref{eq:FT[E],FT[H]} into Eq.~\eqref{eq:boundary_condition_flat},
\begin{align}
  \begin{cases}{}
    \vec{u}_x \cdot
    \left(
      \vec{E}_{\mathbf{k},+\frac{d}{2}-0} - \vec{E}_{\mathbf{k},+\frac{d}{2}+0}
    \right)
    = 0,
    &
    \vec{u}_y \cdot
    \left(
      \vec{E}_{\mathbf{k},+\frac{d}{2}-0} - \vec{E}_{\mathbf{k},+\frac{d}{2}+0}
    \right)
    = 0,
    \\
    \vec{u}_x \cdot
    Z_0
    \left(
      \vec{H}_{\mathbf{k},+\frac{d}{2}-0} - \vec{H}_{\mathbf{k},+\frac{d}{2}+0}
    \right)
    = 0,
    &
    \vec{u}_y \cdot
    Z_0
    \left(
      \vec{H}_{\mathbf{k},+\frac{d}{2}-0} - \vec{H}_{\mathbf{k},+\frac{d}{2}+0}
    \right)
    = 0.
  \end{cases}
\end{align}
and use Eqs.~(\ref{eq:E_expanded}, \ref{eq:H_expanded}) to obtain
\begin{align}
  \begin{cases}{}
    \displaystyle{\sum_{\lambda,\sigma}} 
    \vec{u}_x \cdot \vec{e}_{\lambda,\mathbf{k}}^{\sigma\mathrm{v}}
    e^{i\phi_{\mathbf{k}}^{\sigma \mathrm{v}}}
    Z_{\lambda,\mathbf{k}}^\mathrm{v}
    H_{\lambda,\mathbf{k}}^{\sigma \mathrm{v}}
    =
    \displaystyle{\sum_{\lambda,\sigma}} 
    \vec{u}_x \cdot \vec{e}_{\lambda,\mathbf{k}}^{\sigma\mathrm{m}}
    e^{i\phi_{\mathbf{k}}^{\sigma \mathrm{m}}}
    Z_{\lambda,\mathbf{k}}^\mathrm{m}
    H_{\lambda,\mathbf{k}}^{\sigma \mathrm{m}}
    \\
    \displaystyle{\sum_{\lambda,\sigma}} 
    \vec{u}_y \cdot \vec{e}_{\lambda,\mathbf{k}}^{\sigma\mathrm{v}}
    e^{i\phi_{\mathbf{k}}^{\sigma \mathrm{v}}}
    Z_{\lambda,\mathbf{k}}^\mathrm{v}
    H_{\lambda,\mathbf{k}}^{\sigma \mathrm{v}}
    =
    \displaystyle{\sum_{\lambda,\sigma}} 
    \vec{u}_y \cdot \vec{e}_{\lambda,\mathbf{k}}^{\sigma\mathrm{m}}
    e^{i\phi_{\mathbf{k}}^{\sigma \mathrm{m}}}
    Z_{\lambda,\mathbf{k}}^\mathrm{m}
    H_{\lambda,\mathbf{k}}^{\sigma \mathrm{m}}
    \\
    \displaystyle{\sum_{\lambda,\sigma}} 
    \vec{u}_x \cdot \vec{h}_{\lambda,\mathbf{k}}^{\sigma\mathrm{v}}
    e^{i\phi_{\mathbf{k}}^{\sigma \mathrm{v}}}
    Z_0
    H_{\lambda,\mathbf{k}}^{\sigma \mathrm{v}}
    =
    \displaystyle{\sum_{\lambda,\sigma}} 
    \vec{u}_x \cdot \vec{h}_{\lambda,\mathbf{k}}^{\sigma\mathrm{m}}
    e^{i\phi_{\mathbf{k}}^{\sigma\mathrm{m}}}
    Z_0
    H_{\lambda,\mathbf{k}}^{\sigma \mathrm{m}}
    \\
    \displaystyle{\sum_{\lambda,\sigma}} 
    \vec{u}_y \cdot \vec{h}_{\lambda,\mathbf{k}}^{\sigma\mathrm{v}}
    e^{i\phi_{\mathbf{k}}^{\sigma\mathrm{v}}}
    Z_0
    H_{\lambda,\mathbf{k}}^{\sigma \mathrm{v}}
    =
    \displaystyle{\sum_{\lambda,\sigma}} 
    \vec{u}_y \cdot \vec{h}_{\lambda,\mathbf{k}}^{\sigma\mathrm{m}}
    e^{i\phi_{\mathbf{k}}^{\sigma\mathrm{m}}}
    Z_0
    H_{\lambda,\mathbf{k}}^{\sigma \mathrm{m}}
  \end{cases}
  \label{eq:trans_mat_eq_pre}
\end{align}
where we set the propagating phase factor 
$\phi_{\mathbf{k}}^{\sigma\tau} = \sigma K_\mathbf{k}^\tau d/2$.
Note that modes which are different in the reciprocal vector $\mathbf{k}$ do not talk to each other and thus we can independently perform the calculation for each $\mathbf{k}$.

\begin{figure}[htbp]
  \centering
  \includegraphics[width=\linewidth]
  {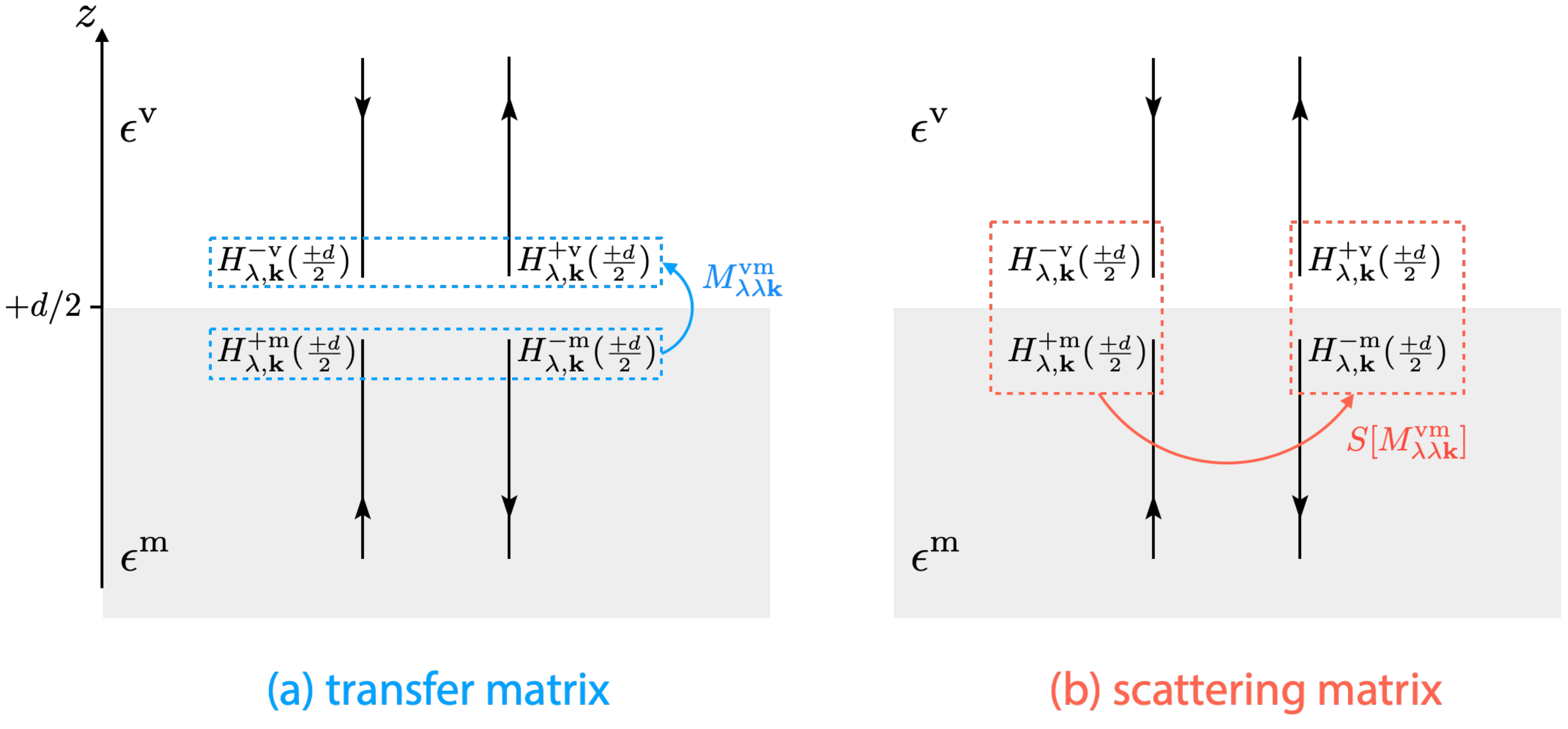}
  \caption{
    Transfer matrix and scattering matrix.
    The transfer matrix associates field amplitudes at one point with the ones at another point, while the scattering matrix connects the incoming field amplitudes and the outgoing field amplitudes.
    The scattering matrix can be computed from the transfer matrix \eqref{eq:sca_trans}.
  }
  \label{fig:sca_trans}
\end{figure}

We arrange Eq.~\eqref{eq:trans_mat_eq_pre} in a matrix-vector form,
\begin{align}
  \begin{pmatrix}
    M_{pp\mathbf{k}}^\mathrm{v} & M_{ps\mathbf{k}}^\mathrm{v} \\
    M_{sp\mathbf{k}}^\mathrm{v} & M_{ss\mathbf{k}}^\mathrm{v}
  \end{pmatrix}
  \begin{pmatrix}
    C_\mathbf{k}^\mathrm{v} & 0\\
    0 & C_\mathbf{k}^\mathrm{v}
  \end{pmatrix}
  \begin{pmatrix}
    H_{p,\mathbf{k}}^{\ssp \mathrm{v}}\\
    H_{p,\mathbf{k}}^{\ssm \mathrm{v}}\\
    H_{s,\mathbf{k}}^{\ssp \mathrm{v}}\\
    H_{s,\mathbf{k}}^{\ssm \mathrm{v}}
  \end{pmatrix}
  = 
  \begin{pmatrix}
    M_{pp\mathbf{k}}^\mathrm{m} & M_{ps\mathbf{k}}^\mathrm{m} \\
    M_{sp\mathbf{k}}^\mathrm{m} & M_{ss\mathbf{k}}^\mathrm{m}
  \end{pmatrix}
  \begin{pmatrix}
    C_\mathbf{k}^\mathrm{m} & 0\\
    0 & C_\mathbf{k}^\mathrm{m}
  \end{pmatrix}
  \begin{pmatrix}
    H_{p,\mathbf{k}}^{\ssp \mathrm{m}}\\
    H_{p,\mathbf{k}}^{\ssm \mathrm{m}}\\
    H_{s,\mathbf{k}}^{\ssp \mathrm{m}}\\
    H_{s,\mathbf{k}}^{\ssm \mathrm{m}}
  \end{pmatrix},
  \label{eq:trans_mat_eq_pre_matrix}
\end{align}
where the matching and propagating phase matrices read
\begin{align}
  M_{pp\mathbf{k}}^\tau 
  &=
  \begin{pmatrix}
  \vec{u}_x \cdot \vec{e}_{p,\mathbf{k}}^{\ssp\tau}
  Z_{p,\mathbf{k}}^\tau
  &
  \vec{u}_x \cdot \vec{e}_{p,\mathbf{k}}^{\ssm\tau}
  Z_{p,\mathbf{k}}^\tau
  \\
  \vec{u}_y \cdot \vec{h}_{p,\mathbf{k}}^{\ssp\tau} 
  Z_0
  &
  \vec{u}_y \cdot \vec{h}_{p,\mathbf{k}}^{\ssm\tau}
  Z_0
  \end{pmatrix},
  \quad
  M_{ps\mathbf{k}}^\tau
  =
  \begin{pmatrix}
  \vec{u}_x \cdot \vec{e}_{s,\mathbf{k}}^{\ssp\tau}
  Z_{s,\mathbf{k}}^\tau
  &
  \vec{u}_x \cdot \vec{e}_{s,\mathbf{k}}^{\ssm\tau}
  Z_{s,\mathbf{k}}^\tau
  \\
  \vec{u}_y \cdot \vec{h}_{s,\mathbf{k}}^{\ssp\tau} 
  Z_0
  &
  \vec{u}_y \cdot \vec{h}_{s,\mathbf{k}}^{\ssm\tau}
  Z_0
  \end{pmatrix},
  \\
  M_{sp\mathbf{k}}^\tau
  &=
  \begin{pmatrix}
  \vec{u}_x \cdot \vec{h}_{p,\mathbf{k}}^{\ssp\tau}
  Z_0
  &
  \vec{u}_x \cdot \vec{h}_{p,\mathbf{k}}^{\ssm\tau}
  Z_0
  \\
  \vec{u}_y \cdot \vec{e}_{p,\mathbf{k}}^{\ssp\tau}
  Z_{p,\mathbf{k}}^\tau
  &
  \vec{u}_y \cdot \vec{e}_{p,\mathbf{k}}^{\ssm\tau}
  Z_{p,\mathbf{k}}^\tau
  \end{pmatrix},
  \quad
  M_{ss\mathbf{k}}^\tau
  =
  \begin{pmatrix}
  \vec{u}_x \cdot \vec{h}_{s,\mathbf{k}}^{\ssp\tau}
  Z_0
  &
  \vec{u}_x \cdot \vec{h}_{s,\mathbf{k}}^{\ssm\tau}
  Z_0
  \\
  \vec{u}_y \cdot \vec{e}_{s,\mathbf{k}}^{\ssp\tau}
  Z_{s,\mathbf{k}}^\tau
  &
  \vec{u}_y \cdot \vec{e}_{s,\mathbf{k}}^{\ssm\tau}
  Z_{s,\mathbf{k}}^\tau
  \end{pmatrix},
  \\
  C_\mathbf{k}^\tau
  &= 
  \begin{pmatrix}
    e^{i\phi_{\mathbf{k}}^{\ssp\tau}} & 0 \\
  0 & e^{i\phi_{\mathbf{k}}^{\ssm\tau}} 
  \end{pmatrix}.
  \label{eq:C}
\end{align}
The translational invariance in the $x$ and $y$ directions allows us to focus on the in-plane situation ($k_y = 0$) without loss of generality.
In the in-plane case,
we have
\begin{align}
  \vec{u}_x \cdot \vec{e}_{s,\mathbf{k}}^{\sigma\tau} 
  = 0,
  \quad
  \vec{u}_y \cdot \vec{e}_{p,\mathbf{k}}^{\sigma\tau}
  = 0,
  \\
  \vec{u}_x \cdot \vec{h}_{p,\mathbf{k}}^{\sigma\tau} 
  = 0,
  \quad
  \vec{u}_y \cdot \vec{h}_{s,\mathbf{k}}^{\sigma\tau}
  = 0,
\end{align}
and the off-diagonal elements of the matching matrix vanishes,
\begin{align}
  M_{ps\mathbf{k}}^\tau = M_{sp\mathbf{k}}^\tau = 0.
\end{align}
This means that TE and TM modes are decoupled.
We can write
\begin{align}
  \begin{pmatrix}
    H_{p,\mathbf{k}}^{\ssp\mathrm{v}}(\frac{+d}{2})\\
    H_{p,\mathbf{k}}^{\ssm\mathrm{v}}(\frac{+d}{2})
  \end{pmatrix}
  &= 
  M_{pp\mathbf{k}}^{\mathrm{v}^{-1}}
  M_{pp\mathbf{k}}^\mathrm{m} 
  \begin{pmatrix}
    H_{p,\mathbf{k}}^{\ssp\mathrm{m}}(\frac{+d}{2})\\
    H_{p,\mathbf{k}}^{\ssm\mathrm{m}}(\frac{+d}{2})
  \end{pmatrix}
  \equiv M_{pp\mathbf{k}}^{\mathrm{vm}}
  \begin{pmatrix}
    H_{p,\mathbf{k}}^{\ssp\mathrm{m}}(\frac{+d}{2})\\
    H_{p,\mathbf{k}}^{\ssm\mathrm{m}}(\frac{+d}{2})
  \end{pmatrix}
  \label{eq:transfer_upper_p}
  \\
  \begin{pmatrix}
    H_{s,\mathbf{k}}^{\ssp\mathrm{v}}(\frac{+d}{2})\\
    H_{s,\mathbf{k}}^{\ssm\mathrm{v}}(\frac{+d}{2})
  \end{pmatrix}
  &= 
  M_{ss\mathbf{k}}^{\mathrm{v}^{-1}}
  M_{ss\mathbf{k}}^\mathrm{m} 
  \begin{pmatrix}
    H_{s,\mathbf{k}}^{\ssp\mathrm{m}}({\frac{+d}{2}})\\
    H_{s,\mathbf{k}}^{\ssm\mathrm{m}}({\frac{+d}{2}})
  \end{pmatrix}
  \equiv 
  M_{ss\mathbf{k}}^\mathrm{vm} 
  \begin{pmatrix}
    H_{s,\mathbf{k}}^{\ssp\mathrm{m}}({\frac{+d}{2}})\\
    H_{s,\mathbf{k}}^{\ssm\mathrm{m}}({\frac{+d}{2}})
  \end{pmatrix},
  \label{eq:transfer_upper_s}
\end{align}
where the amplitude of the field that is labeled by $(\lambda,\mathbf{k},\sigma,\tau)$ and evaluated at $z$ is defined by
$
H_{\lambda,\mathbf{k}}^{\sigma \tau}(z)
\equiv
H_{\lambda,\mathbf{k}}^{\sigma \tau}
e^{i\sigma K_{\mathbf{k}}^{\tau}z}
$, 
and we introduce a transfer matrix for the single interface by
$
M_{\lambda\lambda\mathbf{k}}^\mathrm{\tau_1\tau_2} 
\equiv 
M_{\lambda\lambda\mathbf{k}}^{\tau_1^{\hspace{.1em}-1}}
M_{\lambda\lambda\mathbf{k}}^{\tau_2}
$,
which associates the modal amplitudes at one point with another point [see \figref{fig:sca_trans}(a)].
In the current case,
Eqs.~(\ref{eq:transfer_upper_p}, \ref{eq:transfer_upper_s}) associate the field amplitudes in the lower medium and those in the upper medium.

We can rearrange Eqs.~(\ref{eq:transfer_upper_p}, \ref{eq:transfer_upper_s}) into another form,
\begin{align}
  \begin{pmatrix}
    H_{\lambda,\mathbf{k}}^{\ssp\mathrm{v}}({\frac{+d}{2}})\\
    H_{\lambda,\mathbf{k}}^{\ssm\mathrm{m}}({\frac{+d}{2}})
  \end{pmatrix}
  &=
  S[M_{\lambda\lambda\mathbf{k}}^\mathrm{vm}]
  \begin{pmatrix}
    H_{\lambda,\mathbf{k}}^{\ssm\mathrm{v}}({\frac{+d}{2}})\\
    H_{\lambda,\mathbf{k}}^{\ssp\mathrm{m}}({\frac{+d}{2}})
  \end{pmatrix}
  \qquad
  (\lambda = p,s),
  \label{eq:scattering_upper}
\end{align}
where $S[\bullet]$ corresponds to the scattering matrix, 
whose elements are calculated as following \cite{markos2008wave}:
\begin{align}
  S
  \left[
    \begin{pmatrix}
      M_{\ssp\ssp} & M_{\ssp\ssm}\\
      M_{\ssm\ssp} & M_{\ssm\ssm}
    \end{pmatrix}
  \right]
  =
  \frac{1}{M_{\ssm\ssm}}
  \begin{pmatrix}
    M_{\ssp\ssm} & M_{\ssp\ssp} M_{\ssm\ssm} - M_{\ssp\ssm} M_{\ssm\ssp}\\
    1 & -M_{\ssm\ssp}
  \end{pmatrix}.
  \label{eq:sca_trans}
\end{align}
Note that the left-hand side of Eq.~\eqref{eq:scattering_upper} contains the outgoing field amplitudes from the interface, while the right-hand side contains the incoming field amplitudes.
Therefore, the scattering matrix $S[\bullet]$ connects the incoming and outgoing amplitudes [see \figref{fig:sca_trans}(b)].
In other words, the matrix elements correspond to the reflection and transmission coefficients of the interface.

From the scattering matrix,
we can find not only the reflection and transmission coefficients but also the localised modes at the interface.
Such modes are provided by non-trivial solutions to Eq.~(\ref{eq:scattering_upper}) with no incoming waves
[i.e.~$
H_{\lambda,\mathbf{k}}^{\ssm\mathrm{v}}({\frac{+d}{2}}) 
= H_{\lambda,\mathbf{k}}^{\ssp\mathrm{m}}({\frac{+d}{2}}) 
= 0
$].
The necessary and sufficient condition for the existence of the solutions is given by
\begin{align}
  \frac{1}{\det S[M_{\lambda\lambda\mathbf{k}}^\mathrm{vm}]} = 0,
\end{align}
which leads to the corresponding dispersion relations of the localised modes.

For the TM mode,
we can explicitly write the matching matrices,
\begin{align}
  M_{pp\mathbf{k}}^\tau 
  &= 
  Z_0
  \operatorname{sgn} (\omega)
  \operatorname{sgn} (k_x)
  \begin{pmatrix}
    \frac{+K_{\mathbf{k}}^\tau}{\epsilon^\tau k_0}
    &
    \frac{-K_{\mathbf{k}}^\tau}{\epsilon^\tau k_0}
    \\
    +1 & +1
  \end{pmatrix},
  \quad
  M_{ss\mathbf{k}}^\tau
  =
  Z_0 \operatorname{sgn} (k_x)
  \begin{pmatrix}
  \frac{+K_{\mathbf{k}}^\tau}{\sqrt{|K_{\mathbf{k}}^\tau|^2 + {k_\parallel}^2}}
  &
  \frac{-K_{\mathbf{k}}^\tau}{\sqrt{|K_{\mathbf{k}}^\tau|^2 + {k_\parallel}^2}}
  \\
  \frac{k_0}{\sqrt{|K_{\mathbf{k}}^\tau|^2 + {k_\parallel}^2}}
  &
  \frac{k_0}{\sqrt{|K_{\mathbf{k}}^\tau|^2 + {k_\parallel}^2}}
  \end{pmatrix}.
  \label{eq:M_k}
\end{align}
and compute the determinant,
\begin{align}
  \frac{1}{\det S[M_{pp\mathbf{k}}^\mathrm{vm}]} = 0
  \quad
  \Leftrightarrow
  \quad 
  \xi_{\mathbf{k}}^{\mathrm{mv}}
  \equiv
  -\frac{K_{\mathbf{k}}^\mathrm{m}/\epsm}{K_{\mathbf{k}}^\mathrm{v}/\epsv} 
  = 1. 
  \label{eq:xi_single}
\end{align}
This equation is the dispersion relation for the localised TM mode at the interface known as the surface plasmon mode.

On the other hand,
for the TE mode,
the determinant gives
\begin{align}
  \frac{1}{\det S[M_{ss\mathbf{k}}^\mathrm{vm}]} = 0
  \quad
  \Leftrightarrow
  \quad 
  K_{\mathbf{k}}^\mathrm{v} + K_{\mathbf{k}}^\mathrm{m} = 0.
  \label{eq:no_SPP_by_s}
\end{align}
However, we have $\operatorname{Im} (K_{\mathbf{k}}^\mathrm{v} + K_{\mathbf{k}}^\mathrm{m}) > 0$ by the definition of the wavenumber in the $z$ direction \eqref{eq:K},
and Eq.~\eqref{eq:no_SPP_by_s} does not hold.
This is why there is no localised TE mode at the vacuum-metal interface.

\begin{figure}[htbp]
  \centering
  \includegraphics[width=.6\linewidth]
  {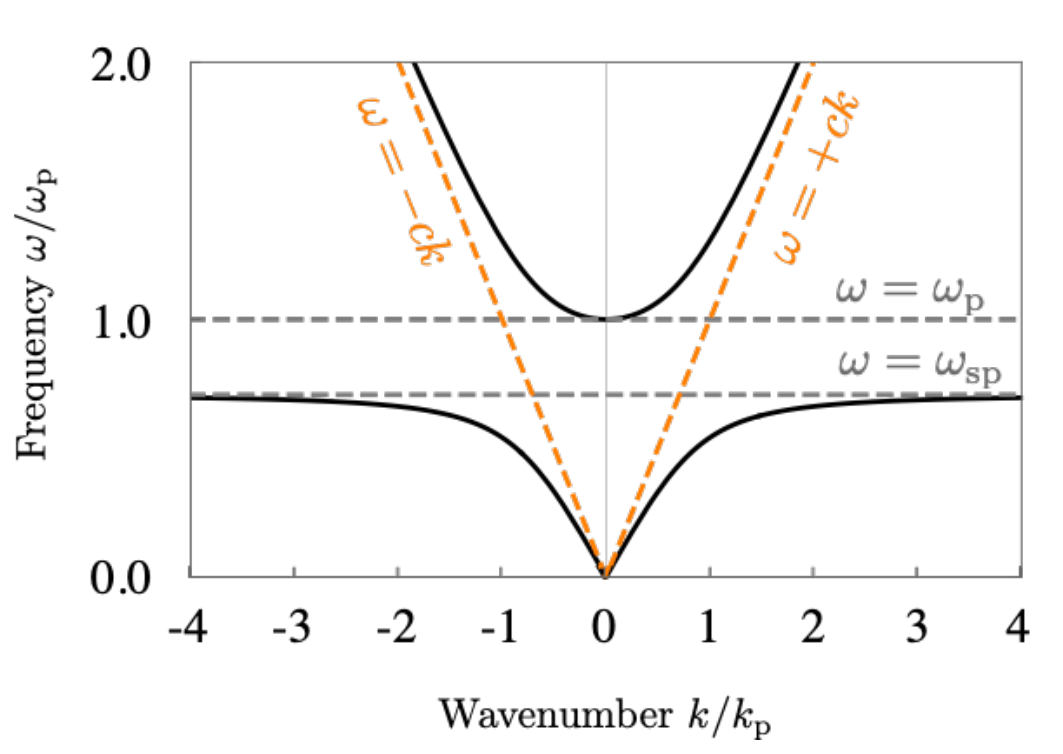}
  \caption{
    Dispersion relations for the bulk mode inside the metal (upper black line) and the surface plasmon mode (lower black line).
    We use the gold plasma frequency $\omega_\mathrm{p} = 2\pi \times 2.068 \times 10^{15}\ [\mathrm{rad \cdot s^{-1}}] \approx 8.55\ [\mathrm{eV}]$ and the vacuum permittivity $\epsv = 1$.
    We have defined $k_p \equiv \omega_\mathrm{p}/c \approx 2\pi \times 6.898\ [\mathrm{rad\cdot\mu m^{-1}}]$.
  }
  \label{fig:dispersion_single}
\end{figure}

In the case of a single interface,
we can solve the implicit dispersion relation \eqref{eq:xi_single} for $\omega$ and obtain the explicit form,
\begin{align}
  \omega 
  = \sqrt{
    \omega_\mathrm{sp}^2 + c^2 k_\parallel^2 
    \pm \sqrt{\omega_\mathrm{sp}^4 + c^4 k_\parallel^4}
  },
  \label{eq:single_omega}
\end{align}
where $\omega_\mathrm{sp} \equiv \omega_\mathrm{p}/\sqrt{2}$ is the frequency of surface plasma oscillation.
In \figref{fig:dispersion_single},
the dispersion relation \eqref{eq:single_omega} is plotted.
The gray dashed lines correspond to the frequency of bulk plasma oscillation $\omega = \omega_\mathrm{p}$ and that of surface plasma oscillation $\omega = \omega_\mathrm{sp}$.
The surface plasma frequencies can be obtained by taking the non-relativistic (quasistatic) limit ($c \rightarrow \infty$) on the lower branch,
\begin{align}
  \omega_\mathrm{sg}
  = \sqrt{
    \omega_\mathrm{sp}^2 + c^2 k_\parallel^2 
    - \sqrt{\omega_\mathrm{sp}^4 + c^4 k_\parallel^4}
  }
  \approx \sqrt{
    \omega_\mathrm{sp}^2 + c^2 k_\parallel^2 - c^2 k_\parallel^2
  } = \omega_\mathrm{sp}.
  \label{eq:omega_sp}
\end{align}

The positive sign inside the square root in the dispersion relation \eqref{eq:single_omega} represents a bulk mode that propagates inside the metal at a high frequency where the metal behaves as a dielectric medium ($\omega > \omega_\mathrm{p}$). 
The orange line is the light line $\omega = c|k|$.
In contrast, the negative sign corresponds to the surface plasmon mode ($\omega < \omega_\mathrm{sp}$).
The dispersion relation of the surface plasmon mode is below the light line and the surface plasma frequency.
This is why the surface plasmon mode cannot escape from the interface to either the vacuum or metal side.

\subsection{Double interfaces (thin film)}
\begin{figure}[htbp]
  \centering
  \includegraphics[width=.5\linewidth]
  {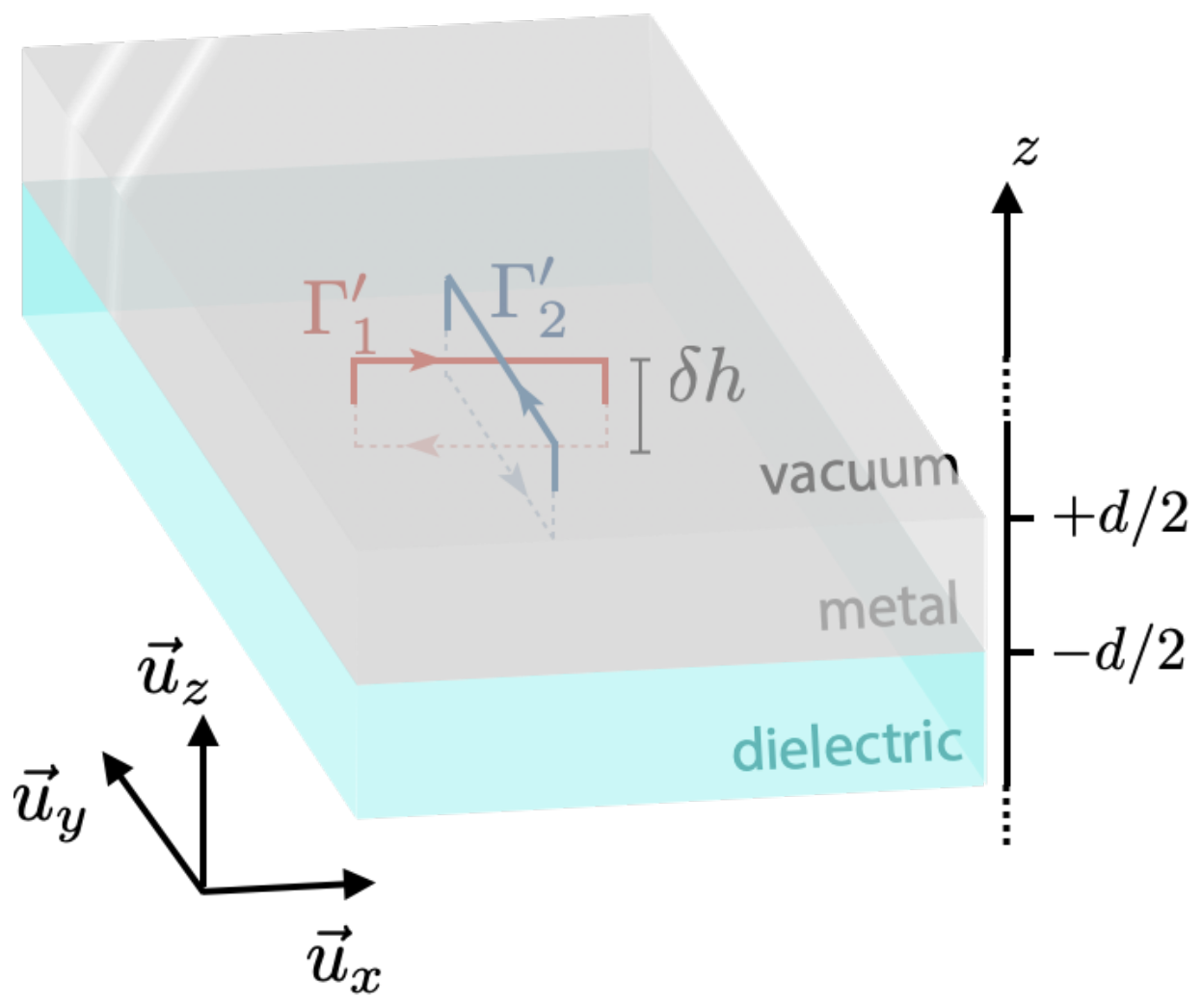}
  \caption{
    Two integration paths $\Gamma_{1,2}'$ enclosing the lower interface of the film,
    which gives Maxwell's boundary conditions at $z=-d/2$.
  }
  \label{fig:metal_film}
\end{figure}

Let us consider what happens if we introduce another interface.
The configuration is shown in \figref{fig:metal_film}.
The upper semi-infinite space is filled with a dielectric medium, and the lower semi-infinite space is another dielectric.
The sandwiched region is filled with a metallic medium.
The permittivity is given by
\begin{align}
  \epsilon
  &=
  \begin{cases}{}
    \epsv
    & (z > +d/2),
    \\
    \epsm
    & (|z| < d/2),
    \\
    \epsi
    & (z < -d/2).
  \end{cases}
\end{align}

Here, we have another interface at $z=-d/2$ in addition to the one at $z=+d/2$ that we calculated in the previous section.
By following the same procedure as we did for the upper interface,
we can derive the boundary conditions for the lower interface.
The integral along the path shown in \figref{fig:metal_film} gives
\begin{align}
  \begin{cases}{}
  \vec{u}_x \cdot
  \left(
    \vec{E}_{\mathbf{k},-\frac{d}{2}-0} - \vec{E}_{\mathbf{k},-\frac{d}{2}+0}
  \right)
  = 0,
  &
  \vec{u}_y \cdot
  \left(
    \vec{E}_{\mathbf{k},-\frac{d}{2}-0} - \vec{E}_{\mathbf{k},-\frac{d}{2}+0}
  \right)
  = 0,
  \\
  \vec{u}_x \cdot
    Z_0
  \left(
    \vec{H}_{\mathbf{k},-\frac{d}{2}-0} - \vec{H}_{\mathbf{k},-\frac{d}{2}+0}
  \right)
  = 0,
  &
  \vec{u}_y \cdot
    Z_0
  \left(
    \vec{H}_{\mathbf{k},-\frac{d}{2}-0} - \vec{H}_{\mathbf{k},-\frac{d}{2}+0}
  \right)
  = 0.
  \end{cases}
  \label{eq:metal-dielectric}
\end{align}

Let us focus on the TM mode.
Substituting Eqs.~(\ref{eq:E_expanded}, \ref{eq:H_expanded}) into Eq.~\eqref{eq:metal-dielectric} and arranging in the matrix-vector form,
we can obtain
\begin{align}
  M_{pp\mathbf{k}}^\mathrm{m} C_\mathbf{k}^{\mathrm{m}^{-1}}
  \begin{pmatrix}
    H_{p,\mathbf{k}}^{\ssp\mathrm{m}}\\
    H_{p,\mathbf{k}}^{\ssm\mathrm{m}}
  \end{pmatrix}
  &= 
  M_{pp\mathbf{k}}^\mathrm{i} C_\mathbf{k}^{\mathrm{i}^{-1}}
  \begin{pmatrix}
    H_{p,\mathbf{k}}^{\ssp\mathrm{i}}\\
    H_{p,\mathbf{k}}^{\ssm\mathrm{i}}
  \end{pmatrix}.
  \label{eq:transfer_lower_p}
\end{align}
This equation is responsible for the lower interface.

Since we have two interfaces in the current configuration,
we should simultaneously solve two matrix equations responsible (\ref{eq:transfer_lower_p}, \ref{eq:transfer_upper_p}).
Eliminating $H_{p,\mathbf{k}}^{\pm \mathrm{m}}$ from Eqs.~(\ref{eq:transfer_upper_p}, \ref{eq:transfer_lower_p}), we can obtain the transfer matrix equation,
\begin{align}
  C_\mathbf{k}^{\mathrm{m}^{-1}}
  M_{pp\mathbf{k}}^{\mathrm{m}^{-1}}
  M_{pp\mathbf{k}}^\mathrm{v} 
  C_\mathbf{k}^{\mathrm{v}}
  \begin{pmatrix}
    H_{p,\mathbf{k}}^{\ssp\mathrm{v}}\\
    H_{p,\mathbf{k}}^{\ssm\mathrm{v}}
  \end{pmatrix}
  &= 
  C_\mathbf{k}^{\mathrm{m}}
  M_{pp\mathbf{k}}^{\mathrm{m}^{-1}}
  M_{pp\mathbf{k}}^\mathrm{i} 
  C_\mathbf{k}^{\mathrm{i}^{-1}}
  \begin{pmatrix}
    H_{p,\mathbf{k}}^{\ssp\mathrm{i}}\\
    H_{p,\mathbf{k}}^{\ssm\mathrm{i}}
  \end{pmatrix},
  \\
  \begin{pmatrix}
    H_{p,\mathbf{k}}^{\ssp\mathrm{v}}(\frac{+d}{2})\\
    H_{p,\mathbf{k}}^{\ssm\mathrm{v}}(\frac{+d}{2})
  \end{pmatrix}
  &=
  M_{pp\mathbf{k}}^{\mathrm{vm}}
  C_\mathbf{k}^{\mathrm{m}}
  C_\mathbf{k}^{\mathrm{m}}
  M_{pp\mathbf{k}}^\mathrm{mi} 
  \begin{pmatrix}
    H_{p,\mathbf{k}}^{\ssp\mathrm{i}}(\frac{-d}{2})\\
    H_{p,\mathbf{k}}^{\ssm\mathrm{i}}(\frac{-d}{2})
  \end{pmatrix},
  \label{eq:trans_mat_eq_film}
\end{align}
where 
$
M_{pp\mathbf{k}}^{\mathrm{vm}}
C_\mathbf{k}^{\mathrm{m}}
C_\mathbf{k}^{\mathrm{m}}
M_{pp\mathbf{k}}^\mathrm{mi} 
$
corresponds to the transfer matrix that relates the field amplitudes at $z=+d/2$ and those at $z=-d/2$.
Remind that $C_\mathbf{k}^{\mathrm{m}}$ contains a half of the propagating phases \eqref{eq:C} so that $C_\mathbf{k}^{\mathrm{m}} C_\mathbf{k}^{\mathrm{m}}$ has the total propagating phases.

We can rearrange the transfer matrix equation \eqref{eq:trans_mat_eq_film} to have the scattering matrix equation as in the single interface case, 
\begin{align}
  \begin{pmatrix}
    H_{p,\mathbf{k}}^{\ssp\mathrm{v}}(\frac{+d}{2})\\
    H_{p,\mathbf{k}}^{\ssm\mathrm{i}}(\frac{-d}{2})
  \end{pmatrix}
  &=
  S[
  M_{pp\mathbf{k}}^{\mathrm{vm}}
  C_\mathbf{k}^{\mathrm{m}}
  C_\mathbf{k}^{\mathrm{m}}
  M_{pp\mathbf{k}}^\mathrm{mi}
  ]
  \begin{pmatrix}
    H_{p,\mathbf{k}}^{\ssm\mathrm{v}}(\frac{+d}{2})\\
    H_{p,\mathbf{k}}^{\ssp\mathrm{i}}(\frac{-d}{2})
  \end{pmatrix}.
  \label{eq:sca_mat_eq_film}
\end{align}
The scattering matrix $S[\bullet]$ is calculated by Eq.~\eqref{eq:sca_trans}.

Substituting the explicit forms of the matching and propagating phase matrices (\ref{eq:C}, \ref{eq:M_k}),
we can derive the dispersion relation for the surface plasmon modes in the thin film,
\begin{align}
  \frac{1}{\det S[
  M_{pp\mathbf{k}}^{\mathrm{vm}}
  C_\mathbf{k}^{\mathrm{m}}
  C_\mathbf{k}^{\mathrm{m}}
  M_{pp\mathbf{k}}^\mathrm{mi}]} = 0
  \quad
  \Leftrightarrow
  \quad
  \cfrac{1 - \xi_{\mathbf{k}}^\mathrm{mv}}{1 + \xi_{\mathbf{k}}^\mathrm{mv}} 
  \cdot
  \cfrac{1 - \xi_{\mathbf{k}}^\mathrm{mi}}{1 + \xi_{\mathbf{k}}^{\mathrm{mi}}} 
  &= e^{-2\operatorname{Im} K_{\mathbf{k}}^\mathrm{m} d}.
  \label{eq:xi_vmi}
\end{align}
where $\xi_{\mathbf{k}}^{\tau_1\tau_2}$ is given by Eq.~\eqref{eq:xi_single}.
If we substitute $\epsi = \epsv$,
we can recover a film standing in free space \cite{maier2007plasmonics},
\begin{align}
  \cfrac{1 - \xi_{\mathbf{k}}^\mathrm{mv}}{1 + \xi_{\mathbf{k}}^\mathrm{mv}} 
  &= \pm e^{-\operatorname{Im} K_{\mathbf{k}}^\mathrm{m} d}.
  \label{eq:xi_vmv}
\end{align}
Taking the infinite thickness limit in the dispersion relation \eqref{eq:xi_vmv},
i.e.~$d\rightarrow \infty$,
we can recover the single interface result \eqref{eq:xi_single}.
This implies the dispersion relation \eqref{eq:xi_vmv} consists of two dispersion relations of surface plasmon polaritons.

\begin{figure}[htbp]
  \centering
  \includegraphics[width=.9\linewidth]
  {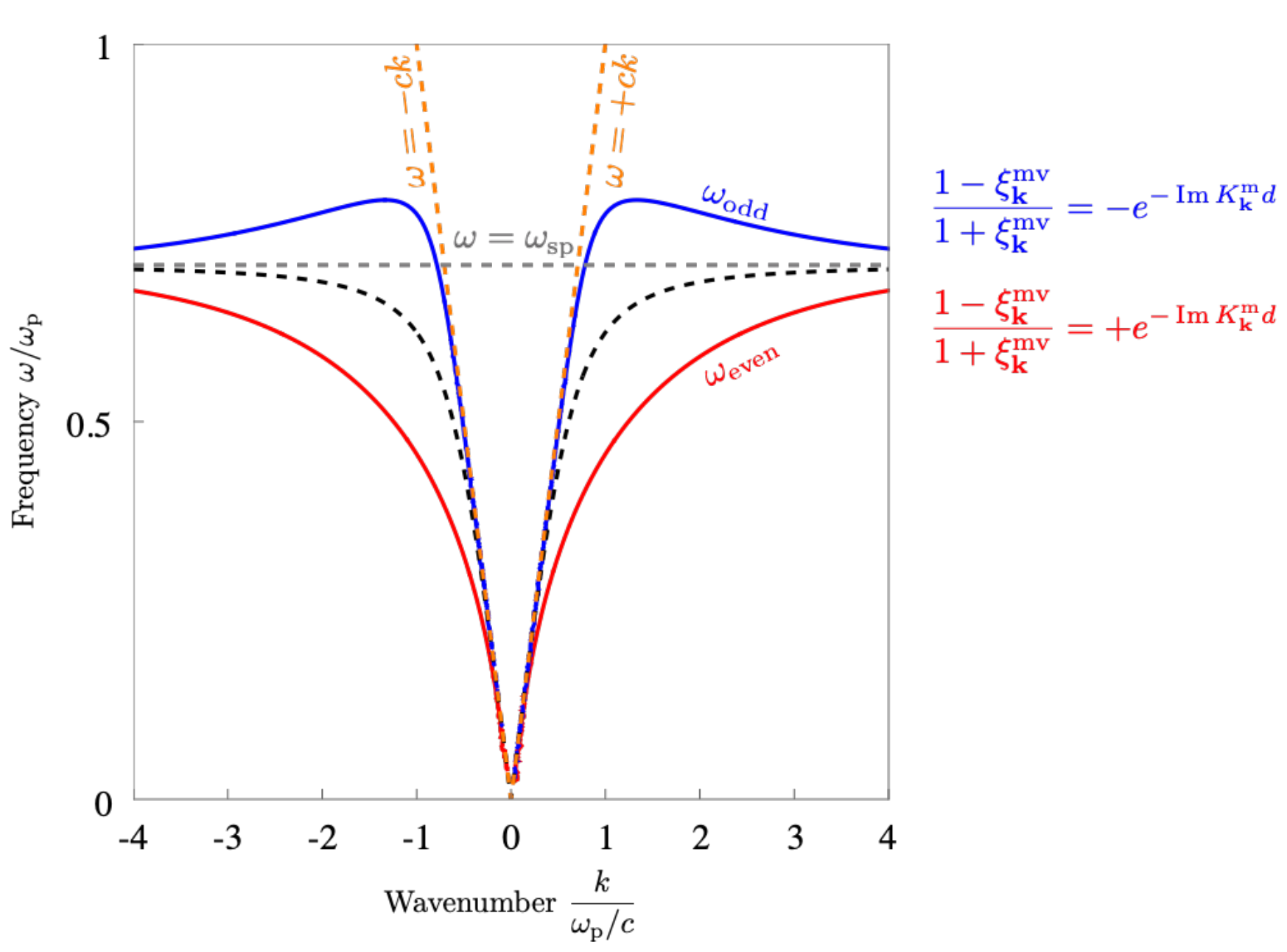}
  \caption{
    Dispersion relation of surface plasmon modes in the thin metallic film
    We use the gold plasma frequency $\omega_\mathrm{p} = 2\pi \times 2.068 \times 10^{15}\ [\mathrm{rad \cdot s^{-1}}]$,
    the vacuum permittivity $\epsv = 1$,
    and the film thickness is 
    $d = 0.1 \times 2\pi c/\omega_\mathrm{p} \approx 100\ [\mathrm{nm}]$.
  }
  \label{fig:dispersion_vmv}
\end{figure}

In \figref{fig:dispersion_vmv},
the dispersion relation \eqref{eq:xi_vmv} is plotted.
The black dashed line is the dispersion relation of the surface plasmon mode at the single interface \eqref{eq:xi_single}.
The orange and grey dashed lines are the light line $\omega=\pm ck$ and the surface plasma frequency $\omega = \omega_\mathrm{sp}$.

We can find there are two branches, red and blue lines in \figref{fig:dispersion_vmv}.
The upper and lower branches correspond to taking $+$ and $-$ on the right-hand side of Eq.~\eqref{eq:xi_vmv}.
They result from the level repulsion of the two surface plasmon modes on the two interfaces.
If the two interfaces are far apart (i.e.~$d$ is large),
the two modes do not interact with each other and are degenerate.
In contrast,
if we bring the two interfaces in close proximity,
they are hybridised and result in the level repulsion,
forming odd and even states.
The upper (lower) branch in \figref{fig:dispersion_vmv} is called an odd (even) mode according to the electric charge density pattern in the film \cite{maier2007plasmonics}.
Since both branches are below the light line $\omega = c |k|$,
they are surface modes.
Note that we cannot obtain the explicit form of the dispersion relation $\omega = \omega(k)$ in this case, unlike the single interface and need to plot the implicit equation.

The scattering matrix calculation via transfer matrices,
which is shown in this chapter,
can straightforwardly be extended for multiple interface calculations.

\chapter{Casimir effect}
\label{ch:casimir}
\section{Casimir effect: radiation force originating from the zero-point field fluctuation}
\label{sec:casimir}

The Casimir effect is the attraction of two neutral interfaces induced by the zero-point fluctuation (i.e.~the uncertainty principle) of electromagnetic field, which is initially discovered by Casimir in 1948 \cite{casimir1948attraction} motivated by his previous work with Polder \cite{casimir1948influence}.
The van der Waals effect is closely related to the Casimir effect.
As the two interfaces come closer to each other, the relativistic retardation effect on the attractive force is smaller and becomes negligible.
In this limit, the attractive force is called the van der Waals force \cite{bordag2009advances,klimchitskaya2015casimir}.

\begin{figure}[htbp]
  \centering
  \includegraphics[width=.8\linewidth]
  {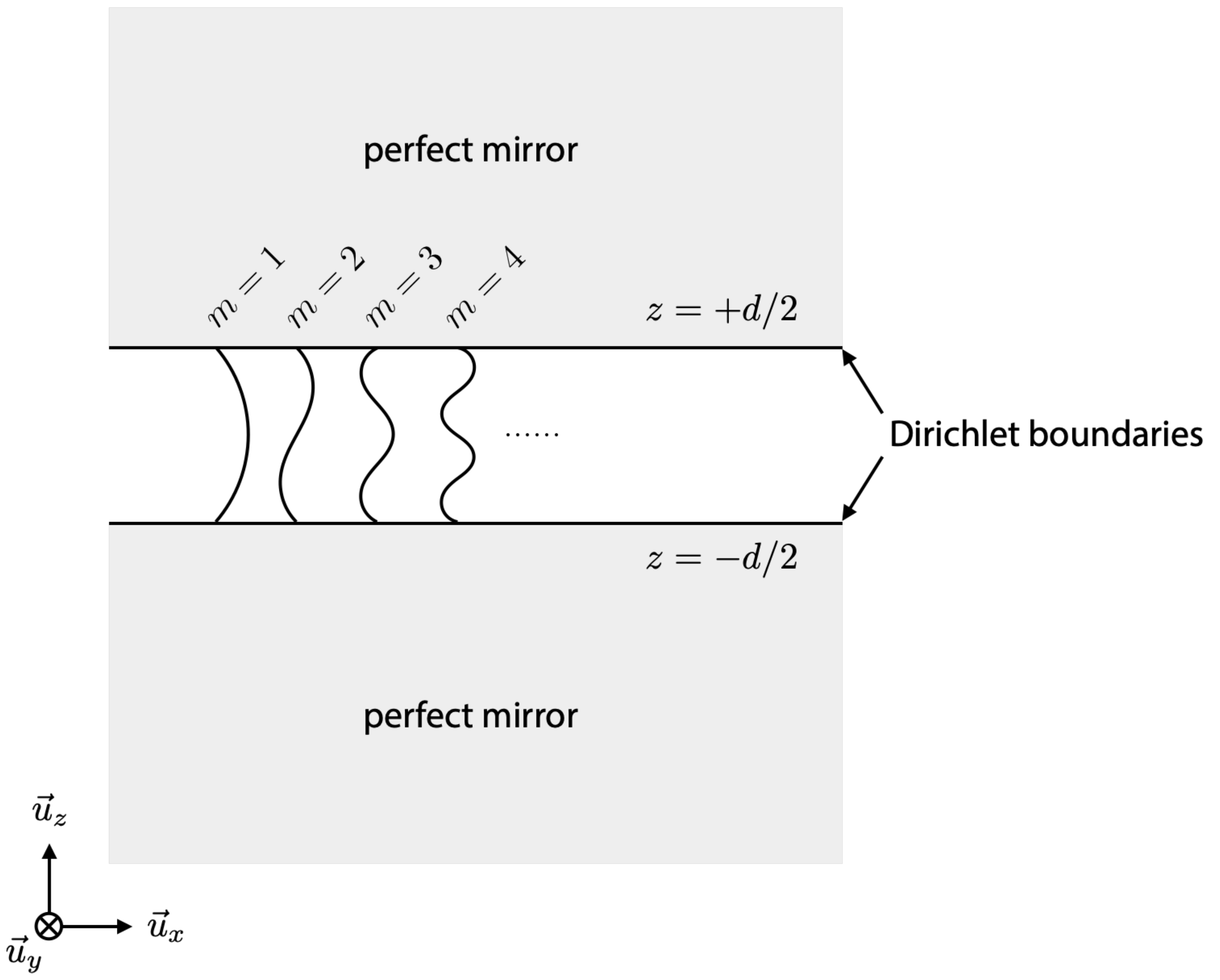}
  \caption{
    Electromagnetic fields originating from the vacuum fluctuation in the gap between two mirrors.
    Since the reflection of the fields at the mirrors is the fixed end one, they are standing waves and called normal modes.
    Each mode has the zero-point energy $\hbar \omega/2$.
    The Casimir energy is the sum of the zero-point energies \eqref{eq:casimir_sum}.
  }
  \label{fig:two_mirrors}
\end{figure}

In his work \cite{casimir1948attraction},
Casimir discussed a pair of planar mirrors separated by a vacuum gap with a width $d$ shown in \figref{fig:two_mirrors}.
Since he assumed that the mirrors are perfect ones,
they can be represented by the Dirichlet boundary conditions.
The fields arising due to quantum fluctuation between the mirrors are standing waves and called normal modes \cite{maier2007plasmonics},
whose wavenumber in the $z$ direction is given by
\begin{align}
  K_\mathbf{k}(d) = \frac{m \pi}{d}
  \quad
  (m = 1,2,\cdots),
  \label{eq:standing_wave_condition}
\end{align}
where $K_\mathbf{k}$ is the radiation wavenumber in the $z$ direction in the vacuum region.
The corresponding frequency is given by 
\begin{align}
  \omega_{\kpara}(d)
  = c\sqrt{{\kpara}^2 + \{K_\mathbf{k}(d)\}^2},
\end{align}
where $\kpara = \sqrt{{k_x}^2 + {k_y}^2}$ is the parallel component of the radiation wavenumber.
Note that the system under consideration possesses the translational invariance in the $x$ and $y$ direction,
and we can work in the reciprocal space,
where the field is labeled by the parallel component of the wave vector $\vec{k}_\parallel = k_x \vec{u}_x + k_y \vec{u}_y$.
The system also has the rotational invariance about the $z$ axis that allow us to give the eigenfrequency in terms of the absolute value $\kpara = \sqrt{{k_x}^2 + {k_y}^2}$.
Each electromagnetic eigenmode labeled by $\vec{k}_\parallel$ is equivalent to a harmonic oscillator with an eigenfrequency $\omega_{\kpara}(d)$,
i.e.~a bosonic excitation,
and hence it has finite energy $\hbar \omega_{\kpara}(d)/2$ even at the absolute zero temperature,
which is known as the zero-point fluctuation
\cite{%
  SchiffLeonardI1968Qm
}.
In the present setup,
the total zero-point fluctuation of the fields satisfying the condition \eqref{eq:standing_wave_condition} is given by
\begin{align}
  U(d) 
  &= 
  \sum_{\kpara} \frac{1}{2} \hbar\omega_{\kpara}
  = 
  \hbar \int
  \left[
    \frac{1}{2}c\kpara
    +
    \sum_{m=1}^{\infty}
    c\sqrt{{\kpara}^2 + \{K_\mathbf{k}(d)\}^2}
  \right]
  \mathrm{d}\vec{k}_\parallel.
  \label{eq:casimir_sum}
  \\
  &=
  \hbar \int
  \left[
    \frac{1}{2}c\kpara
    +
  \int 
  c\sqrt{{\kpara}^2 + {k_z}^2}
  \hspace{.1em}
  \frac{d}{\pi} \mathrm{d}k_z
  \right]
  \mathrm{d}\vec{k}_\parallel.
\end{align}
The sum representation in Eq.~\eqref{eq:casimir_sum} is called the Casimir sum.
Casimir calculated the derivative of the Casimir energy with respect to the gap width \cite{casimir1948attraction},
\begin{align}
  F_\mathrm{Cas}
  &=
  -\frac{\partial U(d)}{\partial d}
  = 
  -\frac{\partial}{\partial d}
  \iint 
  c\sqrt{{\kpara}^2 + {k_z}^2}
  \hspace{.1em}
  \frac{d}{\pi} \mathrm{d}k_z
  \mathrm{d}\vec{k}_\parallel
  =
  -\iint 
  c\sqrt{{\kpara}^2 + {k_z}^2}
  \hspace{.1em}
  \mathrm{d}k_z
  \mathrm{d}\vec{k}_\parallel.
\end{align}
Note that the integrand is positive so that the far right-hand side is negative.
Thus, there is an attractive radiation force between the two mirrors.
Physically speaking,
the electromagnetic mode density and hence the stored energy between the mirrors decreases 
as the gap becomes narrower 
so that shrinking the gap is energetically favourable.

In order to compute the Casimir sum,
we need to evaluate the explicit expression of the dispersion relation 
$\omega=\omega_{\kpara}(d)$;
however,
such explicit expression is not always available,
unlike in the Casimir study,
for example,
when the dielectric dispersion of the mirrors plays a role.
This is why an integral formula proposed by Lifshitz in his seminal work
\cite{lifshitz1956theory}
is useful in practice,
where we can adopt an implicit expression for the dispersion relation instead of the explicit one.
The formula is given through integrations with respect to the parallel component of the wavenumber $\vec{k}_\parallel$ and the frequency $\omega$,
\begin{align}
  U(d) 
  &= 
  \frac{\hbar}{2\pi i}
  \sum_{\lambda=s,p}
  \int_0^{+i\infty} \int
  \ln 
  \Big[
    \frac{1}{1- (r_{\lambda,\mathbf{k}} e^{i K_{\mathbf{k}}^\mathrm{v} d})^2}
  \Big]
  \mathrm{d}\mathbf{k},
  \label{eq:Lifshitz_formula}
\end{align}
where we perform the frequency integration along the imaginary axis.
The reflection coefficients of the mirrors for TE and TM modes are defined as following~\cite{novotny2012principles}:
\begin{align}
  r_{s,\mathbf{k}} 
  &= \frac{\mu^\mathrm{v} K_{\mathbf{k}}^\mathrm{m} - \mu^\mathrm{m} K_{\mathbf{k}}^\mathrm{v}}
  {\mu^\mathrm{v} K_{\mathbf{k}}^\mathrm{m} + \mu^\mathrm{m} K_{\mathbf{k}}^\mathrm{v}},
  \quad
  r_{p,\mathbf{k}} 
  = \frac{\epsilon^\mathrm{v} K_{\mathbf{k}}^\mathrm{m} - \epsilon^\mathrm{m} K_{\mathbf{k}}^\mathrm{v}}
  {\epsilon^\mathrm{v} K_{\mathbf{k}}^\mathrm{m} + \epsilon^\mathrm{m} K_{\mathbf{k}}^\mathrm{v}},
  \label{eq:r_lambda}
\end{align}
where $\epsilon^\mathrm{v,m}$ are the permittivities of the gap region ($|z| < d/2$) and the mirror region $|z| > d/2$,
and $\mu^\mathrm{v,m}$ are the corresponding permeabilities.
The integral \eqref{eq:Lifshitz_formula} is often called Lifshitz formula.
The original derivation of the formula \cite{lifshitz1956theory} was based on the statistical physics point of view as elaborated in his textbook 
\cite{lifshitz1980statistical}.
After the first work,
Lifshitz generalised the formula and calculated the Casimir force acting on a pair of dielectric surfaces, working with Dzyaloshinskii and Pitaevskii  
\cite{%
lifshitz1956theory,%
dzyaloshinskii1961general%
}.

There were several doubts on the Lifshitz theory at that time
\cite{%
  mehra1967temperature,%
  boyer1968quantum,%
  boyer1970quantum%
}
that Schwinger settled to rest by providing an alternative derivation of Lifshitz formula from the field theoretical perspective 
\cite{%
  schwinger1975casimir,%
  schwinger1978casimir%
},
which is suitable in attempts to generalise the Lifshitz formula for non-planar setups 
\cite{%
  golestanian2009casimir,%
  rahi2009scattering,%
  ttira2011lifshitz%
}
as discussed in Sec.~\ref{sec:pfa}.
Lifshitz formula can also be derived by rigorous mode-by-mode summation
\cite{%
  klimchitskaya2000casimir,%
  bordag2011drude,%
  bordag2012electromagnetic%
}
and by using the photonic Green's function and the fluctuation-dissipation theorem
\cite{%
  milton2010casimir,%
  wijnands1997green,%
  parashar2018quantum%
}.
The connection between the Lifshitz formula \eqref{eq:Lifshitz_formula} and the rigorous mode-by-mode summation \eqref{eq:casimir_sum} is discussed in detail in the next section.

\begin{figure}[htbp]
  \centering
  \includegraphics[width=.6\linewidth]
  {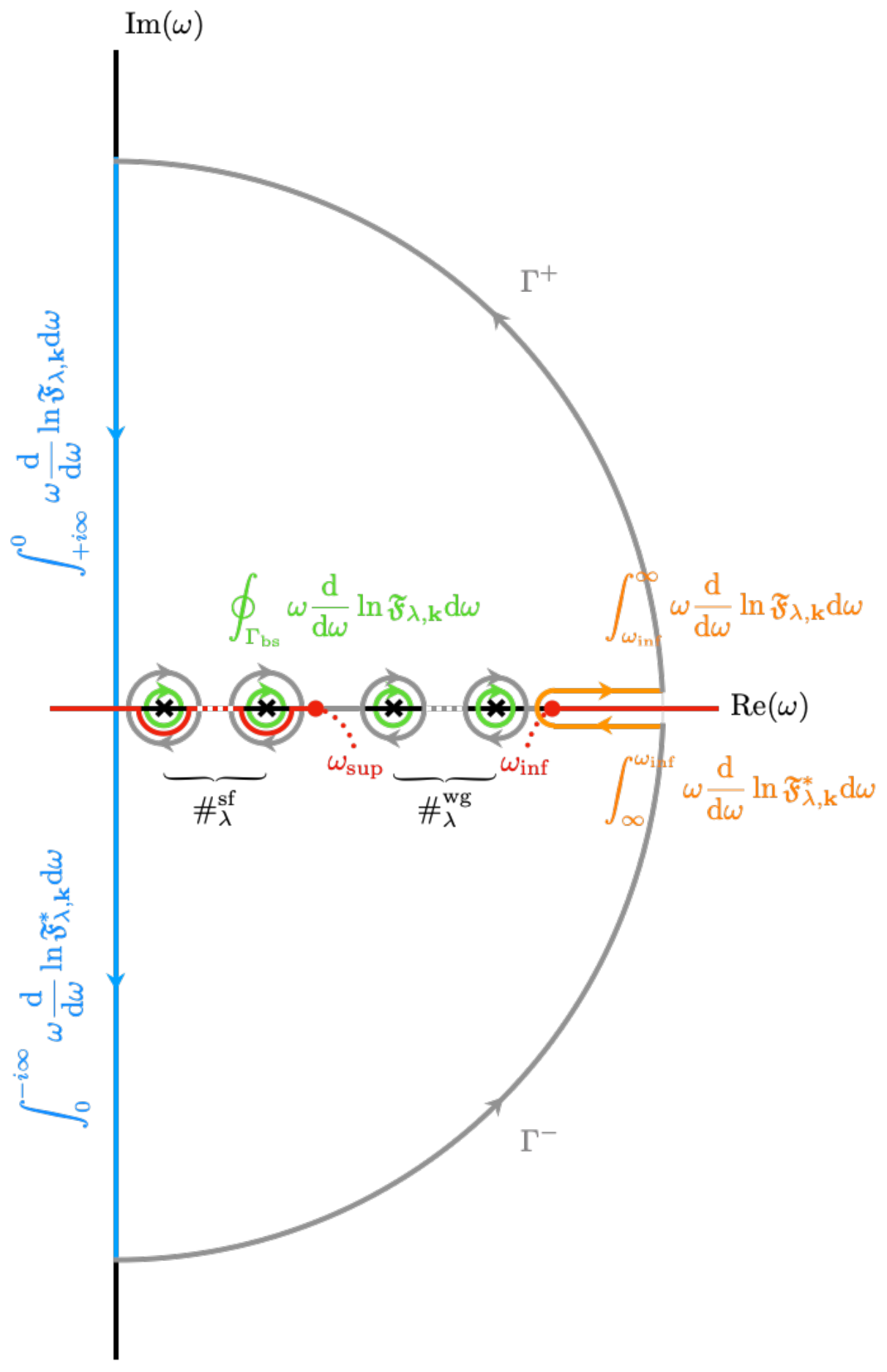}
  \caption{
    The quarter-circle integration contours $\Gamma_{\pm}$ below (above) the real axis on the complex frequency plane.
    Note that we have branch cuts on the real axis represented by the red lines, and poles correspond to the bound states.
    While poles below $\omega_\mathrm{sup}$ correspond to surface modes,
    ones between $\omega_\mathrm{sup}$ and $\omega_\mathrm{inf}$ to waveguide modes.
  }
  \label{fig:contour}
\end{figure}

When the finite temperature effect is taken into consideration, additional poles emerge on the imaginary frequency axis.
They are called Matsubara poles in statistical physics and are equally spaced on the imaginary frequency axis \cite{matsubara1955new}.
When rotating the contour as in \figref{fig:contour}, the contour encounters the Matsubara poles.
This implies not only quantum but also thermal fluctuation does contribute to the Casimir force.
In this case, we have the summation of the corresponding residues instead of the integral
\cite{%
  guerout2014derivation,%
  guerout2016lifshitz%
}.

The Casimir force between structured mirrors has been studied by means of the generalised Lifshitz approach 
\cite{%
davids2010modal,%
lambrecht2008casimir%
}
However, the complementary problem,
the Casimir effect on a slab sandwiched by dielectrics,
has been overlooked until recently 
\cite{%
  klimchitskaya2016casimir,%
  klimchitskaya2017low,%
  klimchitskaya2017casimir,%
  baranov2018contribution,%
  bordag2018free%
}
partly because it is recent studies that suggest the waveguide modes and surface modes should carefully be taken into account 
\cite{%
  intravaia2005surface,%
  intravaia2007role,%
  bordag2011drude,%
  iizuka2019casimir%
}.
In the final part of this chapter,
we discuss how surface corrugation on metallic thin film affects the Casimir effect.

It has been recognised that there are disagreements between the Lifshitz formula and experimental results,
which are called the Casimir puzzle in the metallic systems and the Casimir conundrum in the dielectric systems.
However, it is not the failure of the Lifshitz formula itself but the failure of the local dielectric function $\epsilon(\omega)$ in the high wavenumber regime,
where the parallel wavenumber $k_\parallel$ is comparable with the characteristic length of the surface roughness and/or the lattice constant of the medium as shown in the literature \cite{mostepanenko2021casimir}.
Since the integral in the Lifshitz formula is performed with respect to the parallel component of the wavenumber as well as the frequency,
we should take not only temporal dispersion but also spatial one into consideration.
For large $k_\parallel$,
we cannot neglect the surface roughness and the nonlocal response of dielectrics and metals.
We can improve the accuracy by substituting nonlocal dielectric function $\epsilon(\omega,k_\parallel)$ in the Lifshitz formula.


\section{Casimir sum and Lifshitz integral formula}
Since the poles of the scattering (reflection) coefficients correspond to the frequency of the eigenmodes,
one may expect that the Casimir sum Eq.~\eqref{eq:casimir_sum} can be rewritten in terms of the reflection coefficients.
Following works by Nesterenko and Pirozhenko 
\cite{nesterenko2012lifshitz},
here, I show the two formulae are equivalent.

We begin with writing the general solutions to the Maxwell's equations in terms of Hertz's vector potentials:
\begin{align}
    \vec{\mathcal{E}} 
    &= \nabla \times \nabla \times \vec{\Pi}_p,
    \quad 
    \vec{\mathcal{H}} 
    = \nabla \times 
    \epsilon \frac{\partial \vec{\Pi}_p}{\partial (ct)}
    \quad \qquad
    \mathrm{(TM\ modes)},
    \label{eq:TM_Hertz}
    \\
    \vec{\mathcal{E}} 
    &= -\nabla \times 
    \mu \frac{\partial \vec{\Pi}_s}{\partial (ct)},
    \quad 
    \vec{\mathcal{H}} = \nabla \times \nabla \times \vec{\Pi}_s
    \qquad
    \mathrm{(TE\ modes)},
    \label{eq:TE_Hertz}
\end{align}
where $\vec{\Pi}_{\lambda=s,p}$ are Hertz's potentials for TE and TM modes given by
\begin{align}
  \vec{\Pi}_{\lambda} 
  &= \vec{u}_z e^{i \mathbf{k} \cdot \mathbf{x}} \Phi_{\lambda}
  \quad 
  (\lambda = s,p),
  \label{eq:Hertz_for_TE}
\end{align}
which satisfy the vector Helmholtz equation:
\begin{align}
  \left[\nabla^2 + \epsilon \mu {k_0}^2 \right]
  \vec{\Pi}_\lambda = 0,
  \label{eq:vector_Helmholtz}
\end{align}
and $\Phi_{\lambda,z}$ is called an envelope function,
which are subject to a wave equation of massive Klein-Gordon type,
\begin{align}
  -\frac{\mathrm{d}^2}{\mathrm{d}z^2} \Phi_{\lambda}
  &= \left(\epsilon \mu {k_0}^2 - {k_\parallel}^2 \right)
  \Phi_{\lambda}.
  \label{eq:env_for_TM}
\end{align}

We impose the standard field continuity conditions at $z = -d/2$,
\begin{align}
  \lim_{\delta h \rightarrow 0}
  \int_{-\frac{d}{2}-\delta h}^{-\frac{d}{2}+\delta h}
  \frac{\mathrm{d}(\epsilon \Phi_p)}{\mathrm{d}z} 
  \mathrm{d}z
  &= 0,
  \quad
  \lim_{\delta h \rightarrow 0}
  \int_{-\frac{d}{2}-\delta h}^{-\frac{d}{2}+\delta h}
  \frac{\mathrm{d}^2\Phi_p}{\mathrm{d}z^2}
  \mathrm{d}z
  = 0
  \quad \qquad
  \textrm{(TM modes)},
  \label{eq:boundary_TM_-L}
  \\
  \lim_{\delta h \rightarrow 0}
  \int_{-\frac{d}{2}-\delta h}^{-\frac{d}{2}+\delta h}
  \frac{\mathrm{d}(\mu \Phi_s)}{\mathrm{d}z} 
  \mathrm{d}z
  &= 0,
  \quad
  \lim_{\delta h \rightarrow 0}
  \int_{-\frac{d}{2}-\delta h}^{-\frac{d}{2}+\delta h}
  \frac{\mathrm{d}^2\Phi_s}{\mathrm{d}z^2}
  \mathrm{d}z
  = 0
  \qquad
  \textrm{(TE modes)}.
  \label{eq:boundary_TE_-L}
\end{align}
and at $z = +d/2$,
\begin{align}
  \lim_{\delta h \rightarrow 0}
  \int_{+\frac{d}{2}-\delta h}^{+\frac{d}{2}+\delta h}
  \frac{\mathrm{d}(\epsilon \Phi_p)}{\mathrm{d}z} 
  \mathrm{d}z
  &= 0,
  \quad
  \lim_{\delta h \rightarrow 0}
  \int_{+\frac{d}{2}-\delta h}^{+\frac{d}{2}+\delta h}
  \frac{\mathrm{d}^2\Phi_p}{\mathrm{d}z^2}
  \mathrm{d}z
  = 0
  \quad \qquad
  \textrm{(TM modes)},
  \label{eq:boundary_TM_+L}
  \\
  \lim_{\delta h \rightarrow 0}
  \int_{+\frac{d}{2}-\delta h}^{+\frac{d}{2}+\delta h}
  \frac{\mathrm{d}(\mu \Phi_s)}{\mathrm{d}z} 
  \mathrm{d}z
  &= 0,
  \quad
  \lim_{\delta h \rightarrow 0}
  \int_{+\frac{d}{2}-\delta h}^{+\frac{d}{2}+\delta h}
  \frac{\mathrm{d}^2\Phi_s}{\mathrm{d}z^2}
  \mathrm{d}z
  = 0
  \qquad
  \textrm{(TE modes)}.
  \label{eq:boundary_TE_+L}
\end{align}
Note that we should substitute 
\begin{align}
  \epsilon
  &=
  \begin{cases}{}
    \epsilon^\mathrm{v} 
    &
    (|z| < d/2),
    \\
    \epsilon^\mathrm{m} 
    &
    (|z| > d/2).
  \end{cases}
\end{align}

In the TM case,
we can collect and arrange the conditions (\ref{eq:boundary_TM_-L}, \ref{eq:boundary_TM_+L}) as following:
\begin{align}
  \mathfrak{L}_\mathrm{odd}
  \equiv 
  i \epsilon^\mathrm{v} K_{\mathbf{k}}^\mathrm{m}
  +
  \epsilon^\mathrm{m} K_{\mathbf{k}}^\mathrm{v}
  \tan \left(\frac{K_{\mathbf{k}}^\mathrm{v} d}{2}\right) 
  = 0,
  \label{eq:boundary_asymmetric_TM}
  \\
  \mathfrak{L}_\mathrm{even}
  \equiv 
  i \epsilon^\mathrm{v} K_{\mathbf{k}}^\mathrm{m}
  -
  \epsilon^\mathrm{m} K_{\mathbf{k}}^\mathrm{v}
  \cot \left(\frac{K_{\mathbf{k}}^\mathrm{v} d}{2}\right) 
  = 0.
  \label{eq:boundary_symmetric_TM}
\end{align}
The roots of each equation corresponds to the eigenmodes with odd and even parity.
Here,
we call odd and even according to the parity of the envelope function.
We can rewrite Eqs.~(\ref{eq:boundary_asymmetric_TM}, \ref{eq:boundary_symmetric_TM}) in terms of the reflection coefficient,
\begin{align}
  \mathfrak{L}_\mathrm{odd} = 0
  \quad
  \Leftrightarrow
  \quad
  1 + r_{p,\mathbf{k}} e^{i K_\mathbf{k}^\mathrm{v} d} = 0,
  \label{eq:boundary_asymmetric_ref_TM}
  \\
  \mathfrak{L}_\mathrm{even} = 0
  \quad
  \Leftrightarrow
  \quad
  1 - r_{p,\mathbf{k}} e^{i K_\mathbf{k}^\mathrm{v} d} = 0.
  \label{eq:boundary_symmetric_ref_TM}
\end{align}
merge them into one,
\begin{align}
  \mathfrak{L}_\mathrm{odd} \mathfrak{L}_\mathrm{even}
  &= 0
  \quad
  \Leftrightarrow
  \quad
  \mathfrak{F}_{p,\mathbf{k}}
  \equiv
  1 - (r_{p,\mathbf{k}} e^{iK_{\mathbf{k}}^\mathrm{v}d})^2
  =0.
\end{align}
We can follow the same procedure for the TE mode to obtain
\begin{align}
  \mathfrak{F}_{s,\mathbf{k}}
  \equiv
  1 - (r_{s,\mathbf{k}} e^{iK_{\mathbf{k}}^\mathrm{v}d})^2
  = 0.
\end{align}

Applying the argument principle,
we can write the contributions from bound states,
including surface modes and waveguide modes,
by means of an integral on contours enclosing the corresponding poles on the complex frequency plane instead of the summation,
\begin{align}
  U_\mathrm{bs} (d)
  &= \frac{\hbar}{2} 
  \sum_{\vec{k}_\parallel} \omega_{k_\parallel}^{\mathrm{sf}}(d)
  +
  \frac{\hbar}{2} 
  \sum_{\vec{k}_\parallel} \omega_{k_\parallel}^{\mathrm{wg}}(d)
  \\
  &= \frac{\hbar}{2} 
  \sum_{\lambda=s,p} 
  \int
  \Big(
  \sum_{n=1}^{\num{\lambda}{sf}}
  \omega_{n,\lambda,k_\parallel}^{\mathrm{sf}}(d)
  +
  \sum_{n=1}^{\num{\lambda}{wg}}
  \omega_{n,\lambda,k_\parallel}^{\mathrm{wg}}(d)
  \Big)
  \mathrm{d}\vec{k}_\parallel 
  \\
  &= \frac{\hbar}{4\pi i} \sum_{\lambda=s,p}
  \oint_{\Gamma_\mathrm{bs}}
  \int
  \omega \frac{\mathrm{d}}{\mathrm{d}\omega} 
  \ln \mathfrak{F}_{\lambda,\mathbf{k}}\ 
  \mathrm{d}\vec{k}_\parallel 
  \mathrm{d}\omega
\end{align}
where $\omega_{n;\lambda,k_\parallel}^\mathrm{sf(wg)}(d)$ 
is the frequency of the $n$th surface (waveguide) mode that is $\lambda$ polarised,
$\num{\lambda}{sf(wg)}$ is the number of surface (waveguide) modes,
Note also that we have denoted 
$\mathrm{d}\vec{k}_\parallel 
\equiv 
\mathrm{d}k_x\mathrm{d}k_y/(2 \pi)^2$.

Since the roots of the frequency equation correspond to the bound states,
we can write
\begin{align}
  \mathfrak{F}_{\lambda,\mathbf{k}}
  = \prod_{i=1}^{\num{\lambda}{sf}}
  (\omega - \omega_{i,\lambda,k_\parallel}^\mathrm{sf})
  \prod_{i=1}^{\num{\lambda}{wg}}
  (\omega - \omega_{i,\lambda,k_\parallel}^\mathrm{wg})
  \times \cdots
  \equiv
  \prod_{i=1}^{\num{\lambda}{bs}}
  (\omega - \omega_{i}^\mathrm{bs})
  \times \cdots,
\end{align}
where $\num{\lambda}{bs}$ is the number of the bound states,
and evaluate the derivative with respect to $\omega$,
\begin{align}
  \frac{\mathrm{d}}{\mathrm{d}\omega}
  \ln \mathfrak{F}_{\lambda,\mathbf{k}}
  &= \frac{\mathrm{d}\mathfrak{F}_{\lambda,\mathbf{k}}/\mathrm{d}\omega}{\mathfrak{F}_{\lambda,\mathbf{k}}}
  = 
  \frac{\displaystyle{
      \sum_{l}
      \left[
      \prod_{m\neq l}
      (\omega - \omega_{m}^\mathrm{bs})
    \times \cdots\right]}
    }{\displaystyle{
      \prod_{m=1}^{\num{\lambda}{bs}}
      (\omega - \omega_{m}^\mathrm{bs})
    \times \cdots}
  }.
\end{align}
This is why the residue of the integral kernel at $\omega_{m}^\mathrm{bs}$ is  
\begin{align}
  \underset{\omega_{m}^\mathrm{bs}}
  {\operatorname{Res}}\ 
  \omega
  \frac{\mathrm{d}}{\mathrm{d}\omega}
  \ln \mathfrak{F}_{\lambda,\mathbf{k}}
  &=
  \lim_{\omega \rightarrow \omega_{m}^\mathrm{bs}}
  \frac{\omega\displaystyle{
      \sum_{l}
      \left[
        \prod_{m'\neq l}
        (\omega - \omega_{m'}^\mathrm{bs})
        \times \cdots
      \right]
      }}{\displaystyle{
      \prod_{m'=1}^{\num{\lambda}{bs}}
      (\omega - \omega_{m'}^\mathrm{bs})
      \times \cdots
  }}
  \times
  (\omega - \omega_{m}^\mathrm{bs})
  \\
  &= 
  \frac{\omega_{m}^\mathrm{bs}\displaystyle{
      \sum_{l}
      \left[
        \prod_{m'\neq l}
        (\omega_{m}^\mathrm{bs}-\omega_{m'}^\mathrm{bs})
        \times \cdots
      \right]
      }}{\displaystyle{
      \prod_{m'\neq m}
      (\omega_{m}^\mathrm{bs}-\omega_{m'}^\mathrm{bs})
      \times \cdots
  }}
  =\omega_{m}^\mathrm{bs}.
\end{align}

On the other hand,
the contribution from scattering states having continuous spectrum can be written as
\begin{align}
  U_\mathrm{sc}(d) 
  &= \frac{\hbar}{2} 
  \sum_{\vec{k}_\parallel}
  \omega_{k_\parallel}^{\mathrm{sc}}(d) 
  \\
  &= \frac{\hbar}{2}
  \sum_{\lambda=s,p}
  \int_{\omega_\mathrm{inf}}^\infty
  \int 
  \omega \Delta \rho_{\lambda, \mathbf{k}}(d) 
  \mathrm{d}\vec{k}_\parallel
  \mathrm{d}\omega.
\end{align}
Here, we have introduced the electromagnetic mode density difference,
\begin{align}
  \Delta \rho_{\lambda, \mathbf{k}}(d) 
  \equiv \rho_{\lambda,\mathbf{k}}(d) - \rho_{\lambda,\mathbf{k}}(\infty),
  \label{eq:Delta_rho}
\end{align}
where $\rho_{\lambda,\mathbf{k}}(\infty)$ corresponds to the mode density in free space,
and $\omega_\mathrm{inf}$ is a frequency below which there is no scattering states.
The mathematically rigorous scattering theory provides Krein's theorem 
\cite{%
  aoyama1984casimir,%
  barton1985casimir,%
  barton1985levinson,%
  souma2002local%
},
and the following expression of the mode density difference \eqref{eq:Delta_rho} in terms of the scattering matrix of the system in question (see Appendix \ref{app:Krein}):
\begin{align}
  \Delta \rho_{\lambda, \mathbf{k}}(d) 
  =
  \frac{1}{2\pi i}
  \frac{\mathrm{d}}{\mathrm{d}\omega}
  \ln \det S_{\lambda,\mathbf{k}}
  =
  \frac{1}{2\pi i}
  \frac{\mathrm{d}}{\mathrm{d}\omega}
  \ln 
  \frac{\mathfrak{F}_{\lambda,\mathbf{k}}^*}
  {\mathfrak{F}_{\lambda,\mathbf{k}}},
\end{align}
and thus
\begin{align}
  U_\mathrm{sc}(d)
  &=
  \frac{\hbar}{4\pi i}
  \sum_{\lambda=s,p}
  \int_{\omega_\mathrm{inf}}^\infty
  \int 
  \omega \frac{\mathrm{d}}{\mathrm{d}\omega}
  \ln \frac{\mathfrak{F}_{\lambda,\mathbf{k}}^*}
  {\mathfrak{F}_{\lambda,\mathbf{k}}}
  \mathrm{d}\vec{k}_\parallel
  \mathrm{d}\omega.
\end{align}
Note that this type of calculation can be extended to be applied to the calculation of the Casimir forces on gratings with the help of the modal expansion technique and the scattering theory
\cite{%
  lambrecht2006casimir,%
  lambrecht2008casimir,%
  lambrecht2009theory,%
  davids2010modal%
},
which can be checked by other approaches such as Green's function method \cite{yannopapas2008optical,silveirinha2018fluctuation}.

Summing up the contributions from the bound states and scattering states,
we can obtain
\begin{align}
  U(d)
  &= U_\mathrm{bs}(d) + U_\mathrm{sc}(d) 
  \\
  &= \frac{\hbar}{4\pi i}
  \sum_{\lambda=s,p}
  \int
  \Big[
  \oint_{\Gamma_\mathrm{bs}}
  \omega \frac{\mathrm{d}}{\mathrm{d}\omega} 
  \ln \mathfrak{F}_{\lambda,\mathbf{k}}\ 
  \mathrm{d}\omega
  +
  \int_{\omega_\mathrm{inf}}^\infty
  \omega \frac{\mathrm{d}}{\mathrm{d}\omega}
  \ln \frac{\mathfrak{F}_{\lambda,\mathbf{k}}^*}
  {\mathfrak{F}_{\lambda,\mathbf{k}}}
  \mathrm{d}\omega
  \Big]
  \mathrm{d}\vec{k}_\parallel.
\end{align}
The scattering part can be decomposed into two contributions,
\begin{align}
  \int_{\omega_\mathrm{inf}}^\infty
  \omega \frac{\mathrm{d}}{\mathrm{d}\omega}
  \ln \frac{\mathfrak{F}_{\lambda,\mathbf{k}}^*}
  {\mathfrak{F}_{\lambda,\mathbf{k}}}
  \mathrm{d}\omega
  &=
  \int_{\omega_\mathrm{inf}}^\infty
  \omega \frac{\mathrm{d}}{\mathrm{d}\omega}
  \ln \mathfrak{F}_{\lambda,\mathbf{k}}^*
  \mathrm{d}\omega
  +
  \int_\infty^{\omega_\mathrm{inf}}
  \omega \frac{\mathrm{d}}{\mathrm{d}\omega}
  \ln \mathfrak{F}_{\lambda,\mathbf{k}}
  \mathrm{d}\omega.
\end{align}
Since the complex roots of 
$\mathfrak{F}_{\lambda,\mathbf{k}} = 0\ 
(\mathfrak{F}_{\lambda,\mathbf{k}}^* = 0)
$
lie in the lower (upper) half-plane due to the causality,
we have 
\begin{align}
  \oint_{\Gamma^\ssm}
  \omega \frac{\mathrm{d}}{\mathrm{d}\omega}
  \ln \mathfrak{F}_{\lambda,\mathbf{k}}^*
  \mathrm{d}\omega
  &= 0,
  \\
  \oint_{\Gamma^\ssp}
  \omega \frac{\mathrm{d}}{\mathrm{d}\omega}
  \ln \mathfrak{F}_{\lambda,\mathbf{k}}
  \mathrm{d}\omega
  &= 0,
\end{align}
or
\begin{align}
  \int_{\omega_\mathrm{inf}}^\infty
  \omega \frac{\mathrm{d}}{\mathrm{d}\omega}
  \ln \mathfrak{F}_{\lambda,\mathbf{k}}^*
  \mathrm{d}\omega
  +
  \landdownint
  \omega \frac{\mathrm{d}}{\mathrm{d}\omega}
  \ln \mathfrak{F}_{\lambda,\mathbf{k}}^*
  \mathrm{d}\omega
  &=
  -\int_{-i\infty}^{0}
  \omega \frac{\mathrm{d}}{\mathrm{d}\omega}
  \ln \mathfrak{F}_{\lambda,\mathbf{k}}^*
  \mathrm{d}\omega,
  \\
  \landupint
  \omega \frac{\mathrm{d}}{\mathrm{d}\omega}
  \ln \mathfrak{F}_{\lambda,\mathbf{k}}
  \mathrm{d}\omega
  +
  \int_\infty^{\omega_\mathrm{inf}}
  \omega \frac{\mathrm{d}}{\mathrm{d}\omega}
  \ln \mathfrak{F}_{\lambda,\mathbf{k}}
  \mathrm{d}\omega
  &=
  -\int_{0}^{+i\infty}
  \omega \frac{\mathrm{d}}{\mathrm{d}\omega}
  \ln \mathfrak{F}_{\lambda,\mathbf{k}}
  \mathrm{d}\omega.
\end{align}

Therefore,
we can obtain
\begin{align}
  &\oint_{\Gamma_\mathrm{bs}}
  \omega \frac{\mathrm{d}}{\mathrm{d}\omega}
  \ln \mathfrak{F}_{\lambda,\mathbf{k}}
  \mathrm{d}\omega
  +
  \int_{\infty}^{\omega_\mathrm{inf}}
  \omega \frac{\mathrm{d}}{\mathrm{d}\omega}
  \ln \mathfrak{F}_{\lambda,\mathbf{k}}^*
  \mathrm{d}\omega
  +
  \int_{\omega_\mathrm{inf}}^\infty
  \omega \frac{\mathrm{d}}{\mathrm{d}\omega}
  \ln \mathfrak{F}_{\lambda,\mathbf{k}}
  \mathrm{d}\omega
  \notag \\
  &=
  \int_{0}^{-i\infty}
  \omega \frac{\mathrm{d}}{\mathrm{d}\omega}
  \ln \mathfrak{F}_{\lambda,\mathbf{k}}^*
  \mathrm{d}\omega
  +
  \int_{+i\infty}^{0}
  \omega \frac{\mathrm{d}}{\mathrm{d}\omega}
  \ln \mathfrak{F}_{\lambda,\mathbf{k}}
  \mathrm{d}\omega
\end{align}
Here, we have used the fact that
$\mathfrak{F}_{\lambda,\mathbf{k}}^* = \mathfrak{F}_{\lambda,\mathbf{k}}$
on the branch cut and the real axis for 
$\operatorname{Re} (\omega) < \omega_\mathrm{inf}$.
This is consistent with the fact that there is no scattering states at frequencies below $\omega_\mathrm{inf}$.

Finally, we can obtain the integral along the imaginary frequency axis,
\begin{align}
  U(d)
  &= \frac{\hbar}{4\pi i}
  \sum_{\lambda=s,p}
  \int
  \Big[
  \int_{0}^{-i\infty}
  \omega \frac{\mathrm{d}}{\mathrm{d}\omega}
  \ln \mathfrak{F}_{\lambda,\mathbf{k}}^*
  \mathrm{d}\omega
  +
  \int_{+i\infty}^{0}
  \omega \frac{\mathrm{d}}{\mathrm{d}\omega}
  \ln \mathfrak{F}_{\lambda,\mathbf{k}}
  \mathrm{d}\omega
  \Big]
  \mathrm{d}\vec{k}_\parallel
  \\
  &= 
  \frac{\hbar}{2\pi i}
  \sum_{\lambda=s,p}
  \int_{+i\infty}^0
  \int
  \ln \mathfrak{F}_{\lambda,\mathbf{k}}
  \mathrm{d}\vec{k}_\parallel
  \mathrm{d}\omega
  =
  \frac{\hbar}{2\pi i}
  \sum_{\lambda=s,p}
  \int_0^{+i\infty}
  \int
  \ln (1/\mathfrak{F}_{\lambda,\mathbf{k}})
  \mathrm{d}\mathbf{k}
\end{align}
where we have used 
$\mathfrak{F}_{\lambda,-\mathbf{k}}^* = \mathfrak{F}_{\lambda,\mathbf{k}}$,
and performed the integration by parts.

\chapter{Lowering the Casimir energy by boundary corrugation}
\label{ch:lowering}
\section{Proximity force approximation}
\label{sec:pfa}
The proximity force approximation (PFA) is one of the simplest approximations adopted in the calculation of interaction between non-planar interfaces,
which is originally developed by Dejaguin in the context of colloidal physics 
\cite{derjaguin1957direct,derjaguin1960force}
and applied to the calculation of the Casimir effect.
In the PFA,
a series of planar patches replace non-planar surfaces,
and the interaction energy only between patch pairs in the closest proximity is calculated.
One can finally sum up contributions from all the patch pairs to get the effective interaction energy.
Since the PFA calculation is simple and fast,
it has been utilised to analyse the Casimir force between a sphere and a plate,
which is one of the popular experimental setups in the Casimir physics community
\cite{%
  lamoreaux1997demonstration,%
  mohideen1998precision,%
  roy1999improved%
}.
It has been known that the PFA is accurate in the small curvature limit and in the small separation limit where area elements in the closest proximity dominate the interaction while it fails to fit the experimental measurement of the force between a sphere and a deeply corrugated surface \cite{chan2008measurement}.
Although it works in the analysis of several experimental setups,
it has been considered as a rough approximation with uncontrolled errors partly because the oblique wave contributions and non-parallelism are neglected within the PFA calculation \cite{bordag2009advances}.
However,
in recent studies,
it has been shown that the PFA can be derived from three different approaches as the leading order contribution in the Casimir energy.
In Refs.~\cite{wu2012field,wu2014perturbative},
they apply the semiclassical approximation to the quantum field theory to derive the PFA formula.
Utilising the Feynman diagrams to perform the perturbative expansion of the Casimir energy, they find that the leading order contribution corresponds to the PFA and give a systematic way to obtain the correction.
In Refs.~\cite{bordag2008casimir,teo2011corrections},
the PFA formula is derived from the multiple scattering formalism.
They expanded the Casimir energy in powers of scattering coefficients and showed that leaving only the zeroth-order contribution is equivalent to PFA.

In a series of works 
\cite{%
  fosco2011proximity,%
  fosco2012derivative,%
  fosco2012improved,%
  fosco2013electrostatic,%
  fosco2014derivative%
},
Fosco et al. regarded the Casimir energy as a functional of surface shapes and performed the derivative expansion \cite{voronovich1994small} to obtain the PFA formula at the lowest order and higher-order corrections within the worldline formalism 
\cite{%
  gies2003casimir,%
  gies2006worldline,%
  gies2006quantum,%
  gies2006casimir,%
  corradini2019worldline,%
  dudal2020casimir%
}.
Such expansion is consistent with the studies of the cylinder-plane
\cite{%
  bordag2006casimir,%
  bordag2007generalized,%
  teo2011casimir%
},
sphere-plane
\cite{%
  bordag2008casimir,%
  bordag2010first,%
  teo2011corrections,%
  bimonte2012material,%
  teo2013material%
},
cylinder-cylinder
\cite{%
  teo2011first,%
  bimonte2012casimir%
}
and sphere-sphere 
\cite{%
  bimonte2012casimir,%
  teo2012casimir,%
  bimonte2018beyond%
}
geometries.

\subsection{Worldline formalism for the derivation of PFA}
\label{subsec:worldline}
Here,
I provide a bird's eye view of the essence of 
Refs.~\cite{%
  fosco2011proximity,%
  fosco2012derivative,%
  fosco2012improved,%
  fosco2013electrostatic,%
  fosco2014derivative%
}.
As done by Casimir in his original works,
here,
we simply regard the electromagnetic field as a scalar field subject to the massless Klein-Gordon equation,
\begin{align}
  -\Box \psi_{\mathbf{x},z} = 0,
  \label{eq:massless_KG}
\end{align}
and the Dirichlet boundary conditions at the surfaces of the mirrors,
\begin{align}
  \psi_{\mathbf{x},0} = \psi_{\mathbf{x},a_\mathbf{x}} = 0,
  \label{eq:Dirichlet_bcs}
\end{align}
where we have used the three-component vector notation $\mathbf{x}=(x,y,ict)$,
$\psi_{\mathbf{x},z}$ is a massless scalar field evaluated at $(\mathbf{x},z)$,
and $a_\mathbf{x}$ is the profile of the boundary.
From the view point of quantum field theory,
the Casimir energy can be regarded as the zero temperature limit of a partition function (generating functional of quantum correlation functions) for a massless scalar field subject to Dirichlet boundary conditions,
\begin{align}
  \mathcal{Z}
  &= \int 
  \delta [\psi_{\mathbf{x}, 0}]\ 
  \delta [\psi_{\mathbf{x}, a_\mathbf{x}}]
  e^{-S_0[\psi]}
  \mathcal{D}\psi,
  \label{eq:Z_pre}
\end{align}
where the boundary conditions are encoded into the Dirac delta functions
(i.e., the partition function is finite for configurations such that the scalar field $\phi$ vanishes at $z=0$ and $z=a_\mathbf{x}$),
and its action without the boundary condition is given by
\begin{align}
  S_0[\psi]
  &=
  \frac{1}{2}\int
  \left(
    \frac{\partial \psi_{\mathbf{x},z}}{\partial \mathbf{x}}
    \cdot
    \frac{\partial \psi_{\mathbf{x},z}}{\partial \mathbf{x}}
    +
    \frac{\partial \psi_{\mathbf{x},z}}{\partial z}
    \frac{\partial \psi_{\mathbf{x},z}}{\partial z}
  \right)
  \mathrm{d}\mathbf{x}
  \mathrm{d}z\\
  &=
  \frac{1}{2}\int
  \psi_{\mathbf{x},z}
  \Big(
  -
  \frac{\partial}{\partial \mathbf{x}}
  \cdot
  \frac{\partial}{\partial \mathbf{x}}
  -
  \frac{\partial}{\partial z}
  \frac{\partial}{\partial z}
  \Big)
  \psi_{\mathbf{x},z}
  \mathrm{d}\mathbf{x}
  \mathrm{d}z
  \equiv
  \frac{1}{2}\int
  \psi_{\mathbf{x},z}
  (-\Box)
  \psi_{\mathbf{x},z}
  \mathrm{d}\mathbf{x}
  \mathrm{d}z,
  \label{eq:S0}
\end{align}
where the integral is performed by parts and we have defined d'Alembertian $\Box := \frac{\partial}{\partial \mathbf{x}} \cdot \frac{\partial}{\partial \mathbf{x}} + \frac{\partial}{\partial z}\frac{\partial}{\partial z}$.
We can confirm that the functional derivative of Eq.~\eqref{eq:S0} with respect to $\psi$ yields the massless Klein-Gordon equation \eqref{eq:massless_KG}.
Note that we have introduced an imaginary-time shorthand notation 
$\mathrm{d}\mathbf{x} = \mathrm{d}x\mathrm{d}y\mathrm{d}(ict)$.

Introducing auxiliary scalar fields,
we can represent the delta functions in terms of the functional integral of an exponential function:
\begin{align}
  \delta[\psi_{\mathbf{x},0}]\ 
  \delta[\psi_{\mathbf{x},a_\mathbf{x}}]
  &= 
  \int e^{-S_{\mathrm{bc}}[\psi;\phi^\crT,\phi^\rT]}
  \mathcal{D}\phi^\crT
  \mathcal{D}\phi^\rT,
  \label{eq:delta_func_aux}
  \\
  S_{\mathrm{bc}}[\psi;\phi^\crT,\phi^\rT]
  &= -i\int
  \psi_{\mathbf{x},z}
  \Big(
  \delta(z)
  \phi_{\mathbf{x}}^\crT
  +
  \delta(z-a_\mathbf{x})
  \phi_{\mathbf{x}}^\rT
  \Big)
  \mathrm{d}\mathbf{x} \mathrm{d}z,
  \label{eq:S_bc}
\end{align}
and rewrite Eq.~\eqref{eq:Z_pre} as following:
\begin{align}
  \mathcal{Z}
  &= \int 
  e^{
    -S_0[\psi]
    -S_{\mathrm{bc}}[\psi;\phi^\crT,\phi^\rT]
  }
  \mathcal{D}\psi
  \mathcal{D}\phi^\crT
  \mathcal{D}\phi^\rT.
  \label{eq:Z}
\end{align}

We can formally perform the functional integral over $\psi$ in Eq.~\eqref{eq:Z} in analogy with the $n$-dimensional integral of Gaussian type,
\begin{align}
  \int 
  e^{
    -\frac{1}{2}\bm{x}\cdot A\bm{x} 
    - \bm{x} \cdot \bm{b}
  } \frac{\mathrm{d}\bm{x}}{(2\pi)^{n/2}}
  = 
  e^{\frac{1}{2} \bm{b} \cdot A^{-1} \bm{b}}
  \int 
  e^{
    -\frac{1}{2}\bm{x}\cdot A\bm{x} 
  } \frac{\mathrm{d}\bm{x}}{(2\pi)^{n/2}},
  \label{eq:Gaussian_int}
\end{align}
where we have performed the square completion,
\begin{align}
  -\frac{1}{2}\bm{x}\cdot A\bm{x} - \bm{x} \cdot \bm{b}
  =
  (\sqrt{A} \bm{x} + \sqrt{A}^{\hspace{.2em}-1} \bm{b})
  \cdot
  (\sqrt{A} \bm{x} + \sqrt{A}^{\hspace{.2em}-1} \bm{b})
  + \frac{1}{2} 
  \sqrt{A}^{\hspace{.2em}-1} \bm{b} 
  \cdot 
  \sqrt{A}^{\hspace{.2em}-1} \bm{b}.
  \label{eq:square_completion}
\end{align}
If and only if the operator $A$ is positive definite,
the integral can be evaluated:
\begin{align}
  \int 
  e^{
    -\frac{1}{2}\bm{x}\cdot A\bm{x} 
  } \frac{\mathrm{d}\bm{x}}{(2\pi)^{n/2}},
  = 
  \int 
  e^{
    -\frac{1}{2}\bm{y}\cdot \bm{y} 
  } 
  \sqrt{\frac{1}{\det A}} 
  \frac{\mathrm{d}\bm{y}}{(2\pi)^{n/2}}
  = 
  \sqrt{\frac{1}{\det A}}.
  \label{eq:Gaussian_int_0}
\end{align}
We have changed the integral variable,
$\bm{y} = \sqrt{A}\bm{x}$,
where the matrix square root can be given by the spectral decomposition 
$
\sqrt{A} 
= \sum_i \sqrt{A_i} \bm{a}_i \bm{a}_i^\dagger
$.

By finding the following correspondence:
\begin{align}
  A \rightarrow -\Box,
  \quad 
  \bm{x} \rightarrow \psi
  \quad
  \bm{b} 
  \rightarrow 
  \delta(z)
  \phi_{\mathbf{x}}^\crT
  +
  \delta(z-a_\mathbf{x})
  \phi_{\mathbf{x}}^\rT,
\end{align}
we can factor out a free field contribution,
\begin{align}
  \mathcal{Z}
  &= \int 
  e^{
    - S_0[\psi] 
    - S_\mathrm{bc}[\psi;\phi^\crT,\phi^\rT]
  }
  \mathcal{D} \psi
  \mathcal{D} \phi^\crT
  \mathcal{D} \phi^\rT
  = \mathcal{Z}_0 \int
  e^{-S_\mathrm{eff}[\phi^\crT,\phi^\rT]}
  \mathcal{D} \phi^\crT
  \mathcal{D} \phi^\rT,
  \label{eq:Z_aux}
\end{align}
where the free field contribution is given by 
$
\mathcal{Z}_0
= \int
e^{-S_0[\psi]}
\mathcal{D}\psi
$,
and the effective action is
\begin{align}
  &S_\mathrm{eff}[\phi^\crT,\phi^\rT] \notag\\
  &=\frac{1}{2} \int
  \Big(
  \delta(z)
  \phi_\mathbf{x}^{\crT}
  +
  \delta(z-a_\mathbf{x})
  \phi_\mathbf{x}^{\rT}
  \Big)
  \mathfrak{D}_\mathrm{F}(\mathbf{x}-\mathbf{x}',z-z')
  \Big(
  \delta(z')
  \phi_{\mathbf{x}'}^{\crT}
  +
  \delta(z'-s_{\mathbf{x}'})
  \phi_{\mathbf{x}'}^{\rT}
  \Big)
  \mathrm{d}\mathbf{x} \mathrm{d}\mathbf{x}'
  \mathrm{d}z \mathrm{d}z'.
  \label{eq:S_eff}
\end{align}
The Feynman propagator of the massless Klein-Gordon field can be written by means of Fourier representation,
\begin{align}
  \mathfrak{D}_\mathrm{F}(\mathbf{x}-\mathbf{x}',z-z')
  &=
  \int
  \frac{e^{i\mathbf{k}\cdot(\mathbf{x}-\mathbf{x}')+ik_z(z-z')}}
  {k_z^2 + \mathbf{k}\cdot\mathbf{k}}
  \frac{\mathrm{d}\mathbf{k}\mathrm{d}k_z}{2\pi},
  \label{eq:D_F}
\end{align}
Remind that we introduced 
$
\mathbf{k}=(k_x,k_y,ik_0),\ 
\mathbf{x}=(x,y,ict),\ 
\mathrm{d}\mathbf{k}
:=\mathrm{d}k_x\mathrm{d}k_y\mathrm{d}\omega/(2\pi)^3
$.

By introducing the Dirac bracket notation for the auxiliary scalar fields,
\begin{align}
  &\braket{\mathbf{x}|\phi^\crT} 
  := \phi_\mathbf{x}^\crT,
  \quad
  \braket{\mathbf{x}|\phi^\rT} 
  := \phi_\mathbf{x}^\rT,
  \\
  &\begin{cases}{}
    \Braket{\mathbf{x}|\mathfrak{S}^{\crT\hspace{.1em}\crT}|\mathbf{x}'} 
    := \mathfrak{D}_\mathrm{F}(\mathbf{x}-\mathbf{x}', 0-0),
    &
    \Braket{\mathbf{x}|\mathfrak{S}^{\crT\rT}|\mathbf{x}'} 
    := \mathfrak{D}_\mathrm{F}(\mathbf{x}-\mathbf{x}', 0-a_{\mathbf{x}'}),
    \\ 
    \bra{\mathbf{x}}\mathfrak{S}^{\rT\hspace{.05em}{\crT}}\ket{\mathbf{x}'} 
    := \mathfrak{D}_\mathrm{F}(\mathbf{x}-\mathbf{x}', a_\mathbf{x}-0),
    &
    \Braket{\mathbf{x}|\mathfrak{S}^{\rT\rT}|\mathbf{x}'} 
    := \mathfrak{D}_\mathrm{F}(\mathbf{x}-\mathbf{x}', a_\mathbf{x}-a_{\mathbf{x}'}),
    \label{eq:bracket_notation}
  \end{cases}
\end{align}
we can rewrite the effective action in a simple form,
\begin{align}
  S_\mathrm{eff}[\phi^\crT,\phi^\rT] 
  &= \frac{1}{2}\int
  \begin{pmatrix}
    \bra{\phi^\crT}\\
    \bra{\phi^\rT}
  \end{pmatrix}^\intercal
  \Bigg| \mathbf{x} \Bigg\rangle
  \Bigg\langle \mathbf{x} \Bigg|
  \begin{pmatrix}
    \mathfrak{S}^{\crT\crT} & \mathfrak{S}^{\crT\rT}\\
    \mathfrak{S}^{\rT\hspace{.1em}\crT} & \mathfrak{S}^{\rT\rT} 
  \end{pmatrix}
  \Bigg| \mathbf{x}' \Bigg\rangle
  \Bigg\langle \mathbf{x} \Bigg|
  \begin{pmatrix}
    \ket{\phi^\crT}\\
    \ket{\phi^\rT}
  \end{pmatrix}
  \mathrm{d}\mathbf{x} \mathrm{d}\mathbf{x}'
  \\
  &= \frac{1}{2}
  \braket{\phi|\mathfrak{S}|\phi}
\end{align}
where we have used the completeness relation,
$\int \ket{x}\bra{x} \mathrm{d}\mathbf{x} = 1$,
and defined
\begin{align}
  \ket{\phi}
  := 
  \begin{pmatrix}
    \ket{\phi^\crT}\\
    \ket{\phi^\rT}
  \end{pmatrix},
  \quad
  \mathfrak{S}
  :=
  \begin{pmatrix}
    \mathfrak{S}^{\crT\crT} & \mathfrak{S}^{\crT\rT}\\
    \mathfrak{S}^{\rT\hspace{.1em}\crT} & \mathfrak{S}^{\rT\rT} 
  \end{pmatrix},
  \quad
  \mathcal{D}\phi 
  := \mathcal{D}\phi^\crT \mathcal{D}\phi^\rT
\end{align}

\begin{figure}[htbp]
  \centering
  \includegraphics[width=.6\linewidth]
  {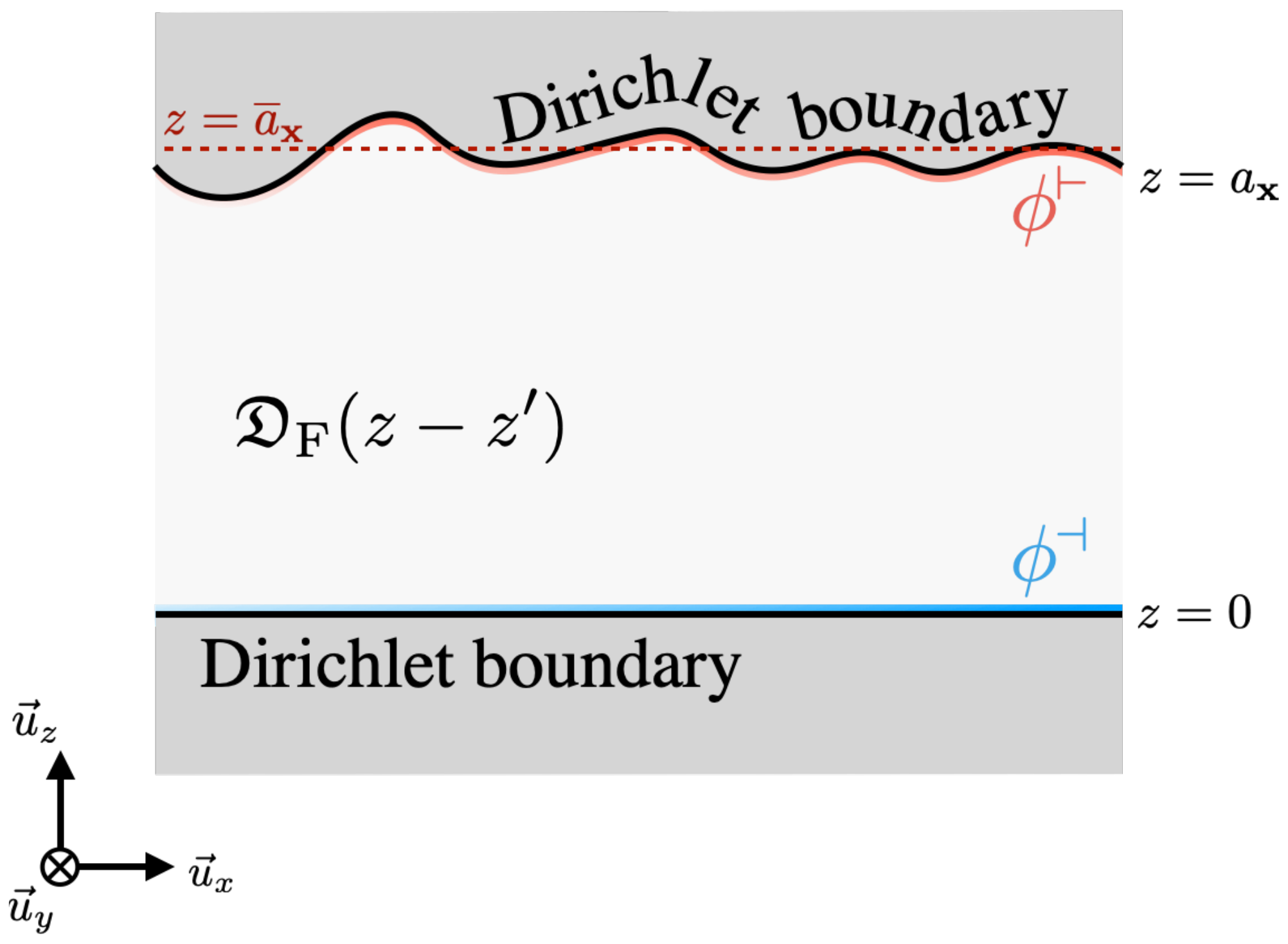}
  \caption{
    Auxiliary fields, $\phi^\rT$ and $\phi^\crT$, lying on Dirichlet boundaries (black lines).
    The lower boundary is flat ($z=0$) while the upper one is structured ($z=a_\mathbf{x}$).
    The averaged part of the profile of the structured boundary is given by $z=\overline{a}_\mathbf{x}$ (dark red dashed line).
    We have a massless Klein--Gordon field which propagates between the boundaries and is represented by the Feynman propagator $\mathfrak{D}_\mathrm{F}$ \eqref{eq:D_F}.
  }
  \label{fig:aux}
\end{figure}

In \figref{fig:aux}, we summarise the current setup.
We have introduced auxiliary fields at the Dirichlet boundaries (\ref{eq:delta_func_aux}, \ref{eq:S_bc}).
The effective action for the auxiliary fields \eqref{eq:S_eff} is given by means of the Feynman propagator (\ref{eq:bracket_notation}) that represents the massless Klein--Gordon field between the boundaries.

\subsection{Expanding the free energy}
\label{subsec:expanding}
In the previous subsection, we derived an effective action in a quadratic form.
Since the effective action is written in a quadratic form, we can evaluate the partition function and its natural logarithm (i.e.~the free energy) that corresponds to the Casimir energy,
\begin{align}
  \mathcal{W} 
  &= -\ln \frac{\mathcal{Z}}{\mathcal{Z}_0}
  = -\ln \int
  e^{-S_\mathrm{eff}[\phi]}
  \mathcal{D}\phi,
  \\
  &= \frac{1}{2} \ln \det \mathfrak{S}
  = \frac{1}{2} \operatorname{tr} \ln \mathfrak{S}
  \label{eq:W_pre}
\end{align}
where we have subtracted the free field contribution $\mathcal{Z}_0$
\footnote{
  Note that the Casimir sum \eqref{eq:casimir_sum} itself is divergent because of the infinite summation $\displaystyle{\sum_{m=1}^{\infty}}[\ldots]$.
  In the derivation of the Lifshitz formula, we subtracted the free space mode density in Eq.~\eqref{eq:Delta_rho} so as to regularise the divergence.
  This subtraction corresponds to what we have done here.
}
and used a matrix identity $\ln \det \mathfrak{S} = \operatorname{tr} \ln \mathfrak{S}$ \cite{withers2010log}.
Note that we have performed the functional integral over $\phi$ in analogy with the multidimensional Gaussian integral \eqref{eq:Gaussian_int_0},
where we should find the following correspondence:
\begin{align}
  A \rightarrow -\mathfrak{S},
  \quad
  \bm{x} \rightarrow \phi.
\end{align}
Although we have obtained the formal expression of the free energy in Eq.~\eqref{eq:W_pre}, we cannot immediately evaluate the expression since the $\mathfrak{S}$ matrix is not diagonalised. 
We should also deal with the functional trace or determinant.

In the following, we approximately evaluate the free energy \eqref{eq:W_pre} and find the PFA expression.
Firstly, we split the surface profile into two parts:
the averaged part shown in \figref{fig:aux} and a deviation part,
\begin{align}
  a_\mathbf{x} = \overline{a}_\mathbf{x} + \Delta a_\mathbf{x},
\end{align}
and then expand the $\mathfrak{S}$ matrix in powers of the deviation parts $\Delta a$,
\begin{align}
  \mathfrak{S} 
  = \mathfrak{S}_{0} 
  + \mathfrak{S}_{1} 
  + \mathfrak{S}_{2} 
  + \cdots.
\end{align}
Accordingly, the free energy can be expanded as following:
\begin{align}
  \mathcal{W} 
  &= \frac{1}{2} \operatorname{tr} \ln
  \left[
    \mathfrak{S}_0
    (1 + \mathfrak{S}_0^{-1} \mathfrak{S}_1 + \mathfrak{S}_0^{-1} \mathfrak{S}_2 + \cdots)
  \right]
  \\
  &= \mathcal{W}_{0}
  + \mathcal{W}_{1} 
  + \mathcal{W}_{2} 
  + \cdots,
\end{align}
This corresponds to perturbative expansion in which the variation of the boundary profile $\Delta a$ is regarded as a perturbation.
The zero-th order term that does not contain $\Delta a$ can be evaluated as following:
\begin{align}
  \mathcal{W}_{0} 
  &= \frac{1}{2}
  \operatorname{tr} \ln \mathfrak{S}_{0}
  = \frac{1}{2}
  \ln \det 
  \begin{pmatrix}
    \mathfrak{S}_0^{\crT\crT} & \mathfrak{S}_0^{\crT\rT}\\
    \mathfrak{S}_0^{\rT\hspace{.1em}\crT} & \mathfrak{S}_0^{\rT\rT} 
  \end{pmatrix}
  \\
  &= \frac{1}{2}
  \ln \det \left[
    1-
    (\mathfrak{S}_0^{\crT\crT})^{-1}
    (\mathfrak{S}_0^{\crT\rT})
    (\mathfrak{S}_0^{\rT\rT})^{-1}
    (\mathfrak{S}_0^{\rT\hspace{.1em}\crT})
  \right]
  - \frac{1}{2}
  \ln \left( 
    \det \mathfrak{S}_0^{\crT\crT}
    \det \mathfrak{S}_0^{\rT\rT}
  \right)
  \label{eq:W_0_pre}
  \\
  &\equiv W_0 + \textrm{`self term'},
\end{align}
Note that $\mathfrak{S}_0^{\crT\crT}$, $\mathfrak{S}_0^{\rT\rT}$ and hence the second term in Eq.~\eqref{eq:W_0_pre} are independent of $\overline{a}_\mathbf{x}$,
which correspond to self terms.
This is why we subtract the second term in Eq.~\eqref{eq:W_0_pre} and evaluate the effective free energy $W_0$ below.

Inserting the identity operators,
$1 = \int \ket{\mathbf{k}}\bra{\mathbf{k}} \mathrm{d}\mathbf{k}$,
gives the Fourier transform of the $\mathfrak{S}_0$ operator,
\begin{align}
  \braket{\mathbf{x}|\mathfrak{S}_0|\mathbf{x}'}
  &= \iint
  \braket{\mathbf{x}|\mathbf{k}}
  \hspace{-.2em}
  \braket{\mathbf{k}|\mathfrak{S}_0|\mathbf{k'}}
  \hspace{-.2em}
  \braket{\mathbf{k}'|\mathbf{x}'}
  \mathrm{d}\mathbf{k} 
  \mathrm{d}\mathbf{k}',
  \\
  &= \iint
  e^{i\mathbf{k}\cdot\mathbf{x}}
  \braket{\mathbf{k}|\mathfrak{S}_0|\mathbf{k'}}
  e^{-i\mathbf{k}'\cdot\mathbf{x}'}
  \mathrm{d}\mathbf{k} 
  \mathrm{d}\mathbf{k}'.
  \label{eq:FT_S}
\end{align}
Comparing Eq.~\eqref{eq:FT_S} with the Fourier representation of the Feynman propagator \eqref{eq:D_F},
we can find the $\mathfrak{S}_0$ matrix is diagonal in the $\mathbf{k}$ space,
\begin{align}
  \mathfrak{S}_0 
  &=
  \int
    \Bigg| \mathbf{k} \Bigg\rangle
    \frac{1}{2|\mathbf{k}|}
    \begin{pmatrix}
      1 & e^{-|\mathbf{k}|\overline{a}_\mathbf{x}}\\
      e^{-|\mathbf{k}|\overline{a}_\mathbf{x}} & 1
    \end{pmatrix}
    \Bigg\langle \mathbf{k} \Bigg|
  \mathrm{d}\mathbf{k},
\end{align}
and obtain
\begin{align}
    W_0 
    &= \frac{1}{2} \operatorname{tr}
    \ln \left[
      1-
      (\mathfrak{S}_0^{\crT\crT})^{-1}
      (\mathfrak{S}_0^{\crT\rT})
      (\mathfrak{S}_0^{\rT\rT})^{-1}
      (\mathfrak{S}_0^{\rT\hspace{.1em}\crT})
    \right]
    = \frac{1}{2} \operatorname{tr} 
    \ln \left[
      \int
        (1 - e^{-2 |\mathbf{k}| \overline{a}_\mathbf{x}})
      \ket{\mathbf{k}}
      \hspace{-.2em}
      \bra{\mathbf{k}}
      \mathrm{d}\mathbf{k}
    \right],
    \\
    &= \frac{1}{2}\operatorname{tr} 
    \int
    \left[
    \ln 
    \left(
      1 - e^{-2 |\mathbf{k}| \overline{a}_\mathbf{x}}
    \right)
    \right]
    \ket{\mathbf{k}}
    \hspace{-.2em}
    \bra{\mathbf{k}}
    \mathrm{d}\mathbf{k}.
    \label{eq:PFA_pre0}
\end{align}
In Eq.~\eqref{eq:PFA_pre0},
we can straightforwardly evaluate the functional trace,
\begin{align}
  W_0
    &= \int 
    \Bigg\langle \mathbf{x} \Bigg|\ 
      \frac{1}{2} \int
      \ln \left(
        1 - e^{-2 |\mathbf{k}| \overline{a}_\mathbf{x}}
      \right)
      \ket{\mathbf{k}}
      \hspace{-.2em}
      \bra{\mathbf{k}}
      \mathrm{d}\mathbf{k}\ 
    \Bigg| \mathbf{x} \Bigg\rangle
    \mathrm{d}\mathbf{x}
    = \frac{1}{2} \iint 
    \ln \left(
      1 - e^{-2 |\mathbf{k}| \overline{a}_\mathbf{x}}
    \right)
    \braket{\mathbf{x}|\mathbf{k}}
    \hspace{-.2em}
    \braket{\mathbf{k}|\mathbf{x}}
    \mathrm{d}\mathbf{k}
    \mathrm{d}\mathbf{x},
    \notag \\
    &= \int \left[
      \frac{1}{2} 
      \int \ln \left(
        1 - e^{-2 |\mathbf{k}| \overline{a}_\mathbf{x}}
      \right)
      \mathrm{d}\mathbf{k}
    \right]
    \mathrm{d}\mathbf{x}.
\end{align}

We have had two variables, $\overline{a}$ and $\Delta a$, so far.
What we would like to have is an expression that contains only $a$ instead of $\overline{a}$ and $\Delta a$.
We can formally substitute $a$ into $\overline{a}$ in order to have such expression.
The zeroth-order term becomes
\begin{align}
  W_0
  &= 
  \int
  \left[
    \frac{1}{2} \int
    \ln
    (1-e^{-2 |\mathbf{k}| a_\mathbf{x}})
    \mathrm{d}\mathbf{k}
  \right]
  \mathrm{d}\mathbf{x}.
  \label{eq:PFA_pre}
\end{align}
The quantity inside the square bracket corresponds to the Lifshitz formula,
which returns the zero-point energy per unit area stored in a vacuum gap with width $a_\mathbf{x}$ between two perfect mirrors.
Therefore,
Eq.~\eqref{eq:PFA_pre} means that we can evaluate the Casimir energy between the flat surface $z=0$ and the corrugated one $z=a_\mathbf{x}$ by the following procedures:\\
(1) Compute the Casimir energy between planar patches separated by a distance $a_\mathbf{x}$.\\
(2) Sum up the contributions from each patch pair at $\mathbf{x}$.\\
As we have seen the equivalence of the Lifshitz formula and the Casimir sum in the previous chapter,
we can substitute the Casimir sum into the square bracket in Eq.~\eqref{eq:PFA_pre} to obtain
\begin{align}
  W_0
  = \int
  \left[
    \frac{\hbar}{2}
    \sum_{\vec{k}_\parallel}
    \omega_{k_\parallel}(a_\mathbf{x})
  \right]
  \mathrm{d}\mathbf{x}.
  \label{eq:PFA}
\end{align}
This is nothing but the PFA formula \cite{bordag2009advances}.

\section{Casimir-induced instabilities}
We are interested in whether the very shallow corrugation at one interface of a metallic thin film further lowers the Casimir energy and leads to the structural instability of the film,
thereby we regard the corrugation as perturbation and safely use the proximity force approximation.

When performing the proximity force calculation,
it is convenient to compute the Casimir sum directly.
On the other hand,
Lifshitz formula is better to be extended if we have the scattering matrix of the corrugated system.

\begin{figure}[htbp]
  \centering
  \includegraphics[width=.6\linewidth]
  {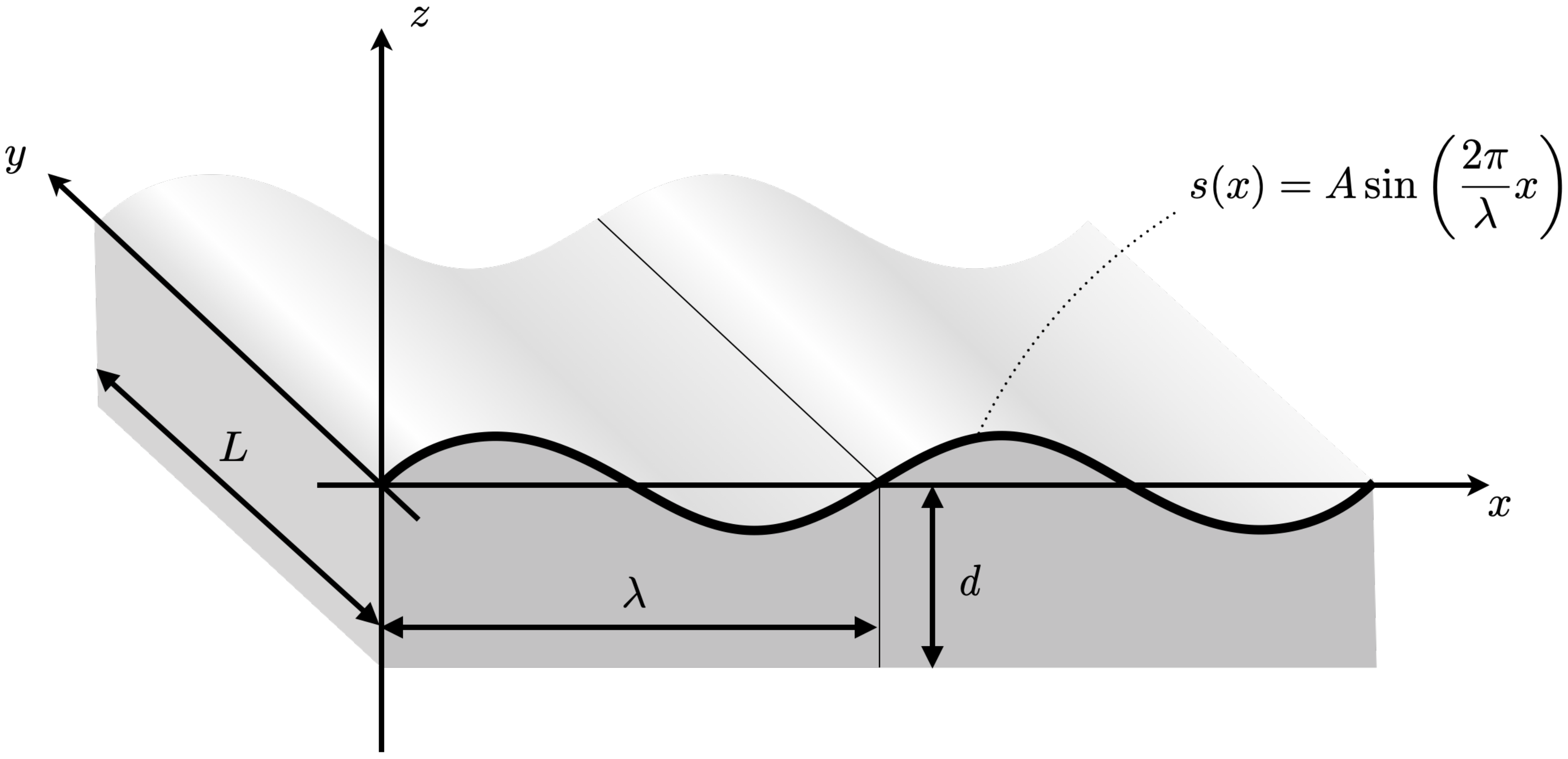}
  \caption{
    Schematic image of a corrugated metallic film.
  }
  \label{fig:corrugated_plate}
\end{figure}

In \figref{fig:corrugated_plate}, the setup studied in the following is shown.
We consider a metallic film whose interface is corrugated.
Here, we consider the corrugation of sinusoidal type,
whose profile reads
\begin{align}
  a(x) &= A \sin \left(\frac{2\pi}{\lambda} x \right), \label{eq:corrugation} \\
  a'(x) &= \left(\frac{2\pi}{\lambda}A\right) \cos \left(\frac{2\pi}{\lambda} x \right), 
  \label{eq:corrugation_derivative}
\end{align}
where $A$ and $\lambda$ are the corrugation depth and period, respectively.

In this section,
we investigate whether the corrugation reduces the Casimir energy or not.
As we have seen in Sec.~\ref{sec:sca_trans_mat},
the dispersion relation is split into two branches if two interfaces are close to each other.
We can expect the change of the zero-point energy due to this splitting.
Let us evaluate it here within the quasistatic approximation 
(see Appendix \ref{app:quasistatic} for this approximation),
\begin{align}
  \omega_\mathrm{odd}
  &\simeq \omega_\mathrm{sp}\sqrt{1 + e^{-\kpara d}},
  \quad
  \omega_\mathrm{even}
  \simeq \omega_\mathrm{sp}\sqrt{1 - e^{-\kpara d}}.
  \label{eq:omega_quasistatic}
\end{align}
The change of the zero-point energy can be evaluated as following:
\begin{align}
  \Delta U
  &=
  \int_{-\infty}^\infty \int_{-\infty}^\infty 
  \Bigg[
    \Bigg(
      \cfrac{\hbar \omega_\mathrm{odd}}{2} 
      -
      \cfrac{\hbar \omega_\mathrm{sp}}{2}
    \Bigg) 
    +
    \Bigg(
    \cfrac{\hbar \omega_\mathrm{even}}{2}
    -
    \cfrac{\hbar \omega_\mathrm{sp}}{2} 
    \Bigg) 
  \Bigg]
  \mathrm{d}k_x \mathrm{d}k_y,
  \\
  &= \frac{\hbar}{2}
  \int_0^{2\pi} \int_0^\infty
  (\omega_\mathrm{odd} + \omega_\mathrm{even} - 2\omega_\mathrm{sp})
  \kpara \mathrm{d}\kpara
  \mathrm{d}\theta,
  \label{eq:DU_pre}
  \\
  &\simeq 
  \cfrac{\hbar \omega_\mathrm{sp}}{2}
  \int_0^{2\pi} \int_0^\infty 
  \Bigg(
    \cfrac{e^{-\kpara d}}{2} 
    - \cfrac{e^{-\kpara d}}{2} 
    - \cfrac{e^{-2\kpara d}}{8}
  \Bigg)
  \kpara \mathrm{d}\kpara 
  \mathrm{d} \theta,\\
  &= -\frac{\pi \hbar \omega_\mathrm{sp}}{32 d^2}.
  \label{eq:DU}
\end{align}
Note that we have changed the integral variables, 
$(k_x, k_y) = (k\cos \theta, k\sin \theta)$,
and the corresponding measure is
\begin{align}
  \mathrm{d}k_{x,y}
  &= \cfrac{\partial k_{x,y}}{\partial k} \mathrm{d}k + \cfrac{\partial k_{x,y}}{\partial \theta} \mathrm{d}\theta,
  \quad 
  \mathrm{d}k_x\mathrm{d}k_y 
  = 
  \Bigg(
    \cfrac{\partial k_{x}}{\partial k}\cfrac{\partial k_{y}}{\partial \theta} 
    -
    \cfrac{\partial k_{y}}{\partial k} \cfrac{\partial k_{x}}{\partial \theta} 
  \Bigg)
  \mathrm{d}k \mathrm{d}\theta
  = k\mathrm{d}k\mathrm{d}\theta.
\end{align}
Note also that we have expanded the square root up to the second order in $e^{-\kpara d}$,
\begin{align}
  \sqrt{1 \pm e^{-\kpara d}} 
  \simeq 
  1 \pm \cfrac{e^{-\kpara d}}{2} - \cfrac{e^{-2\kpara d}}{8}.
\end{align}
Equation \eqref{eq:DU} implies the splitting of the surface plasmon modes lowers the zero-point energy.
This is because the plasmon dispersion splitting is asymmetric due to the second-order time derivative in the wave equation \cite{ding2021casimir}.
When we apply the Fourier transform, we obtain $\omega^2$ from the second order time derivative,
which results in the square-root dispersion,
\begin{align}
  \omega_\mathrm{sg} 
  \simeq \omega_\mathrm{sp}
  \rightarrow
  \omega_\mathrm{odd,even} 
  \simeq \omega_\mathrm{sp}\sqrt{1 \pm e^{-\kpara d}}. 
\end{align}
By expanding the square root,
we can indeed find that the integrand in Eq.~\eqref{eq:DU_pre} is negative for each $\kpara$, 
\begin{align}
  &(\omega_\mathrm{odd} - \omega_\mathrm{sg})
  +
  (\omega_\mathrm{even} - \omega_\mathrm{sg})
  \simeq
  \omega_\mathrm{sp}
  \bigg[
  \Big(\sqrt{1 + e^{-\kpara d}} - 1\Big)
  +
  \Big(\sqrt{1 - e^{-\kpara d}} - 1\Big)
  \bigg]
  \notag \\
  &=
  \frac{1}{2}
  \omega_\mathrm{sp}
  \Bigg(
    \Big[
    (+e^{-\kpara d})
    +
    (-e^{-\kpara d})
  \Big]
    +
    \sum_{n=2}
    \frac{(2n-3)!!}{n!} 
    \bigg(\frac{-1}{2}\bigg)^{n-1}
    \Big[
      (+e^{-\kpara d})^n
      +
      (-e^{-\kpara d})^n
    \Big]
  \Bigg)
  \notag \\
  &= -\omega_\mathrm{sp}
    \sum_{n=1}
    \frac{(4n-3)!!}{2n!} 
    \bigg(\frac{1}{2}\bigg)^{2n-1}
    e^{-2n\kpara d}.
\end{align}
Note that the odd-order terms do not contribute to the change and the even-order terms are always negative for each $\kpara$.

Since the change of the zero-point energy $\Delta U$ has the inverse square dependence on the film thickness $d$, we can expect further lowering of the energy by introducing the corrugation.
In the following, we use the proximity force approximation and investigate how much the zero-point energy decreases or increases by the corrugation.

\begin{figure}[htbp]
  \centering
  \includegraphics[width=.6\linewidth]
  {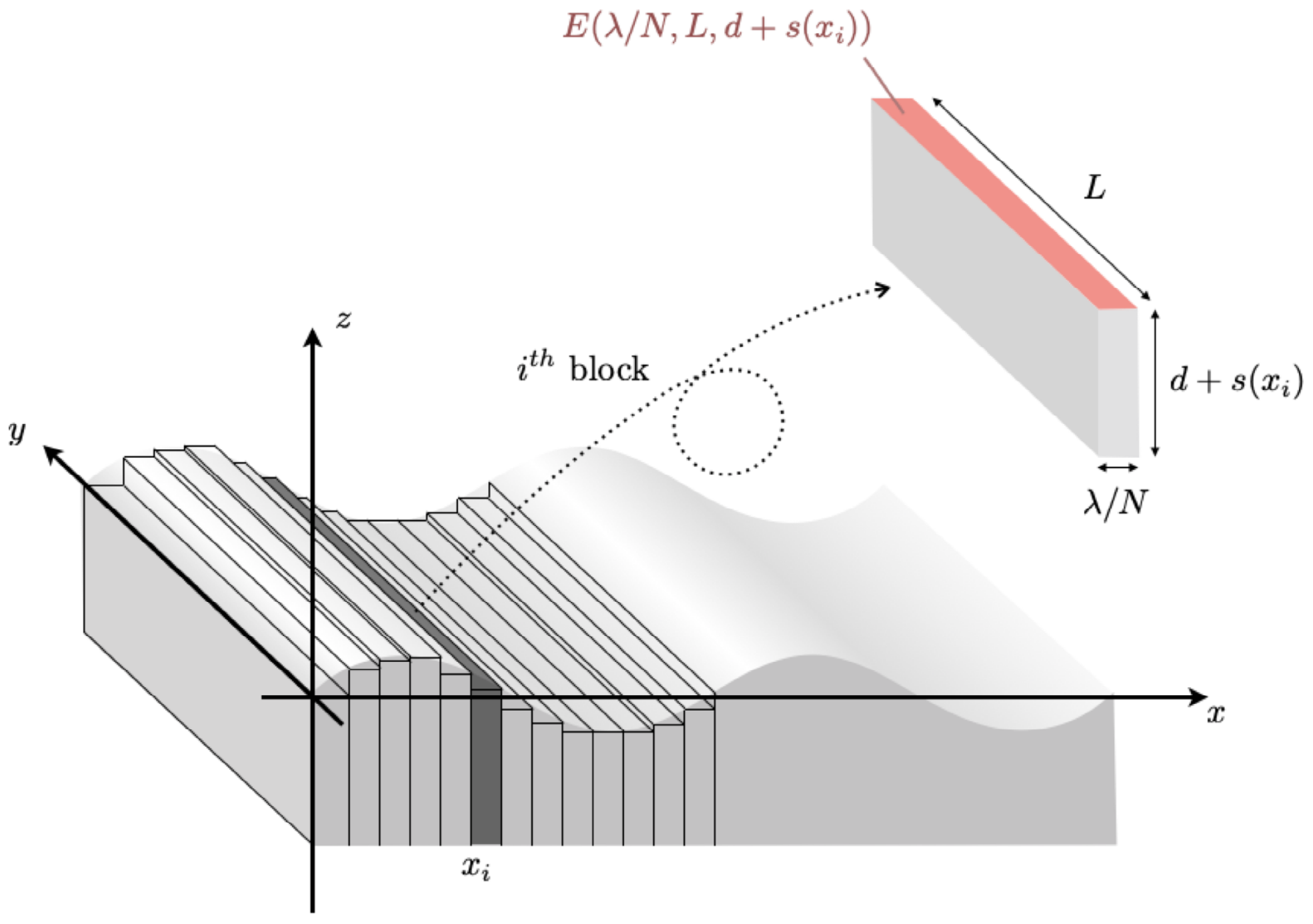}
  \caption{
    Proximity force approximation for the corrugated metallic film.
    One unit cell ($0 \leq x \leq \lambda$) of the film is replaced by $N$ blocks.
    The height of the $i$th box is given by $d+a(x_i)$,
    and the width is $\lambda/N$.
    We evaluate the Casimir energy at each box and then sum up contributions from all boxes.
  }
  \label{fig:pfa}
\end{figure}

Let us replace the corrugated film with a set of blocks as shown in \figref{fig:pfa} and evaluate the Casimir energy at each block.
This replacement corresponds to the proximity force approximation.
Considering the mode density $\rho(\kpara)$ at each block, we can rewrite the summation evaluating the Casimir energy at each block into an integral form,
\begin{align}
  \sum_{\kpara}
  \frac{\hbar\omega_i(\kpara)}{2}
  &\rightarrow
  \frac{\hbar}{2}
  \iint 
  \rho_i(\kpara)
  \omega_i(\kpara)
  \frac{\mathrm{d}k_x}{2\pi} 
  \frac{\mathrm{d}k_y}{2\pi},
  \label{eq:sum->int}
\end{align}
where we defined the mode density $\rho_i(\kpara)$ and the dispersion $\omega_i$ at the $i$th block.

In general, 
the mode density in a box $\rho(\kpara)$ can be evaluated by imposing boundary conditions.
Here, we choose the periodic boundary conditions,
\begin{align}
  \vec{\mathcal{E}}(x=0) &= \vec{\mathcal{E}}(x=L_x),
  \quad
  \vec{\mathcal{H}}(x=0) = \vec{\mathcal{H}}(x=L_y),
  \\
  \vec{\mathcal{E}}(y=0) &= \vec{\mathcal{E}}(y=L_x),
  \quad
  \vec{\mathcal{H}}(y=0) = \vec{\mathcal{H}}(y=L_y),
\end{align}
which discretise the wavenumbers in the $x$ and $y$ directions,
\begin{align}
  k_{x,y}
  &= \cfrac{2\pi n_{x,y}}{L_{x,y}},
  \quad
  n_{x,y} = \pm 1, \pm 2,\cdots, 
  \\
  {\kpara}^2
  &=
  \frac{4\pi^2}{L_x L_y}\left(n_x^2+n_y^2\right).
\end{align}
The number of modes within an infinitesimal wavenumber interval $[\kpara, \kpara+\mathrm{d}\kpara]$ is equivalent to the number of cells with a size $(2\pi/L_x) \times (2\pi/L_y)$ in the annular area,
\begin{align}
  \rho(\kpara) \mathrm{d}\kpara
  &= 
  \frac{\pi (\kpara+\mathrm{d}\kpara)^2 - \pi \kpara^2}{(2\pi/L_x) \times (2\pi/L_y)}
  \simeq 
  \frac{L_x L_y}{2\pi}
  \kpara\mathrm{d}\kpara. 
  \label{eq:mode_density}
\end{align}
We use Eq.~\eqref{eq:mode_density} to evaluate the Casimir energy at each block \eqref{eq:sum->int}.

The zero-point energy per unit cell ($0 \leq x \leq \lambda,\ 0 \leq y \leq L$) can be evaluated by summing up the contribution from each block,
\begin{align}
  U_\mathrm{sp}[a(x), d] 
  &= \sum_{i=0}^{N-1} 
  \frac{\hbar}{2}
  \iint
  \rho_i(\kpara)
  (\omega_{i;\hspace{.2em} \mathrm{odd}}
  +\omega_{i;\hspace{.2em} \mathrm{even}})
  \frac{\mathrm{d}k_x}{2\pi}
  \frac{\mathrm{d}k_y}{2\pi}
  \label{eq:sum_pfa}
  \\
  &= 
  \sum_{i=0}^{N-1} 
  \frac{\hbar \omega_\mathrm{sp}}{2}
  \frac{\lambda}{N}L 
  \int_0^{\infty} 
  \left( \sqrt{1+e^{-\kpara(d+a(x_i))}} + \sqrt{1-e^{-\kpara(d+a(x_i))}} \right)
  \kpara \mathrm{d}\kpara,
  \\
  &= \frac{\hbar \omega_\mathrm{sp}}{2}L
  \int_{0}^{\lambda}
  \int_0^{\infty} 
  \left( \sqrt{1+e^{-k(d+a(x))}} + \sqrt{1-e^{-k(d+a(x))}} \right)
  \kpara \mathrm{d}\kpara
  \mathrm{d}x,
\end{align}
where $\omega_{i;\hspace{.2em} \mathrm{odd,even}}$ are the odd and even mode dispersion in the $i$th block.

Subtracting the zero-point energy in the case of no corrugation, we define the change of zero-point energy in a unit cell,
\begin{align}
  \Delta U_\mathrm{sp}
  &=
  U_\mathrm{sp}[a(x),d] - U_\mathrm{sp}[0,d].
  \label{eq:DU_sp}
\end{align}
Let us expand the energy in powers of the corrugation depth $A$,
\begin{align}
  \Delta U_\mathrm{sp} 
  &= 
  \Gamma_\mathrm{sp}^{(1)} A
  + \Gamma_\mathrm{sp}^{(2)} A^2 
  + \cdots.
  \label{eq:DU_expanded}
\end{align}
The first order coefficient vanishes, $\Gamma_\mathrm{sp}^{(1)} = \partial U_\mathrm{sp}/\partial A = 0$, because our system configuration is invariant under $A \mapsto -A$.
Thus, the leading contribution is the second order, 
\begin{align}
  \Gamma_\mathrm{sp}^{(2)} 
  &= \frac{\partial^2 U_\mathrm{sp}}{\partial A^2} \Bigg|_{A=0}, \\
  &= -\hbar \omega_\mathrm{sp}\int_0^{\infty} 
  \left[ 
    \kpara^3\left( \frac{e^{-\kpara d}}{2(1 - e^{-\kpara d})} + 1 \right)
    \frac{e^{-\kpara d}}{2 \sqrt{1 - e^{-\kpara d}}} 
    \int_0^{\lambda} 
    \sin^2 \left(\frac{2 \pi x}{ \lambda} \right) 
    \mathrm{d}x
  \right] 
  \mathrm{d}\kpara, 
  \\
  &= -\frac{\hbar \omega_\mathrm{sp}}{2}
  \int_0^{\infty}
  \left[
    \kpara^3 \left( \frac{e^{-\kpara d}}{2(1 - e^{-\kpara d})} + 1 \right) 
    \frac{e^{-\kpara d}}{2 \sqrt{1 - e^{-\kpara d}}}
  \right]
  \mathrm{d}\kpara.
\end{align}
This is the Casimir effect contribution to the decrease of system energy by introducing corrugation.

On the other hand, introducing the corrugation raises the surface tension energy.
The increase of the surface tension energy in a unit cell is given by
\begin{align}
  \Delta U_\mathrm{sf}[a(x)] 
  = \gamma_\mathrm{sf} L \Delta S[a(x)],
\end{align}
where we have introduced the arc length
\begin{align}
  S[a(x)]
  &= 
  \int_0^\lambda
  \sqrt{1 + \{a'(x)\}^2}
  \mathrm{d}x,
  \\
  \Delta S[a(x)]
  &=
  S[a(x)] - S[0],
\end{align}
so that $L \times \Delta S[a(x)]$ is the increase of the surface area in a unit cell,
and $\gamma_\mathrm{sf}$ is the surface tension coefficient. 
Let us expand the surface tension energy change in powers of the corrugation depth $A$ as we did for the zero-point energy,
\begin{align}
  \Delta U_\mathrm{sf}[a(x)] 
  &=
  \Gamma_\mathrm{sf}^{(1)} A
  + \Gamma_\mathrm{sf}^{(2)} A^2
  + \cdots.
\end{align}
Again, the first order coefficient vanishes, $\partial U_\mathrm{sf}/\partial A = 0$,
and the leading term is the second order,
\begin{align}
  \Gamma_\mathrm{sf}^{(2)} 
  &= \frac{\partial^2 U_\mathrm{sf}}{\partial A^2} \Bigg|_{A=0}
  = \gamma_\mathrm{sf}L \frac{\partial^2 S}{\partial A^2} \Bigg|_{A=0}
  = \frac{4 \pi^2}{\lambda^2}\gamma_\mathrm{sf}L 
  \int_0^\lambda
  \cos^2 \left(\frac{2 \pi x}{\lambda}\right)
 \mathrm{d}x
  = \frac{2 \pi^2}{\lambda} \gamma_\mathrm{sf} L.
\end{align}

\begin{figure}[htbp]
  \centering
  \includegraphics[width=\linewidth]
  {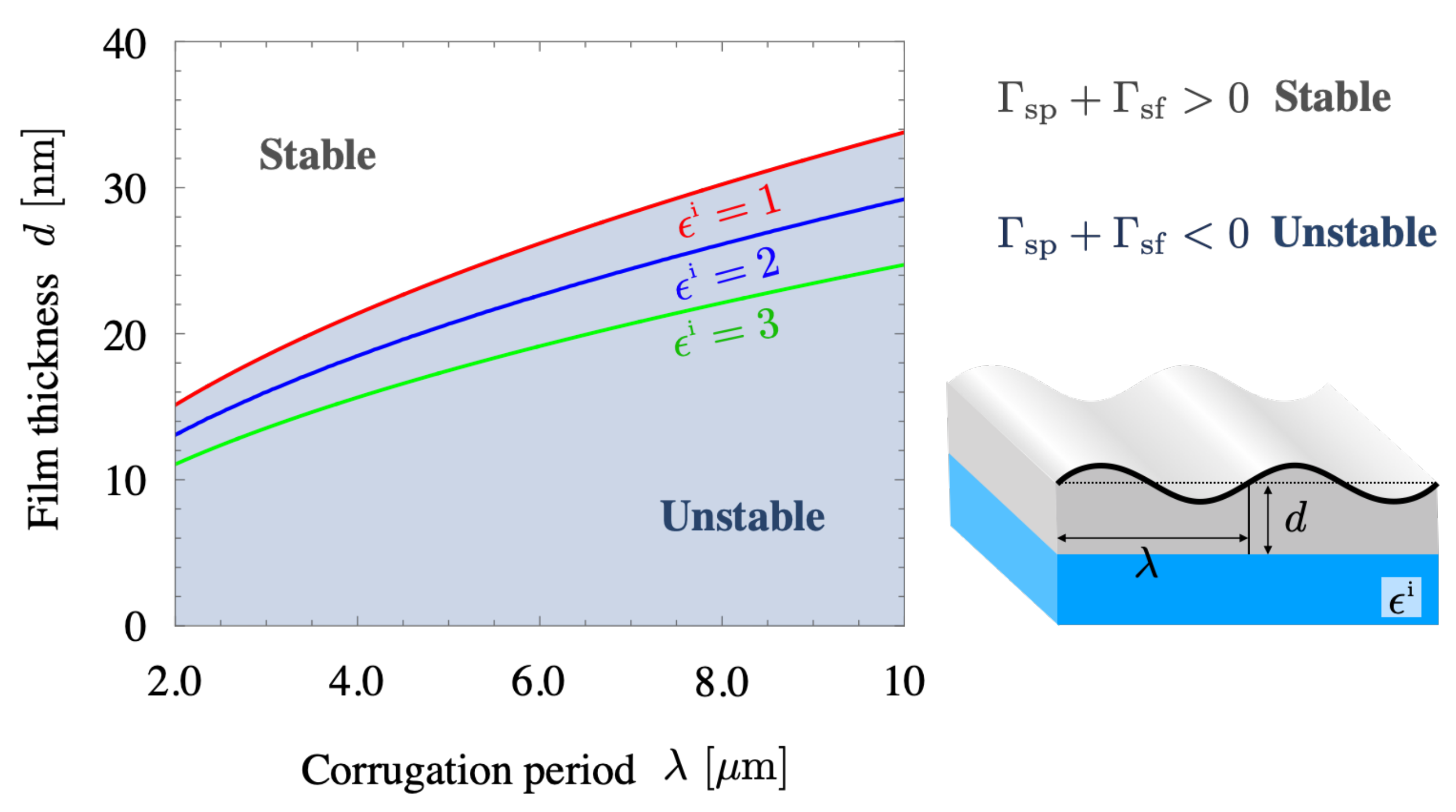}
  \caption{
    The dependence of $\Gamma_\mathrm{sp}^{(2)} + \Gamma_\mathrm{sf}^{(2)} = 0$ lines for various substrates 
    ($\epsilon^{\scriptscriptstyle \mathrm{i}} = 1.0, 2.0, 3.0$)
    on the thickness $d$ and corrugation period $\lambda$.
    The inset is our corrugated film located on a substrate with a permittivity $\epsilon^{\scriptscriptstyle \mathrm{i}}$.
  }
  \label{fig:stable_unstable}
\end{figure}

Therefore, the net change of film energy is given by
$
\Gamma_\mathrm{sp}^{(2)}
+
\Gamma_\mathrm{sf}^{(2)}
$.
In \figref{fig:stable_unstable}, regions where this quantity is positive and negative are shown.
We adopt mercury as the film material. 
The plasma frequency of mercury 
$\omega_\mathrm{p} = 6.83\ \mathrm{eV}$ 
and the surface tension coefficient 
$\gamma_\mathrm{sf} = 27.6\ \mathrm{meV/{\AA}^2}$ are taken from \cite{kim1995mercury}.
In the positive region,
introducing the corrugation raises the net energy of the film so that the film is robust to the surface corrugation and stable as it is.
On the contrary, in the negative region, the corrugation lowers the net energy.
This means the energy decrease as the corrugation depth $A$ increase so that the film is structurally unstable.
The zero line $\Gamma_\mathrm{sp}^{(2)} + \Gamma_\mathrm{sf}^{(2)} = 0$ represents the critical thickness.
Once the film crosses this line, the corrugated surface is preferred over the flat one.

\chapter{Conclusion of Part I}
\label{ch:conclusion_casimir}
To sum up, we investigated the structural instabilities of a metallic film induced by the Casimir effect in Part I.
The Casimir effect is attraction of two interfaces induced by the zero-point fluctuations of electromagnetic fields.
We began with the \textit{microscopic} Maxwell--Heaviside equations that determine the dynamics of electromagnetic fields with electric polarisation and magnetisation.
Since we are interested in the metallic films where electric response dominates the magnetic one,
we adopted the Drude--Lorentz model,
which is a simple model often used in the plasmonics community to describe the electric response of an medium.
From the Drude--Lorentz model,
we obtained the relationship between the electric polarisation and the electric field and wrote closed systems of equations,
the \textit{macroscopic} Maxwell--Heaviside equations.

We used the transfer matrix approach to calculate the light scattering (reflection and transmission) by the metallic film.
The transfer matrix is derived from the macroscopic Maxwell--Heaviside equations and has information of the field propagation and boundary conditions.
Multiplying $n$ transfer matrices together, we can calculate the reflection from a system with $n$ interfaces.
We calculated the poles of the reflection coefficient to find eigenmodes in the thin film system.
The degenerated plasmon modes localised at upper and lower interfaces of the film are hybridised, forming even and odd parity plasmon modes.

According to the Casimir prescription,
we summed up the zero-point energy of each plasmon mode,
whose eigenfrequency is evaluated in the modal analysis,
to find net zero-point energy in the system.
We revealed that decreasing the film thickness causes the lowering of the net zero-point energy.
The amount of the energy decrease has the inverse square dependence on the thickness.
From this result, one would expect further lowering of the zero-point energy by introducing small corrugation on the film surface.
The energy decrease makes the growth of the corrugation energetically favorable.
On the other hand, introducing corrugation increases the surface area of the film, and thus the surface tension energy,
which makes the growth of the corrugation energetically unfavorable.
We used the proximity force approximation to evaluate the zero-point energy and compared it with the surface tension energy.
We found a critical thickness above which the zero-point contribution overcomes the surface tension contributions,
making the growth of the corrugation energetically favorable.
This means the film becomes unstable with small corrugation if it is thinner than the critical thickness. 


\part{\v{C}erenkov radiation from a dynamical grating}
\chapter{\v{C}erenkov radiation}
\label{ch:cerenkov}
\section{Light emission from a swift particle}
\label{sec:light_emission}
\begin{figure}[htbp]
  \centering
  \includegraphics[width=.8\linewidth]{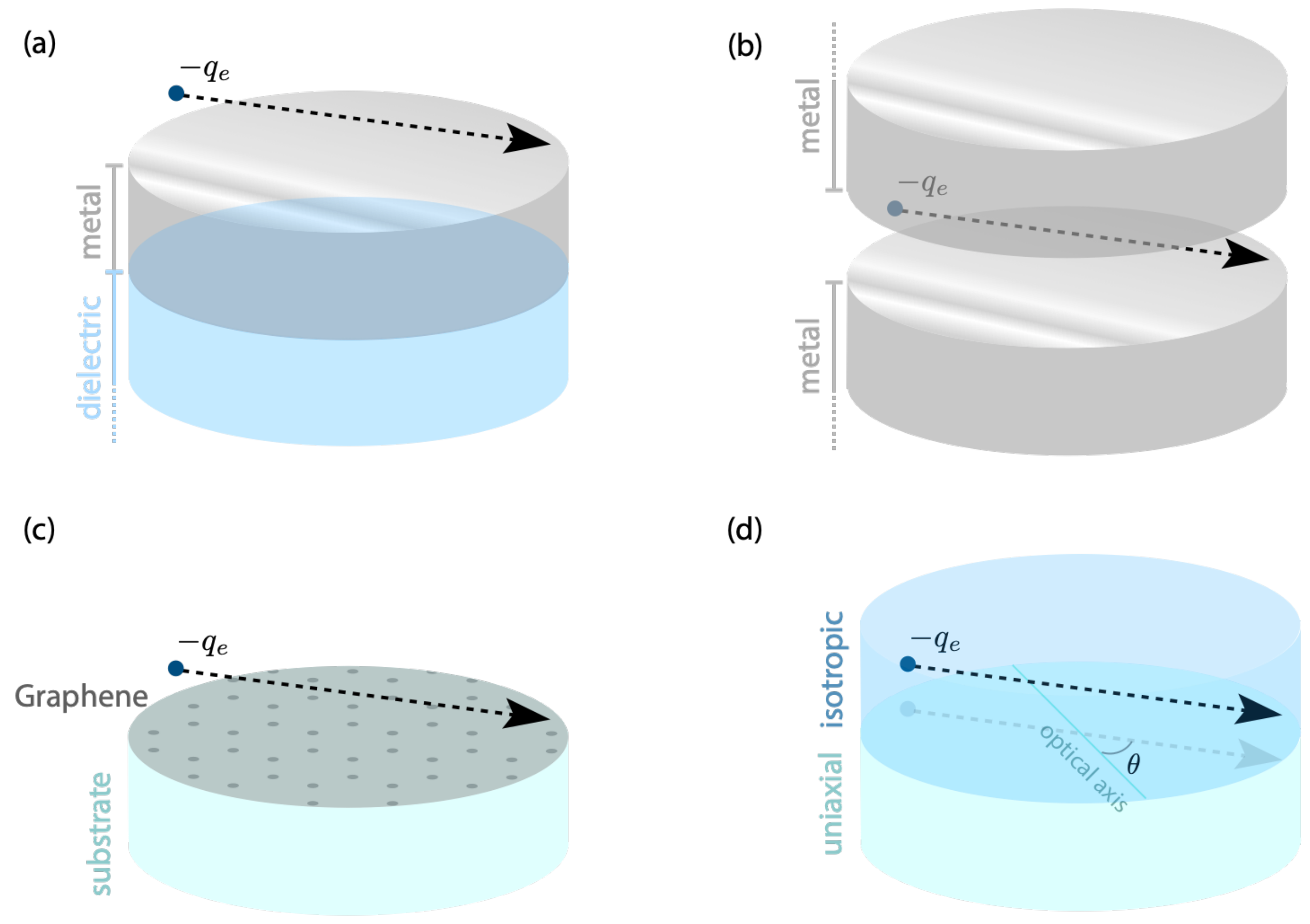}
  \caption{
    Surface modes emission by charged particles travelling near surfaces and interfaces due to the \v{C}erenkov mechanism has extensively been discussed.
    (a) Surface plasmon on a metallic film, (b) Surface plasmon in a Fabry-P\'{e}rot cavity, (c) Graphene plasmon and (d) Dyakonov wave can be generated by swift electrons.
  }
  \label{fig:surface-cerenkov}
\end{figure}

\v{C}erenkov radiation is radiation emitted by a charged particle travelling at a velocity greater than the phase velocity of electromagnetic waves in the medium.
It was first observed in a refractive medium by \v{C}erenkov in 1934 \cite{cherenkov1934visible} and then theoretically interpreted by Frank and Tamm in 1937 \cite{frank1937coherent}. 
\v{C}erenkov, Frank and Tamm shared the Nobel prize in physics 1958 
\footnote{
  Vavilov, who supervised \v{C}erenkov and was the co-discoverer of the radiation, was not included because he died in 1951.
  The radiation is also called Vavilov--\v{C}erenkov radiation to give him credit for the co-discovery 
  \cite{%
    frank1984vavilov,%
    ginzburg1996radiation%
  }.
}.
Since it was first observed, \v{C}erenkov radiation has been observed in various systems:
in photonic crystals and metamaterials,
the threshold of the radiation is modified due to the photonic band structure and negative refractive index
\cite{%
  antipov2008observation,%
  xi2009experimental,%
  de2010optical,%
  veselago1967electrodynamics,%
  pendry1994energy,%
  lu2003vcerenkov,%
  szczepkowicz2020frequency%
}.
Analogous radiation in the presence of surfaces and interfaces is called Smith--Purcell radiation
\cite{%
  smith1953visible,%
  pendry1994energy,%
  de2004boundary,%
  szczepkowicz2020frequency%
}.
In the presence of surfaces or interfaces, there are also emissions of surface waves such as surface plasmon polaritons 
\cite{%
  de2004boundary,%
  liu2012surface,%
  tao2016reverse%
},
graphene plasmons
\cite{%
  fares2019quantum%
}
and Dyakonov waves 
\cite{%
  hao2020surface%
}
can be emitted by the same mechanism (\figref{fig:surface-cerenkov}).
Even uncharged particles emit \v{C}erenkov radiation if it is polarised
\cite{%
  frank1984vavilov,%
  ginzburg1996radiation,%
  milton2020self%
},
which is an origin of the quantum friction
\cite{%
  pendry1997shearing,%
  pendry1998can,%
  pendry2010quantum,%
  manjavacas2010vacuum,%
  maghrebi2013quantum,%
  milton2016reality,%
  farias2019motion%
}.
In the nonlinear regime,
\v{C}erenkov radiation is emitted from moving light foci, which induce polarisation in media
\cite{%
  auston1984cherenkov,%
  bakunov2009cherenkov,%
  smith2016steerable%
}
or optical solitons in optical fibres
\cite{%
  cao1994soliton,%
  akhmediev1995cherenkov,%
  chang2010highly%
}
and micro resonators
\cite{%
  brasch2016photonic,%
  cherenkov2017dissipative,%
  vladimirov2018effect%
}.

Thanks to its availability in various systems, \v{C}erenkov radiation is applied in the broad research areas: it has been utilised for the cosmic ray detection at super-Kamiokande and pioneered the high energy physics \cite{fukuda2003super} and for an optical frequency comb and a tunable and broadband light source in metrology and spectroscopy
\cite{%
  tu2009optical,%
  skryabin2010colloquium,%
  brasch2016photonic%
}.

Since nothing can physically travel faster than light in vacuum, \v{C}erenkov emission in vacuum calls for a background field such as external strong electromagnetic and Chern--Simons fields,
by which the speed of light in a vacuum effectively decreases,
\cite{%
  kaufhold2007vacuum,%
  macleod2019cherenkov,%
  lee2020cherenkov,%
  artemenko2020quasiclassical%
}.
Alternatively, we can make charge density or polarisation pattern on a surface virtually travel faster than light to trigger \v{C}erenkov radiation as in the case of moving light foci
\cite{%
  bakunov2009cherenkov,%
  smith2016steerable%
}.
Inspired by these studies, we discuss making use of a spatiotemporally modulated surface (a kind of dynamical metasurfaces) to generate a \v{C}erenkov source virtually travelling faster than light in the following chapters.

\section{Frank--Tamm formula}
\label{sec:frank-tamm}

Frank and Tamm studied the radiation from a charged particle travelling at a constant velocity in 1937,
which is also known as the Tamm problem \cite{frank1937coherent}.
They calculated the radiation power by integrating the Poynting vector over a cylinder enclosing the particle trajectory,
\begin{align}
  W
  &=
  \oiint
  \left(
    \int_{-\infty}^{+\infty}
    \vec{\mathcal{E}}_t \times \vec{\mathcal{H}}_t
    \mathrm{d}t
  \right)
  \cdot
  \mathrm{d}\vec{S},
  \\
  &=  
  8\pi\oiint
  \left(
    \int_{0}^{+\infty}
    \frac{1}{2}
    \operatorname{Re}
    \left[
      \vec{E}_\omega^*
      \times
      \vec{H}_\omega
    \right]
    \mathrm{d}\omega
  \right)
    \cdot
  \mathrm{d}\vec{S},
  \label{eq:W}
\end{align}
where the quantity in the parentheses represents the time average.
Note that the integrand in Eq.~\eqref{eq:W} is equivalent to the conventional definition of the time-averaged Poynting vector \cite{%
  jackson1999classical,%
  berry2009optical,%
  bekshaev2011internal,%
  bliokh2014extraordinary,%
  bekshaev2015transverse%
}
The electromagnetic field from an electron moving with a constant velocity $\beta c\ (0 \leq \beta \leq 1)$ in a medium with a refractive index $n$ is given in terms of the vector and scalar potentials in the cylindrical coordinate system via the Lorentz boost 
\cite{%
  pendry1994energy,%
  szczepkowicz2020frequency%
},
\begin{align}
  \vec{A}_\omega
  &= -\frac{q_e \mu_0}{(2\pi)^2}
  \mathcal{K}_0
  \left(
    i\frac{\omega}{c}r\sqrt{n^2 - \frac{1}{\beta^2}}
  \right)
  e^{-i (\omega/\beta c) z}
  \vec{u}_z
  \\
  \Phi_\omega
  &= \frac{c}{n^2 \beta}
  \vec{u}_z \cdot \vec{A}_\omega,
\end{align}
where $\mathcal{K}_0$ is the modified Bessel function of the second kind.
Using the asymptotics of the modified Bessel function in the far field region
($r\rightarrow \infty$),
\begin{align}
  \mathcal{K}_0(ix)
  &\sim
  \sqrt{\frac{\pi}{2x}}
  e^{-i (x + \pi/4)},
  \quad
  x = \frac{\omega}{c}r\sqrt{n^2 - \frac{1}{\beta^2}},
\end{align}
we can evaluate the electric and magnetic fields,
\begin{align}
  \vec{E}_\omega
  &= 
  -i\omega \vec{A}_\omega 
  -
  \left(
    \vec{u}_r \frac{\partial}{\partial r} 
    +
    \vec{u}_z \frac{\partial}{\partial z}
  \right)
  \Phi_\omega
  \\
  \vec{H}_\omega 
  &= \mu_0^{-1} 
  \left(
    \vec{u}_r \frac{\partial}{\partial r} 
    +
    \vec{u}_z \frac{\partial}{\partial z}
  \right)
  \times 
  \vec{A}_\omega
  = \mu_0^{-1} 
  \vec{u}_r 
  \times
  \frac{\partial \vec{A}_\omega}{\partial r}, 
\end{align}
and the time-averaged power flow in the radial direction,
\begin{align}
  \vec{u}_r
  \cdot
  \frac{1}{2}
  \operatorname{Re}
  \left[
    \vec{E}_\omega^*
    \times
    \vec{H}_\omega
  \right]
  = 
  \cfrac{\mu_0 q_e^2\omega}{64 \pi^3 r_c}
  \Bigg[
    1 - 
    \left(
      \frac{c/n}{\beta c}
    \right)^2
  \Bigg]
  \Theta(\beta c - c/n),
\end{align}
where the quantity inside the square brackets on the left-hand side becomes a pure imaginary number if $\beta c < c/n$,
and thus we have the Heaviside unit step function $\Theta(\beta c - c/n)$ on the right-hand side.
This step function results in the fact that the \v{C}erenkov radiation is emitted if and only if the particle velocity exceeds the speed of light in the medium ($\beta c > c/n$).
The propagation direction of the radiation is given by the \v{C}erenkov angle,
\begin{align}
  \theta_\mathrm{\check{C}R}
  \equiv 
  \tan^{-1} 
  \frac{(\omega/c)\sqrt{n^2 - \beta^{-2}}}{\omega/(\beta c)}
  = \tan^{-1} \sqrt{\frac{(\beta c)^2}{(c/n)^2} - 1}
  \label{eq:theta_CR}
\end{align}

\begin{figure}[htbp]
  \centering
  \includegraphics[width=.6\linewidth]{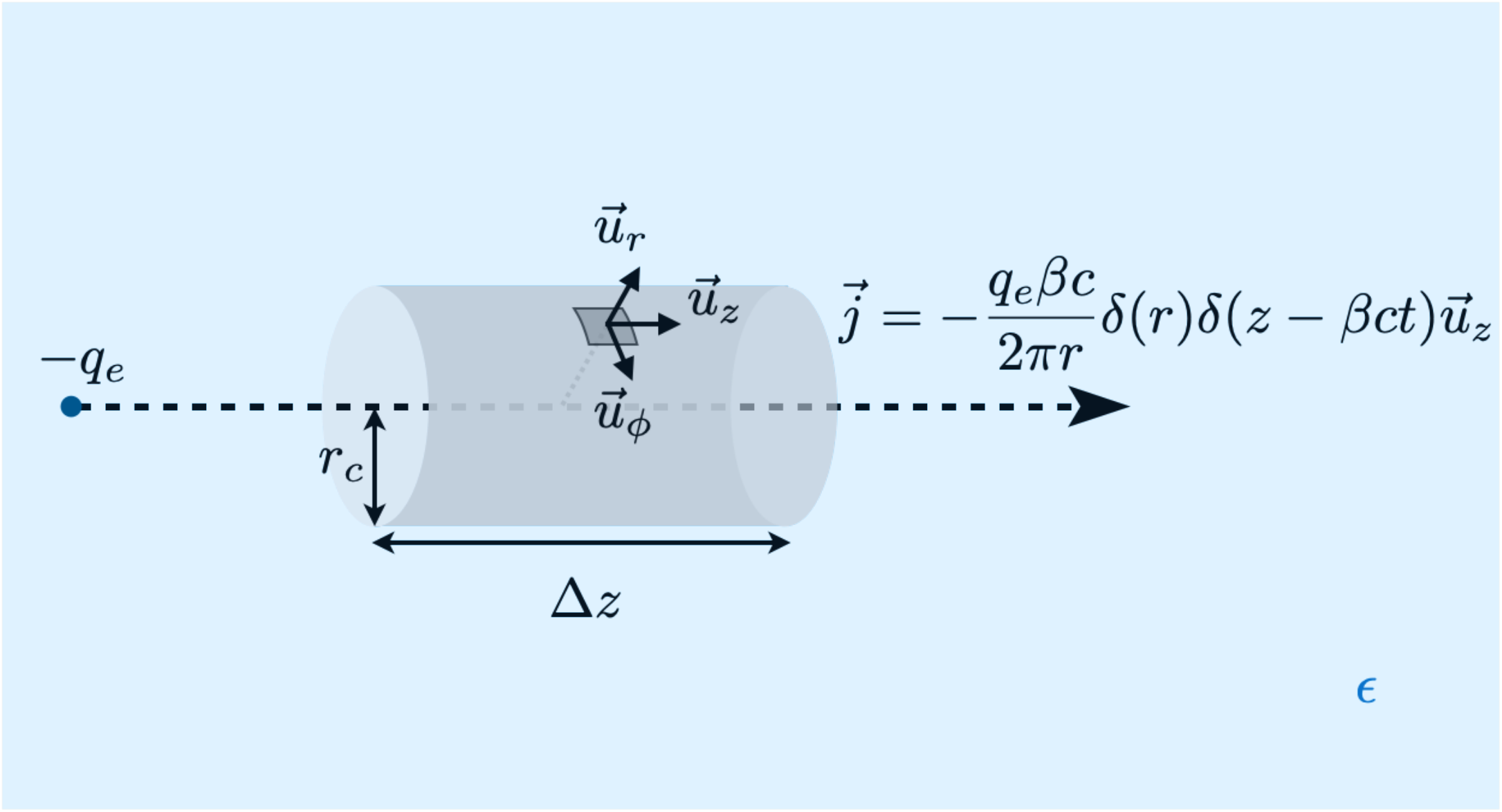}
  \caption{
    Cylinder with a radius $r_c$ and a length $\Delta z$,
    which encloses the electron trajectory in a refractive medium ($n = \sqrt{\epsilon}$).
    The swift electron can be regarded as an effective source current that induces the vector and scalar potentials at the cylinder,
    and the corresponding electromagnetic wave carries energy out through the cylinder \eqref{eq:DW/Dz}.
  }
  \label{fig:Frank-Tamm}
\end{figure}
The power flow through a cylinder with a radius $r_c$ and a length $\Delta z$ shown in \figref{fig:Frank-Tamm} can be evaluated by performing the surface integral in Eq.~\eqref{eq:W},
\begin{align}
  \Delta W
  &=
  8\pi\int_0^{+\infty}
  \frac{1}{2}
  \operatorname{Re}
  \left[
    \vec{E}_\omega^*
    \times
    \vec{H}_\omega
  \right]
  \cdot
  (2\pi r_c \Delta z) \vec{u}_r
  \mathrm{d}\omega
  \\
  \frac{\Delta W}{\Delta z}
  &= \int_0^{+\infty}
  \frac{\mu_0 q_e^2 \omega}{4\pi}
  \cdot
  \Theta(\beta c - c/n)
  \cdot
  \sin^2 \theta_\mathrm{\check{C}R}
  \hspace{.1em} 
  \mathrm{d}\omega
  \label{eq:DW/Dz}
\end{align}
The integrand in Eq.~\eqref{eq:DW/Dz} is equivalent to the Frank--Tamm formula 
\cite{%
  frank1937coherent,%
  jackson1999classical%
},
which gives the power of radiation emitted from the electron within a frequency interval $[\omega, \omega + \mathrm{d}\omega]$ while it travels the unit length.
From the Frank--Tamm formula \eqref{eq:DW/Dz},
we can see that \v{C}erenkov radiation is emitted if and only if the swift electron travels faster than light in the medium ($\beta c > c/n$),
and the radiation power is proportional to the radiation frequency $\omega$ and is quadratically dependent on the radiation direction, $\tan \theta_\mathrm{\check{C}R}$.
If we take the dispersion of the medium into consideration [i.e.~$n \rightarrow n(\omega)$],
the radiation becomes band-limited due to the unit step function $\Theta(\beta c - c/n(\omega))$.
The extension of this type of radiation power calculation for negative index media was firstly discussed by Veselago \cite{veselago1967electrodynamics}, later refined by Lu et al. 
\cite{%
  lu2003vcerenkov,%
  wu2007left,%
  duan2008cherenkov%
},
where they also examined the effects of dissipation and anisotropy.
Their theoretical studies have been confirmed in various configurations 
(see, for example, 
\cite{%
  antipov2008observation,%
  xi2009experimental,%
  de2010optical%
}).

\chapter{Differential formalism for a dynamical grating}
\label{ch:df}
The spatiotemporal modulation of bulky media has been extensively studied both experimentally and theoretically
\cite{%
  sounas2017non,%
  caloz2018electromagnetic,%
  shaltout2019spatiotemporal,%
  galiffi2019broadband,%
  huidobro2019fresnel%
}.
They found that temporal modulation breaks the reciprocity of the systems and results in novel phenomena such as light amplification and Fresnel drag,
which do not have static counterparts.
Keeping these studies in mind,
we analyse the light scattering at the spatiotemporally modulated surfaces in this chapter.

One popular way to calculate the light scattering is to use the dyadic Green's function in the system in question.
We can give the perturbative expansion of the scattered field and thus obtain the scattering matrix
\cite{%
  hill1981integral,%
  goedecke1988scattering,%
  pendry1998can,%
  Yurkin_2007%
}.

Another approach is the boundary matching method.
In this approach,
the input and scattered fields are expanded in a series of eigenmodes in the media,
where we can unambiguously define the scattered field amplitudes,
and then Maxwell's field continuity conditions are imposed at the boundaries.
Those conditions form an equation system that we can invert to obtain the scattering matrix.
One of the most popular examples is the calculation of Mie scattering,
where the continuity conditions are imposed at the surface of a spherical particle,
and the Mie coefficients correspond to the scattered field amplitudes
\cite{mie1908beitrage}.

Here, we consider the boundary matching at a dynamically deformed boundary with the help of a differential formalism.
This formalism is firstly proposed by Chandezon et al. in order to analyse arbitrary structured dielectric surfaces,
\cite{%
  chandezon1980new,%
  chandezon1982multicoated,%
  li1994multilayer%
}.
In these works
the structured boundary is mapped to flat one by a global coordinate translation scheme,
and thus, we can straightforwardly impose the boundary conditions.
This is why differential formalism has been applied to calculate not only dielectric surfaces but also dispersive, lossy, anisotropic surfaces
\cite{%
  barnes1995photonic,%
  harris1995differential,%
  barnes1996physical,%
  harris1996conical,%
  kitamura2013hermitian,%
  murtaza2017study%
}.
It is worth noting that a similar formalism is proposed by Johnson et al. \cite{johnson2002perturbation},
and extensively used in the context of optomechanics in order to evaluate the optomechanical interaction energy 
(see, for example, \cite{eichenfield2009optomechanical, chan2012optimized}).

The differential formalism can handle both shallow and deep structures;
however, we need to adopt local coordinate distortion instead of the global coordinate translation as studied in a series of studies
\cite{%
  li1996improvement,%
  li1996use,%
  xu2014simple,%
  xu2017numerical,%
  xu2020numerical,%
  shcherbakov2013efficient,%
  shcherbakov20153d,%
  shcherbakov2017generalized,%
  shcherbakov2018direct,%
  shcherbakov2019curvilinear%
}.
That is because the local transformation affects only a bounded region near the structured surface and enables us to apply a simple criterion to select the incoming and outgoing fields.
Felix et al. also showed that the local transformation drastically improves the convergence 
\cite{%
  essig2010generation,%
  felix2014local%
}.

\begin{figure}[htbp]
  \centering
  \includegraphics[width=.6\linewidth]
  {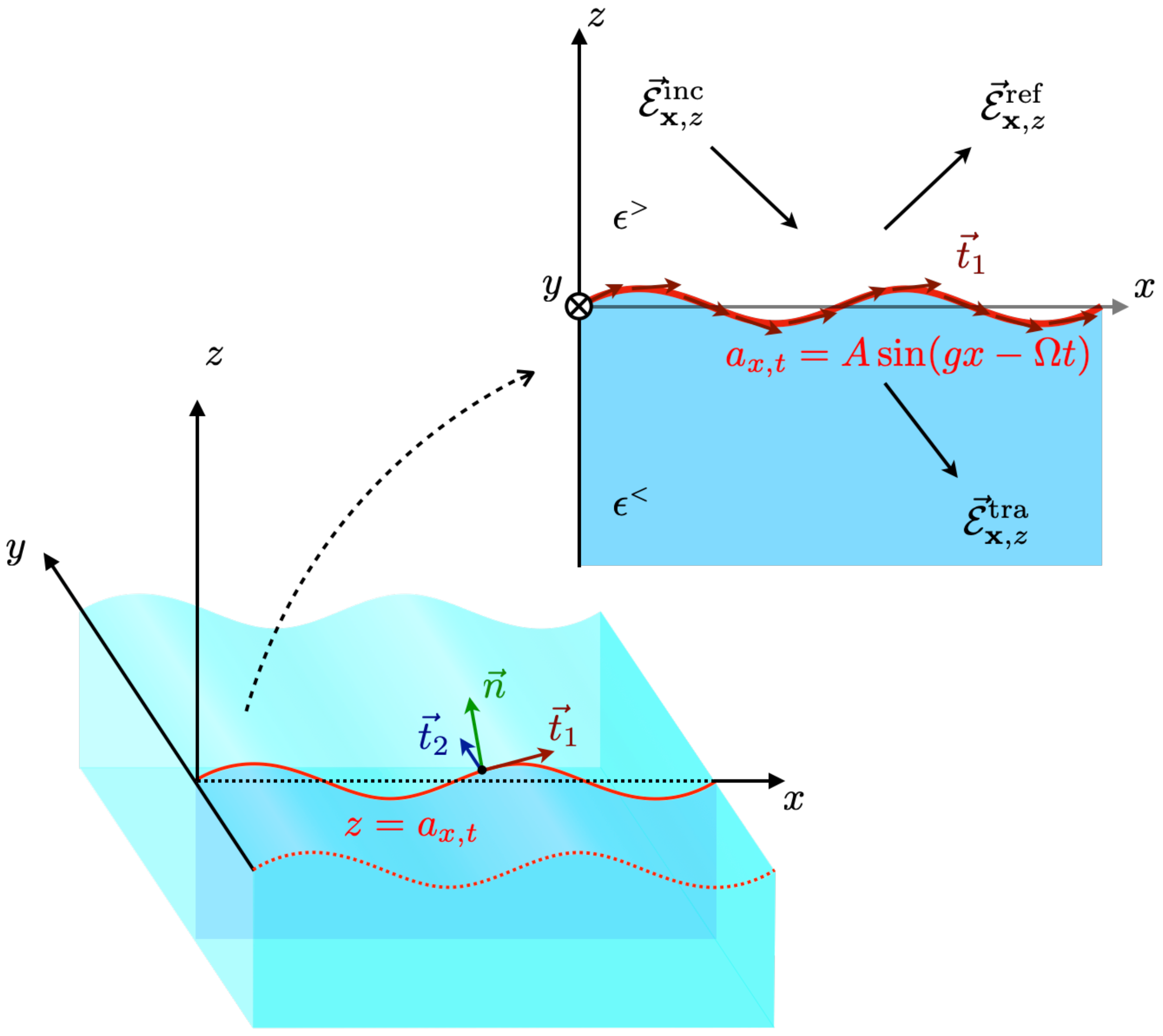}
  \caption{
    Spatiotemporally modulated surface.
    The surface is weakly modulated in space and time.
    The corrugation function is sinusoidal and given by Eq.~\eqref{eq:boundary}.
    We consider the in-plane incidence of an electromagnetic field and analyse the reflection and the transmission at the surface.
    The red and blue arrows $\vec{t}_{1,2}$ are the tangential vectors of the surface at each point.
    The permittivities of medium and lower media are $\epsilon^{\gl}$, 
    respectively.
    The permeability is assumed to be unity in both media, $\mu^{\gl} = 1$.
  }
  \label{fig:sketch}
\end{figure}

In \figref{fig:sketch}, 
the schematic image of the system we consider in this chapter is shown.
The system is composed of two dielectric media characterised by their permittivities $\epsilon^\gl$.
Note that the permittivities of our dielectric media are regarded as constants.
Note also that we assume the permeability is unity everywhere.
The boundary is dynamically corrugated, and the profile is given by $z=a_\mathbf{x}$.
Let us consider one of the simplest profiles,
\begin{align}
  a_\mathbf{x} 
  = 
  A \sin(\mathbf{q}\cdot\mathbf{x})
  = A \sin \left[g\left(x - \frac{\Omega}{g} t\right)\right].
  \label{eq:boundary}
\end{align}
Here, we have introduced three-component vectors $\mathbf{q} = ( g, 0, -i\Omega/c)$ and $\mathbf{x} = (x, y, -ict)$,
where $A$, $g$ and $\Omega$ are the strength of the modulation, the spatial and temporal frequencies of the modulation, respectively.
The surface itself is not physically moving in the $x$ direction,
but its profile is shifting at the phase velocity $\Omega/g$ that can exceed the speed of light.
Note that the surface oscillates in the $z$ direction, 
and the maximum velocity in the $z$ direction is given by $A\Omega/2\pi$, 
which cannot exceed the speed of light ($A < 2\pi c/\Omega$).
The permittivity distribution can be written in terms of the surface profile,
\begin{align}
  \epsilon_{\mathbf{x},z}
  &= 
  \epsilon^\ssl
  \Theta(a_\mathbf{x}-z) 
  +
  \epsilon^\ssg
  \Theta(z-a_\mathbf{x})
  = 
  \alpha
  \Theta(a_\mathbf{x}-z)
  +
  \epsilon^\ssg,
\end{align}
where $\Theta(z)$ is Heaviside's unit step function,
and 
$\alpha \equiv (\epsilon^\ssl-\epsilon^\ssg)$
is the permittivity difference.
The tangential and normal vectors of the surface are calculated by calculating the space derivatives of the surface profile,
\begin{align}
  \begin{cases}{}
    \vec{t}_{1}
    =
    \cfrac{\vec{u}_x + a_\mathbf{x}' \vec{u}_z}{\sqrt{1+{a_\mathbf{x}'}^2}},
    \quad
    \vec{t}_2
    =
    \vec{u}_y.
    \\
    \vec{n} 
    = 
    \vec{t}_1 \times \vec{t}_2
    =
    \cfrac{-a_\mathbf{x}' \vec{u}_x + \vec{u}_z}{\sqrt{1+{a_\mathbf{x}'}^2}},
  \end{cases}
  \label{eq:t_1,2,n}
\end{align}
where $a_\mathbf{x}' = (\partial/\partial x)a_\mathbf{x}$ is the space derivative of the boundary profile in the $x$ direction,
and $\vec{u}_{x,y,z}$ are the unit vectors in the $x, y$ and $z$ directions.

\section{Field expansion}
\label{sec:field_exp}
In order to proceed with the scattering calculation, we expand the electric and magnetic fields above and below the boundary as we analysed reflection and transmission at a thin film in Sec.~\ref{sec:sca_trans_mat}.
Here, we assign the incident, reflected and transmitted fields as following:
\begin{align}
  \begin{cases}{}
    \vec{\mathcal{E}}{}_{\mathbf{x},z}^\mathrm{inc} 
    = \displaystyle{\int_{\mathbf{k}}}
    e^{i\mathbf{k}\cdot \mathbf{x}}
    \vec{E}{}_{\mathbf{k},z}^{\Inc},
    \quad
    \vec{\mathcal{H}}{}_{\mathbf{x},z}^\mathrm{inc} 
    = \displaystyle{\int_{\mathbf{k}}}
    e^{i\mathbf{k}\cdot \mathbf{x}}
    \vec{H}{}_{\mathbf{k},z}^{\Inc}, 
    & 
    (z \geq a_\mathbf{x}),
    \vspace{.5em}
    \\
    \vec{\mathcal{E}}{}_{\mathbf{x},z}^\mathrm{ref} 
    = \displaystyle{\int_{\mathbf{k}}}
    e^{i\mathbf{k}\cdot \mathbf{x}}
    \vec{E}{}_{\mathbf{k},z}^{\Refl},
    \quad
    \vec{\mathcal{H}}{}_{\mathbf{x},z}^\mathrm{ref} 
    = \displaystyle{\int_{\mathbf{k}}}
    e^{i\mathbf{k}\cdot \mathbf{x}}
    \vec{H}{}_{\mathbf{k},z}^{\Refl}
    &
    (z \geq a_\mathbf{x}),
    \vspace{.5em}
    \\
    \vec{\mathcal{E}}{}_{\mathbf{x},z}^\mathrm{tra} 
    = \displaystyle{\int_{\mathbf{k}}}
    e^{i\mathbf{k}\cdot \mathbf{x}}
    \vec{E}{}_{\mathbf{k},z}^{\Tra},
    \quad
    \vec{\mathcal{H}}{}_{\mathbf{x},z}^\mathrm{tra} 
    = \displaystyle{\int_{\mathbf{k}}}
    e^{i\mathbf{k}\cdot \mathbf{x}}
    \vec{H}{}_{\mathbf{k},z}^{\Tra}
    &
    (z \leq a_\mathbf{x}),
    \vspace{.5em}
  \end{cases}
  \label{eq:E,H_lnc,ref,tra}
\end{align}
The superscripts on the left-hand side,
$\Lambda \in \{\mathrm{inc},\ \mathrm{ref},\ \mathrm{tra}\}$,
are labels which identify modes in the real space and the time domain.
The two superscripts on the right-hand side,
$\sigma \in \{+, -\}$ and $\tau \in \{<, >\}$,
are corresponding labels in the reciprocal space.
While $\sigma$ specifies in which direction the mode propagates,
$\tau$ specifies in which medium the mode lives.
The fields in the real space,
$\vec{\mathcal{E}}_{\mathbf{x},z}^{\Lambda}$ 
and
$\vec{\mathcal{H}}_{\mathbf{x},z}^{\Lambda}$,
are given by means of complex-valued Fourier components,
$\vec{E}{}_{\mathbf{k},z}^{\sigma\tau}$ 
and 
$\vec{H}{}_{\mathbf{k},z}^{\sigma\tau}$.
Note that we have introduced a reciprocal vector $\mathbf{k} = (k_x, k_y, -ik_0)$, where $k_0 \equiv \omega/c$.
Since the electric $\vec{\mathcal{E}}_{\mathbf{x},z}^{{\sigma\tau}}$ and magnetic fields $\vec{\mathcal{H}}_{\mathbf{x},z}^{{\sigma\tau}}$ in the real space are real-valued, each Fourier component satisfies
\begin{align}
  \vec{E}{}_{\mathbf{k},z}^{\sigma\tau}
  &= 
  \vec{E}{}_{-\mathbf{k},z}^{\sigma\tau\cc},
  \quad
  \vec{H}{}_{\mathbf{k},z}^{\sigma\tau}
  = 
  \vec{H}{}_{-\mathbf{k},z}^{\sigma\tau\cc}.
\end{align}
Note also that we employed a shorthand notation,
\begin{align}
  \int_{\mathbf{k}} [\ldots]
  =
  \int 
  [\ldots]
  \mathrm{d}\mathbf{k}.
\end{align}
Recall that we introduced the shorthand notation for the integral measure, $\mathrm{d}\mathbf{k} := \mathrm{d}\omega \mathrm{d}k_x \mathrm{d}k_y/(2\pi)^3$.

Let us summarise the Fourier components of electric and magnetic fields,
\begin{align}
  \begin{cases}{}
    \vec{E}{}_{\mathbf{k},z}^{\sigma\tau}
    =
    e^{i\sigma K_{\mathbf{k}}^\tau z}
    \vec{E}{}_{\mathbf{k},0}^{\sigma \tau},
  &
  \vec{E}{}_{\mathbf{k},0}^{\sigma \tau}
  =
  \displaystyle{\sum_{\lambda=s,p}}
  E_{\lambda,\mathbf{k}}^{\sigma\tau}
  \vec{e}_{\lambda,\mathbf{k}}^{\hspace{.2em}\sigma\tau},
  \vspace{.5em}
  \\
  \vec{H}{}_{\mathbf{k},z}^{\sigma\tau}
  =
  e^{i\sigma K_{\mathbf{k}}^\tau z}
  \vec{H}{}_{\mathbf{k},0}^{\sigma \tau},
  &
  \vec{H}{}_{\mathbf{k},0}^{\sigma \tau}
  =
  \displaystyle{\sum_{\lambda=s,p}}
  H_{\lambda,\mathbf{k}}^{\sigma\tau}
  \vec{h}_{\lambda,\mathbf{k}}^{\hspace{.2em}\sigma\tau}.
  \end{cases}
  \label{eq:E,H_z,alpha^sigmatau}
\end{align}
Remind that the polarisation vectors are given by
\begin{align}
  \vec{e}_{\lambda,\mathbf{k}}^{\sigma\tau}
  &=
  \begin{cases}{}
    \displaystyle{
      \frac{\operatorname{sgn}(\omega) \vec{k}_\mathbf{k}^{\sigma\tau} \times \vec{u}_z}
      {|\vec{k}_\mathbf{k}^{\sigma\tau} \times \vec{u}_z|}
    }
    &
    (\lambda = s),
    \vspace{.5em}
    \\
    \displaystyle{
      \frac{\operatorname{sgn}(\omega) \vec{k}_\mathbf{k}^{\sigma\tau} \times \vec{e}_{s,\mathbf{k}}^{\sigma\tau}}
      {|\vec{k}_\mathbf{k}^{\sigma\tau} \times \vec{e}_{s,\mathbf{k}}^{\sigma\tau}|}
    }
    &
    (\lambda = p),
  \end{cases}
  \tag{\ref{eq:e_lambda}}
\end{align}
which satisfies the transversality condition \eqref{eq:transversality}.
Remind also that we introduced another basis for the magnetic field for convenience,
\begin{align}
    \begin{pmatrix}
      \vec{h}_{s,\mathbf{k}}^{\sigma\tau}
      \\
      \vec{h}_{p,\mathbf{k}}^{\sigma\tau}
    \end{pmatrix}
  =
  \begin{pmatrix}
    0 & +1
    \\
    -1 & 0
  \end{pmatrix}
  \begin{pmatrix}
    \vec{e}_{s,\mathbf{k}}^{\hspace{.2em}\sigma\tau}
    \\
    \vec{e}_{p,\mathbf{k}}^{\hspace{.2em}\sigma\tau}
  \end{pmatrix}.
  \tag{\ref{eq:e_h}}
\end{align}

\section{Boundary matching equations}
\label{sec:boundary}
\begin{figure}[htbp]
  \centering
  \includegraphics[width=.5\linewidth]
  {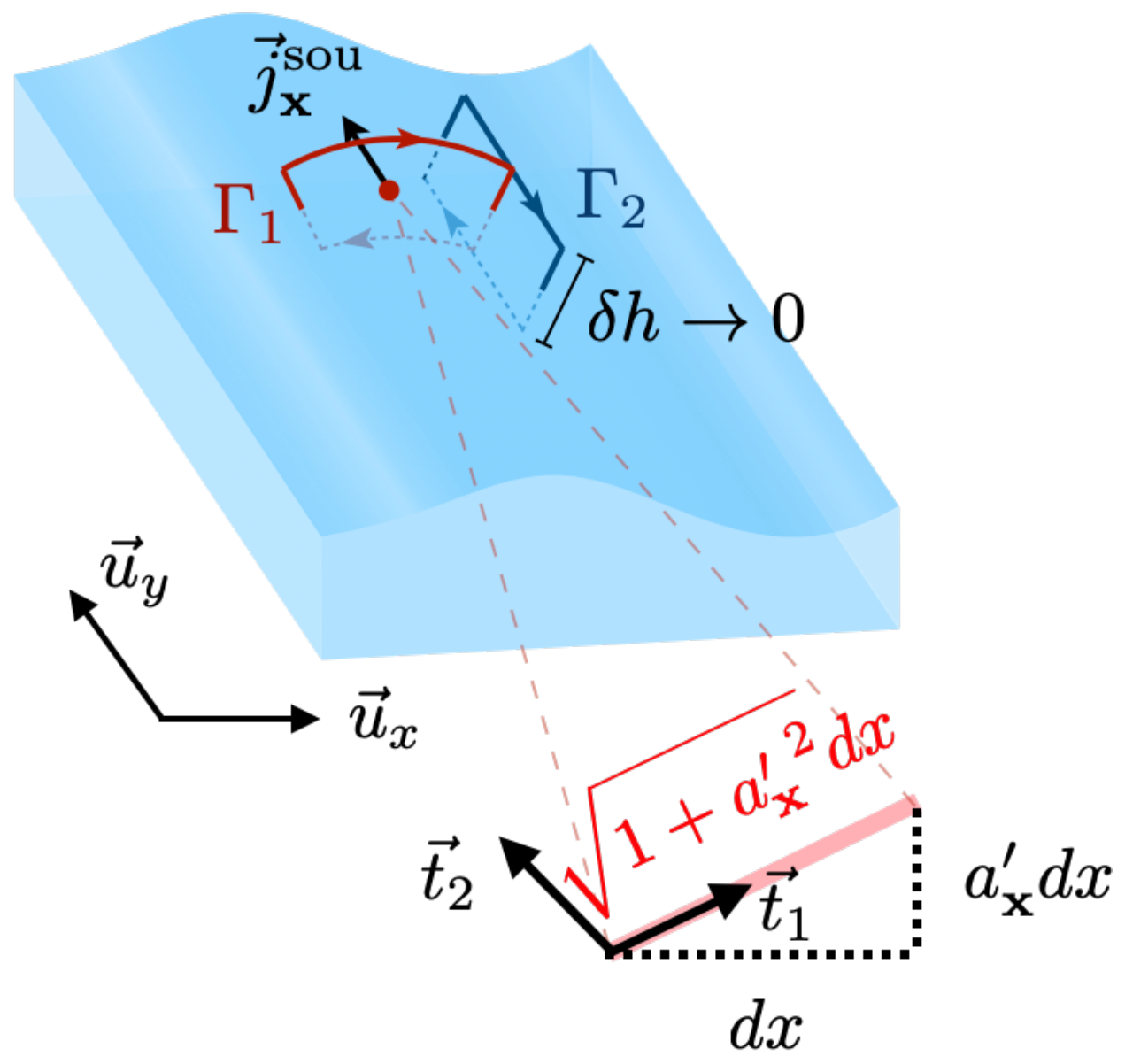}
  \caption{
    Integration paths $\Gamma_{1,2}$ to derive a boundary matching equation.
    Note that $\Gamma_1$ is curved and $\Gamma_2$ is tilted along the boundary unlike the flat surface calculation (\figref{fig:metal_single}).
  }
  \label{fig:boundary_matching}
\end{figure}

Since we have completed the field expansion in the previous section, we are ready to derive the boundary matching equations that associate the scattered field amplitudes with the incident field amplitudes.
In order to find the matching equations, we integrate the Maxwell--Heaviside equations along paths enclosing the boundary as we analysed thin films in Sec.~\ref{sec:sca_trans_mat}.
In \figref{fig:boundary_matching}, one of the paths to find the matching equations is shown.
Note that our boundary is now moving and structured, and we should choose dynamically curved paths.

Integrating the Faraday law \eqref{eq:faraday_macro} along a path $\Gamma_1$ shown in \figref{fig:boundary_matching}, we can obtain the electric field continuity at the boundary,
\begin{align}
  \lim_{\delta h \rightarrow 0}
  \oiint_{\Gamma_1}
  (\nabla \times \vec{\mathcal{E}}_{\mathbf{x},z})
  \cdot 
  \mathrm{d}\vec{S} 
  &=
  -\lim_{\delta h \rightarrow 0}
  \oiint_{\Gamma_1}
  \Bigg(
  \frac{\partial}{\partial t} 
  \mu_0 \vec{\mathcal{H}}_{\mathbf{x},z},
  \Bigg)
  \cdot 
  \mathrm{d}\vec{S} 
  \label{eq:Faraday_Gamma1}
  \\
  \lim_{\delta h \rightarrow 0}
  \oint_{\Gamma_1}
  \vec{\mathcal{E}}_{\mathbf{x},z} 
  \cdot 
  \mathrm{d}\vec{r} 
  &= 0,
\end{align}
where we have used the Stokes' theorem.
The vectorial line element $\mathrm{d} \vec{r}$ is given by means of the tangential vector $\vec{t}_1$ and a scalar element $\sqrt{1+{a_\mathbf{x}'}^2}\mathrm{d}x$.
We can explicitly write an integral equation,
\begin{align}
  \int \mathrm{d}x \sqrt{1+{a_\mathbf{x}'}^2}  
  \vec{t}_1 
  \cdot 
  (\vec{\mathcal{E}}_{\mathbf{x},a_\mathbf{x}+0}
  - 
  \vec{\mathcal{E}}_{\mathbf{x},a_\mathbf{x}-0}) = 0,
  \label{eq:boundary_condition_pre}
\end{align}
where 
$\vec{\mathcal{E}}_{\mathbf{x},a_\mathbf{x}+0}
= \vec{\mathcal{E}}_{\mathbf{x},z}^\mathrm{inc} 
+ \vec{\mathcal{E}}_{\mathbf{x},z}^\mathrm{ref}$
and 
$\vec{\mathcal{E}}_{\mathbf{x},a_\mathbf{x}-0} 
= \vec{\mathcal{E}}_{\mathbf{x},z}^\mathrm{tra}$
are the electric fields evaluated above and below the boundary,
respectively.
Since Eq.~\eqref{eq:boundary_condition_pre} is independent of the choice of the integration interval,
we can obtain a boundary matching equation for the electric field,
\begin{align}
  \eta 
  \vec{t}_1 
  \cdot
  (
  \vec{\mathcal{E}}_{\mathbf{x},a_\mathbf{x}}^\mathrm{inc} 
  +
  \vec{\mathcal{E}}_{\mathbf{x},a_\mathbf{x}}^\mathrm{ref}
  - 
  \vec{\mathcal{E}}_{\mathbf{x},a_\mathbf{x}}^\mathrm{tra}
  ) = 0.
  \label{eq:t_1_E}
\end{align}
where we set the stretching ratio of the line element $\eta = \sqrt{1+{a_\mathbf{x}'}^2}$ for brevity, which has the information of the surface geometry.

Let us consider the matching equation for the magnetic field.
Integrating the Amp\`{e}re--Maxwell equation \eqref{eq:ampere_macro} along the path $\Gamma_1$,
we can obtain 
\begin{align}
  \lim_{\delta h \rightarrow 0}
  \oiint_{\Gamma_1}
  (\nabla \times \vec{\mathcal{H}}_{\mathbf{x},z})
  \cdot 
  \mathrm{d}\vec{S} 
  &=
  \lim_{\delta h \rightarrow 0}
  \oiint_{\Gamma_1}
\Bigg(
  \frac{\partial}{\partial t} 
  \epsilon_{\mathbf{x},z}\epsilon_0
  \vec{\mathcal{E}}_{\mathbf{x},z}
\Bigg)
  \cdot 
  \mathrm{d}\vec{S},
  \\
  \lim_{\delta h \rightarrow 0}
  \oint_{\Gamma_1}
  \mathrm{d}\vec{r} 
  \cdot 
  \vec{\mathcal{H}}_{\mathbf{x},z}
  &=
  +\vec{t}_2
  \cdot
  \vec{j}_\mathbf{x}^\mathrm{sou},
  \label{eq:Ampere-Maxwell_Gamma1}
\end{align}
where we have used Stokes' theorem.
This equation represents the magnetic field discontinuity at the boundary.
Recall that our boundary is moving, and the time derivative of the permittivity distribution results in the Dirac delta function,
$
\partial \epsilon_{\mathbf{x},z}/\partial t 
=
\dot{a}_\mathbf{x}
\alpha
\delta(a_\mathbf{x}-z)
$.
Thus, we have a source term on the right-hand side in Eq.~\eqref{eq:Ampere-Maxwell_Gamma1} as a consequence of the temporal modulation (i.e.~$\dot{a}_\mathbf{x} \neq 0$),
\begin{align}
  \vec{j}_\mathbf{x}^\mathrm{sou}
  =
  \frac{\dot{a}_{\mathbf{x}}}{cZ_0}
  \alpha \vec{\mathcal{E}}_{\mathbf{x},a_\mathbf{x}}^\mathrm{tra}.
  \label{eq:jsou}
\end{align}
Evaluating the left-hand side in Eq.~\eqref{eq:Ampere-Maxwell_Gamma1} as in the case of the electric field, we can obtain the explicit representation of the magnetic discontinuity at the boundary,
\begin{align}
  \eta
  \vec{t}_1 
  \cdot 
  (
  \vec{\mathcal{H}}_{\mathbf{x},a_\mathbf{x}}^\mathrm{inc} 
  +
  \vec{\mathcal{H}}_{\mathbf{x},a_\mathbf{x}}^\mathrm{ref}
  - 
  \vec{\mathcal{H}}_{\mathbf{x},a_\mathbf{x}}^\mathrm{tra}
  ) 
  =
  +\vec{t}_2
  \cdot
  \vec{j}_\mathbf{x}^\mathrm{sou}.
  \label{eq:t_1_H}
\end{align}

From Eq.~\eqref{eq:Faraday_Gamma1} to Eq.~\eqref{eq:t_1_H},
we have considered integration along the path $\Gamma_1$.
Below this, we analogously integrate the equations along another path $\Gamma_2$ enclosing the boundary.
The integration of the Faraday law \eqref{eq:faraday_macro} results in the electric field continuity at the boundary,
\begin{align}
  &\vec{t}_2 \cdot 
  (
  \vec{\mathcal{E}}{}_{\mathbf{x},a_\mathbf{x}}^\mathrm{inc} 
  + 
  \vec{\mathcal{E}}{}_{\mathbf{x},a_\mathbf{x}}^\mathrm{ref} 
  - 
  \vec{\mathcal{E}}{}_{\mathbf{x},a_\mathbf{x}}^\mathrm{tra}
  ) 
  = 0,
  \vspace{.5em}
  \label{eq:t_2_E}
\end{align}
while the Amp\`{e}re--Maxwell law \eqref{eq:ampere_macro} provides the discontinuity of the magnetic field at the boundary,
\begin{align}
  \vec{t}_2 \cdot 
  (
  \vec{\mathcal{H}}{}_{\mathbf{x},a_\mathbf{x}}^\mathrm{inc} 
  +
  \vec{\mathcal{H}}{}_{\mathbf{x},a_\mathbf{x}}^\mathrm{ref} 
  - 
  \vec{\mathcal{H}}{}_{\mathbf{x},a_\mathbf{x}}^\mathrm{tra}
  )
  =
  -\eta
  \vec{t}_1
  \cdot
  \vec{j}_\mathbf{x}^\mathrm{sou}.
  \label{eq:t_2_H}
\end{align}
where we have the source term on the right-hand side as in the integral along $\Gamma_1$.

Let us collect the boundary matching equations,
\begin{align}
  \begin{cases}{}
    \vec{t}_2 \cdot 
    (
    \vec{\mathcal{E}}{}_{\mathbf{x},a_\mathbf{x}}^\mathrm{inc} 
    + 
    \vec{\mathcal{E}}{}_{\mathbf{x},a_\mathbf{x}}^\mathrm{ref} 
    - 
    \vec{\mathcal{E}}{}_{\mathbf{x},a_\mathbf{x}}^\mathrm{tra}
    ) 
    = 0,
    \vspace{.5em}
    \\
    \eta
    \vec{t}_1 \cdot 
    Z_0 (
    \vec{\mathcal{H}}{}_{\mathbf{x},a_\mathbf{x}}^\mathrm{inc} 
    + 
    \vec{\mathcal{H}}{}_{\mathbf{x},a_\mathbf{x}}^\mathrm{ref} 
    -
    \vec{\mathcal{H}}{}_{\mathbf{x},a_\mathbf{x}}^\mathrm{tra}
    )
    =
    +\vec{t}_2
    \cdot
    Z_0 \vec{j}_\mathbf{x}^\mathrm{sou}.
  \end{cases}
  \label{eq:boundary_conditions_s_pol}
  \\
  \begin{cases}{}
    \eta
    \vec{t}_1 \cdot 
    (
    \vec{\mathcal{E}}{}_{\mathbf{x},a_\mathbf{x}}^\mathrm{inc} 
    + 
    \vec{\mathcal{E}}{}_{\mathbf{x},a_\mathbf{x}}^\mathrm{ref} 
    -
    \vec{\mathcal{E}}{}_{\mathbf{x},a_\mathbf{x}}^\mathrm{tra}
    ) = 0,
    \vspace{.5em}
    \\
    \vec{t}_2 \cdot 
    Z_0 (
    \vec{\mathcal{H}}{}_{\mathbf{x},a_\mathbf{x}}^\mathrm{inc} 
    +
    \vec{\mathcal{H}}{}_{\mathbf{x},a_\mathbf{x}}^\mathrm{ref} 
    - 
    \vec{\mathcal{H}}{}_{\mathbf{x},a_\mathbf{x}}^\mathrm{tra}
    )
    =
    -\eta\vec{t}_1
    \cdot
    Z_0 \vec{j}_\mathbf{x}^\mathrm{sou}.
  \end{cases}
  \label{eq:boundary_conditions_p_pol}
\end{align}
Note that the magnetic equations are multiplied by $Z_0$ so that the units of all equations are unified.
These results (\ref{eq:boundary_conditions_s_pol}, \ref{eq:boundary_conditions_p_pol}) are general and can be applied to various smooth profiles as in the Chandezon works
\cite{%
  chandezon1980new,
  chandezon1982multicoated,
  li1994multilayer%
}.
Below this,
we focus on the sinusoidal profile \eqref{eq:boundary} as one of the simplest possible examples.

\section{Reflection and transmission matrices}
\label{sec:ref_tra}
We apply the Fourier transform to the matching equations in the real space and the time domain (\ref{eq:boundary_conditions_s_pol}, \ref{eq:boundary_conditions_p_pol}) and find the corresponding equations in the reciprocal space that associate the scattered modal amplitudes and the incident ones and lead to the reflection and transmission matrices.
For convenience, we define the Fourier transform operator,
\begin{align}
\mathscr{F}[g_\mathbf{x}]_\mathbf{k} 
\equiv 
\int  
g_\mathbf{x}
e^{-i\mathbf{k}\cdot\mathbf{x}}
d\mathbf{x}.
\end{align}
Since our configuration (\ref{fig:sketch}) is of translational invariance in the $y$ direction, our scattering problem is two-dimensional.
This is why we focus on two fundamental polarisations, transverse electric ($s$) and transverse magnetic ($p$) modes.

Here, we calculate the transverse electric mode ($s$ polarisation).
The same procedure can be applied to the transverse magnetic mode ($p$ polarisation) as given in the Appendix \ref{app:p-pol}.
When the incident field is the TE mode, the electric field oscillates in the $y$ direction, 
and we can focus on two equations \eqref{eq:boundary_conditions_s_pol} out of four.
Note that Eqs.~\eqref{eq:boundary_conditions_p_pol} is automatically satisfied.
The tangential components of the electric field,
which appear in Eqs.~\eqref{eq:boundary_conditions_s_pol},
are evaluated by means of inner product,
\begin{align}
  \vec{t}_2 
  \cdot 
  \vec{\mathcal{E}}_{\mathbf{x},a_\mathbf{x}}^{\Lambda}
  &=
  -\int_\mathbf{k}
  e^{i\mathbf{k}\cdot\mathbf{x}}
  \operatorname{sgn}(\omega)
  \frac{k_x}{k_\parallel}
  e^{i\phi_{\mathbf{k}}^{\sigma\tau}\sin \mathbf{q}\cdot\mathbf{x}}
  E_{s,\mathbf{k}}^{\sigma\tau},
  \label{eq:t_2dotE}
\end{align}
where we have defined the propagating phase factor for a mode characterised by a set $\{\sigma, \mathbf{k}\}$ in a medium $\tau$,
$\phi_{\mathbf{k}}^{\sigma \tau}
=\sigma K_{\mathbf{k}}^\tau A$.
We have the exponential of the trigonometric function in Eq.~\eqref{eq:t_2dotE} that can be expanded by using the Jacobi--Anger identity \cite{cuyt2008handbook}, 
\begin{align}
  \vec{t}_2 
  \cdot 
  \vec{\mathcal{E}}_{\mathbf{x},a_\mathbf{x}}^{\Lambda}
  &=
  -\SumInt_{m,\mathbf{k}}
  e^{i\mathbf{k}\cdot\mathbf{x}}
  \frac{k_{x,m}}{k_{\parallel,m}}
  \operatorname{sgn}(\omega_m)
  J_{-m}(\phi_{\mathbf{k}_m}^{\sigma\tau})
  E_{s,\mathbf{k}_m}^{\sigma\tau},
\end{align}
where $J_m$ is the $m\mathrm{th}$ order Bessel function of the first kind.
Note that we have introduced a shorthand notation,
\begin{align}
  \SumInt_{m,\mathbf{k}}
[\ldots]
  = 
  \sum_{m=-\infty}^{+\infty}
  \int_\mathbf{k}
  [\ldots].
  \label{eq:SumInt}
\end{align}
We apply the Fourier transform evaluating the modal amplitude at $\mathbf{k}_l = \mathbf{k} + l\mathbf{q}$,
\begin{align}
  \mathscr{F}
  \left[
    \vec{t}_2 
    \cdot 
    \vec{\mathcal{E}}{}_{\mathbf{x},a_\mathbf{x}}^{\Lambda} 
  \right]_{\mathbf{k}_l}
  =
  \left[
    \mathsf{M}_{\mathbf{k}}^{\sigma\tau}
  \mathbb{E}_{s,\mathbf{k}}^{\sigma\tau}
\right]_l,
\label{eq:F[t_2dotE]}
\end{align}
where we have introduced an electric modal amplitude vector,
\begin{align}
  \mathbb{E}_{\lambda,\mathbf{k}}^{\sigma\tau}
   &=
   \begin{pmatrix}
     \cdots
     &
     E_{\lambda,\mathbf{k}_{-1}}^{\sigma\tau}
     &
     E_{\lambda,\mathbf{k}_{0}}^{\sigma\tau}
     &
     E_{\lambda,\mathbf{k}_{+1}}^{\sigma\tau}
     &
     \cdots
   \end{pmatrix}^\top,
   \label{eq:modal_amp_vec}
\end{align}
whose components correspond to the amplitudes of Floquet replicas \cite{chicone2006ordinary}.
Note that the subscript $l$ on the right-hand side in Eq.~\eqref{eq:F[t_2dotE]} denote the $l$th element.
We also introduced a matrix-vector representation with the coefficient matrix $\mathsf{M}_\mathbf{k}^{\sigma\tau}$ that has the information of the boundary geometry,
\begin{align}
  [\mathsf{M}_{\mathbf{k}}^{\sigma\tau}]_{lm}
  &=
  \frac{k_{x,m}}{k_{\parallel,m}}
  \operatorname{sgn}(\omega_m)
  J_{l-m}(\phi_{\mathbf{k}_m}^{\sigma\tau}).
  \label{eq:M}
\end{align}
Recall that the subscript $m$ represents the $m$th replica (e.g.~$\mathbf{k}_m = \mathbf{k} + m\mathbf{q}$).

We apply the Fourier transform formula \eqref{eq:F[t_2dotE]} to the incident and scattered (reflected and transmitted) fields to obtain a mode matching equation,
\begin{align}
  \mathsf{M}_{\mathbf{k}}^{\Inc}
  \mathbb{E}_{s,\mathbf{k}}^{\Inc}
  +
  \mathsf{M}_{\mathbf{k}}^{\Refl}
  \mathbb{E}_{s,\mathbf{k}}^{\Refl}
  -
  \mathsf{M}_{\mathbf{k}}^{\Tra}
  \mathbb{E}_{s,\mathbf{k}}^{\Tra}
  &= 0.
  \label{eq:t_2_E_FT}
\end{align}
Here, $\mathbb{E}_\mathbf{k}^{\Inc}$ corresponds to the incident modal amplitude vector,
which is known quantities,
while $\mathbb{E}_\mathbf{k}^{\Refl}$ and $\mathbb{E}_\mathbf{k}^{\Tra}$ are the scattered modal amplitude vectors, unknown quantities.
In order to fix these unknowns, we need one more equation, which is derived just below.

Similarly to the electric field calculation above,
we can expand the tangential components appearing in Eqs.~\eqref{eq:boundary_conditions_s_pol} by using the Jacobi--Anger identity,
\begin{align}
  &\eta
  \vec{t}_1
  \cdot
  Z_0
  \vec{\mathcal{H}}_{\mathbf{x},a_\mathbf{x}}^{\Lambda}
  =
  \int_\mathbf{k}
  e^{i\mathbf{k}\cdot\mathbf{x}}
  e^{
    i\sigma K_{\mathbf{k}}^\tau a_\mathbf{x}
  }
  \eta
  \vec{t}_1
  \cdot
  \vec{h}_{s,\mathbf{k}}^{\sigma\tau}
  Z_0
  H_{s,\mathbf{k}}^{\sigma\tau},
  \\
  &=
  \int_\mathbf{k}
  e^{i\mathbf{k}\cdot\mathbf{x}} 
  \frac{\sigma K_\mathbf{k}^\tau k_x - a_\mathbf{x}' {k_\parallel}^2}
  {k_\parallel |k_0|}
  e^{
    i\sigma 
    K_{\mathbf{k}}^\tau
    a_\mathbf{x}
  }
  E_{s,\mathbf{k}}^{\sigma \tau},
  \\
  &=
  \SumInt_{m,\mathbf{k}}
  e^{
    i(\mathbf{k} - m\mathbf{q})
    \cdot\mathbf{x}
  }
  \left(
    \frac{k_x}
    {k_\parallel}
    \frac{\sigma K_\mathbf{k}^\tau}
    {|k_0|}
    -
    \frac{-mg}
    {\sigma K_{\mathbf{k}}^\tau}
    \frac{k_\parallel}
    {|k_0|}
  \right)
  J_{-m}(\phi_{\mathbf{k}}^{\sigma\tau}) 
  E_{s,\mathbf{k}}^{\sigma \tau}.
\end{align}
Note that the magnetic field amplitude can be given by the electric field amplitude.
The modal amplitude in the reciprocal space can be calculated via the Fourier transform,
\begin{align}
  \mathscr{F}
  \left[
  \eta
  \vec{t}_1
  \cdot
  Z_0
  \vec{\mathcal{H}}_{\mathbf{x},a_\mathbf{x}}^{\Lambda}
  \right]_{\mathbf{k}_l}
  =
  \left[
    \mathsf{N}_{\mathbf{k}}^{\sigma\tau}
  \mathbb{E}_{s,\mathbf{k}}^{\sigma\tau}
\right]_l,
\label{eq:F[t_1dotZ_0H]}
\end{align}
where we have adopted the same notation as in Eq.~\eqref{eq:F[t_2dotE]} and defined another coefficient matrix whose element reads
\begin{align}
  [\mathsf{N}_{\mathbf{k}}^{\sigma\tau}]_{lm}
  &=
  \left(
    \frac{k_{x,m}}{k_{\parallel,m}}
    \frac{\sigma K_{\mathbf{k}_m}^\tau}{|k_{0,m}|}
    -
    \frac{(l-m)g}{\sigma K_{\mathbf{k}_m}^\tau}
    \frac{k_{\parallel,m}}{|k_{0,m}|}
  \right)
  J_{l-m}(\phi_{\mathbf{k}_m}^{\sigma\tau}).
  \label{eq:N}
\end{align}
Here, we have two terms because both the $x$ and $z$ components of the fields contribute as the boundary is curved.
In the second term, we have a factor $(l-m)g$ which stems from the space derivative in the stretching ratio $\eta = \sqrt{1+{a_\mathbf{x}'}^2}$.

The source term contribution in \eqref{eq:boundary_conditions_s_pol} can also be mapped to the reciprocal space,
\begin{align}
  \mathscr{F}
  \left[
    \vec{t}_2
    \cdot
    \vec{j}_\mathbf{x}^\mathrm{sou}
  \right]_{\mathbf{k}_l}
  &=
  \left[
    \mathsf{L}_{\mathbf{k}}
    \mathbb{E}_{s,\mathbf{k}}^{\Tra}
  \right]_l.
\end{align}
Since the source current stems from the induced polarisation in the dielectric medium,
it is composed of the electric field in the medium $\mathbb{E}_{s,\mathbf{k}}^{\Tra}$.
Note that we adopted the matrix-vector notation again,
where the coefficient matrix is given by
\begin{align}
  [\mathsf{L}_{\mathbf{k}}]_{lm}
  &=
  \frac{\alpha}{c}
  \frac{(l-m)\Omega}{-K_{\mathbf{k}_m}^{\ssl}}
  \times
  \operatorname{sgn}(\omega_m)
  \frac{k_{x,m}}{k_{\parallel,m}}
  J_{l-m}(\phi_{\mathbf{k}_m}^{\ssm\ssl}).
  \label{eq:L}
\end{align}
Here, we have a factor $(l-m)\Omega$ generated by the time derivative $\dot{a}_\mathbf{x}$ in the source current $\vec{j}_\mathbf{x}^\mathrm{sou}$ \eqref{eq:jsou} as we had $(l-m)g$ in the $\mathsf{N}$ matrix \eqref{eq:N} because of the space derivative $a_\mathbf{x}'$ in the stretching ratio $\eta$.

Finally, we can write the Fourier transform of the magnetic field discontinuity for the TE mode,
\begin{align}
  \mathsf{N}_{\mathbf{k}}^{\Inc}
  \mathbb{E}_{s,\mathbf{k}}^{\Inc}
  +
  \mathsf{N}_{\mathbf{k}}^{\Refl}
  \mathbb{E}_{s,\mathbf{k}}^{\Refl}
  -
  (
  \mathsf{N}_{\mathbf{k}}^{\Tra}
  +
  \mathsf{L}_{\mathbf{k}}
  )
  \mathbb{E}_{s,\mathbf{k}}^{\Tra}
  = 0.
  \label{eq:t_1_H_FT}
\end{align}
We simultaneously solve this equation and the other mode matching equation \eqref{eq:t_2_E_FT} to find the reflection and transmission matrices.

We arrange the two mode matching equations (\ref{eq:t_2_E_FT}, \ref{eq:t_1_H_FT}) in a matrix form,
\begin{align}
  \begin{pmatrix}
    \mathsf{M}_{\mathbf{k}}^{\Refl} 
    &
    -\mathsf{M}_{\mathbf{k}}^{\Tra}
    \\
    \mathsf{N}_{\mathbf{k}}^{\Refl} 
    &
    -(\mathsf{N}_{\mathbf{k}}^{\Tra} 
    +
    \mathsf{L}_{\mathbf{k}} 
    )
  \end{pmatrix}
  \begin{pmatrix}
    \mathbb{E}_{s,\mathbf{k}}^{\Refl}
    \\
    \mathbb{E}_{s,\mathbf{k}}^{\Tra}
  \end{pmatrix}
  = 
  \begin{pmatrix}
    -\mathsf{M}_{\mathbf{k}}^{\Inc}
    \mathbb{E}_{s,\mathbf{k}}^{\Inc}
    \\
    -\mathsf{N}_{\mathbf{k}}^{\Inc} 
    \mathbb{E}_{s,\mathbf{k}}^{\Inc}
  \end{pmatrix},
  \label{eq:s_pol}
\end{align}
and inverting the matrix on the left-hand side gives the reflection and transmission matrices for the TE mode,
\begin{align}
  \begin{pmatrix}
    \mathsf{R}_{s,\mathbf{k}}
    \\
    \mathsf{T}_{s,\mathbf{k}}
  \end{pmatrix}
  &=
  \begin{pmatrix}
    \mathsf{M}_{\mathbf{k}}^{\Refl} 
    &
    -\mathsf{M}_{\mathbf{k}}^{\Tra}
    \\
    \mathsf{N}_{\mathbf{k}}^{\Refl} 
    &
    -(
    \mathsf{N}_{\mathbf{k}}^{\Tra} 
    +
    \mathsf{L}_{\mathbf{k}}
    )
  \end{pmatrix}^{-1}
  \begin{pmatrix}
    -\mathsf{M}_{\mathbf{k}}^{\Inc}
    \\
    -\mathsf{N}_{\mathbf{k}}^{\Inc}
  \end{pmatrix}.
  \label{eq:RTs}
\end{align}

\section{Numerical implementation}
\label{sec:numerical}
In this subsection, we discuss the numerical computation of the reflection and transmission matrices derived above.
Firstly, we confirm the coefficient matrices $\mathsf{M, N}$ and $\mathsf{L}$ are invariant when we scale parameters keeping any two lengths as in the calculation of photonic crystals \cite{sakoda2004optical}.
All of the matrix elements consist of dimensionless numbers such as $K_\mathbf{k}^{\sigma \tau} A$ and $g/K_\mathbf{k}^{\sigma \tau}$.
These quantities are invariant under scaling the modulation depth and the reciprocal vectors,
\begin{align}
    A \mapsto \nu A,
    \quad
    \mathbf{q} \mapsto \nu^{-1} \mathbf{q},
    \quad
    \mathbf{k} \mapsto \nu^{-1} \mathbf{k},
\end{align}
where $\nu > 0$ is the scaling parameter.

In the analytical calculation, the coefficient matrices are of infinite ranks (i.e.~$l,m \in [-\infty,+\infty]$) because they are generated by the Fourier transform.
When we implement the calculation in numerics, we have to truncate the $\mathsf{M}, \mathsf{N}$ and $\mathsf{L}$ matrices to finite rank ones.
Here, we introduce the cutoff number $m_c$ so that the subscripts $l$ and $m$ are finite (i.e.~$l,m \in [-m_c, +m_c]$).
We can justify this truncation as in the conventional Fourier modal method because our boundary profile is differentiable, and the media above and below the boundary is homogeneous
\cite{%
  li1999justification,
  shcherbakov2013efficient%
}.

\begin{figure}[htbp]
  \centering
  \includegraphics[width=.7\linewidth]
  {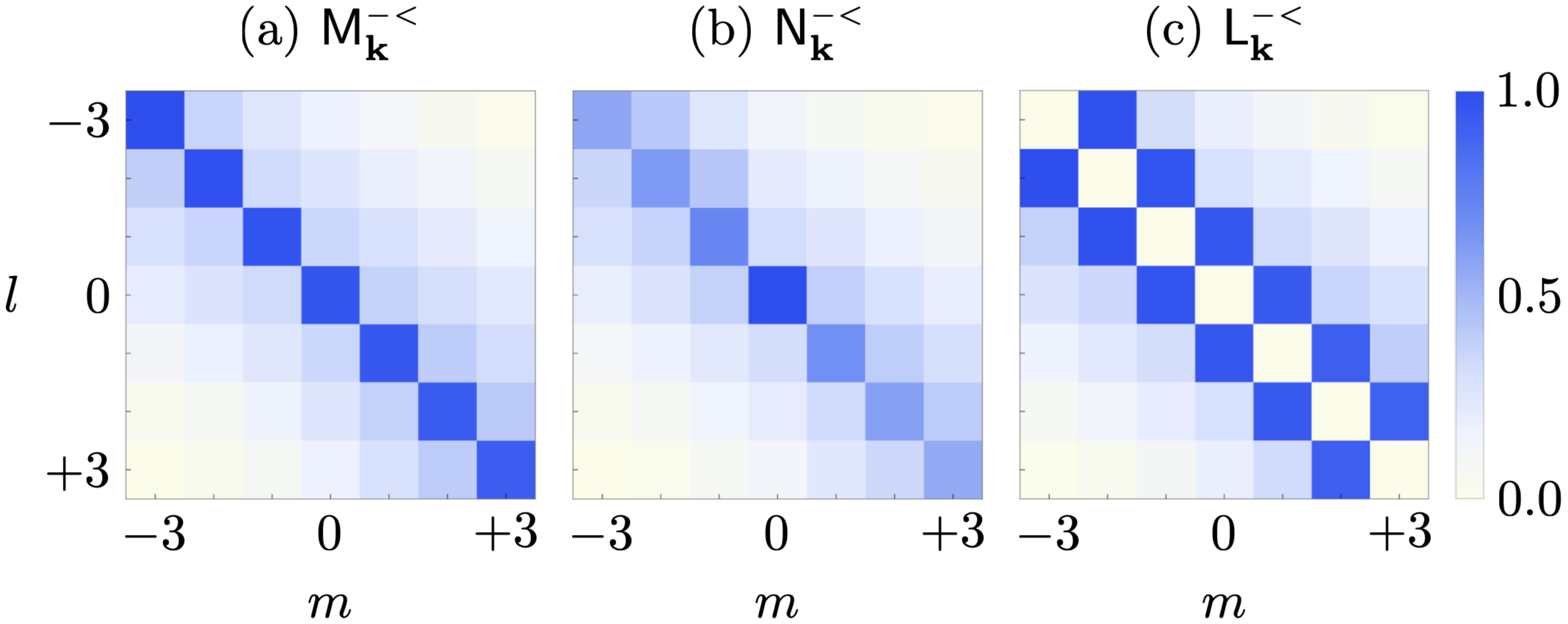}
  \caption{
    Typical distribution of the matrix elements.
    The absolute values of the matrix elements in $\mathsf{M}$, $\mathsf{N}$ and $\mathsf{L}$ are plotted.
    The common colorbar is shown on the right of (c).
    The horizontal and vertical axes are the column and row indices.
    Note that all elements are normalised by the maximum element in each matrix.
    In these plots,
    we use the following parameters:
    $\epsilon^{\ssg} = 1.0$,
    $\epsilon^{\ssl} = 2.25$,
    $A = 10\ \mathrm{[nm]}$,
    $g = 2\pi\ \mathrm{[\mu m^{-1}]}$,
    $\Omega = 0.2gc$,
    $\omega_\mathrm{in} = 0.8 gc$,
    $\theta_\mathrm{in} = 0$,
    $k_y = 0$.
    The cutoff number is $m_c = 3$.
  }
  \label{fig:decay}
\end{figure}
We can also confirm that the truncation works by plotting the matrix elements.
In \figref{fig:decay}, each matrix element of the coefficient matrices is shown.
We can find that the large elements,
which dominates the matrix calculation,
are localised near the diagonal elements.
This implies the truncation, neglection far off-diagonal elements, is justified.

If the grating depth is large compared with other length scales (e.g.~$gA > 1$), large matrix elements do not localise near the diagonal elements any longer.
This phenomenon occurs even in the conventional grating calculation handled by the differential formalism 
\cite{%
  li1996improvement,%
  li1996use,%
  xu2014simple,%
  xu2017numerical,%
  xu2020numerical,%
  shcherbakov2013efficient,%
  shcherbakov20153d,%
  shcherbakov2017generalized,%
  shcherbakov2018direct,%
  shcherbakov2019curvilinear,%
  essig2010generation,%
  felix2014local%
}
as mentioned at the beginning of this chapter.
In that regime, we have to adopt local coordinate distortion instead of the global coordinate translation \eqref{eq:boundary}.

\chapter{Asymmetric diffraction and \v{C}erenkov radiation from dynamical gratings}
\label{ch:superluminal}
\section{Asymmetric diffraction} 
\label{sec:asymmetric}
\begin{figure}[htbp]
  \centering
  \includegraphics[width=.6\linewidth]
  {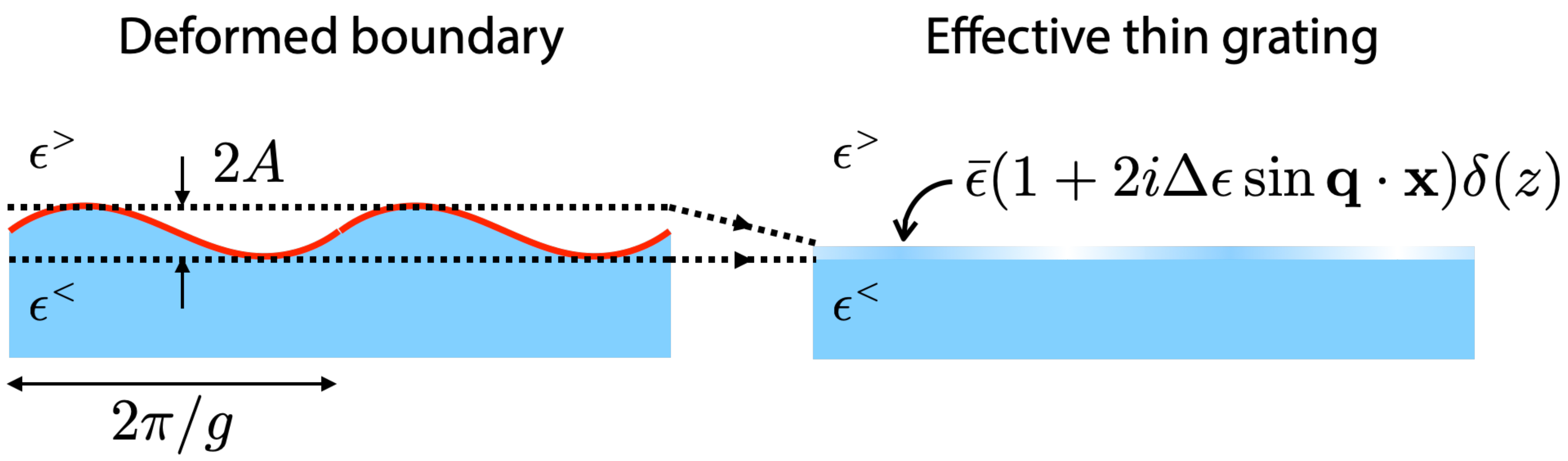}
  \caption{
    The dynamically modulated boundary and the effective thin grating.
    Averaging the permittivity in the $z$ direction within the modulated region ($-A \leq z \leq +A$) generates a very thin grating model.
  }
  \label{fig:homogenisation}
\end{figure}

\begin{figure}[htbp]
  \centering
  \includegraphics[width=.8\linewidth]
  {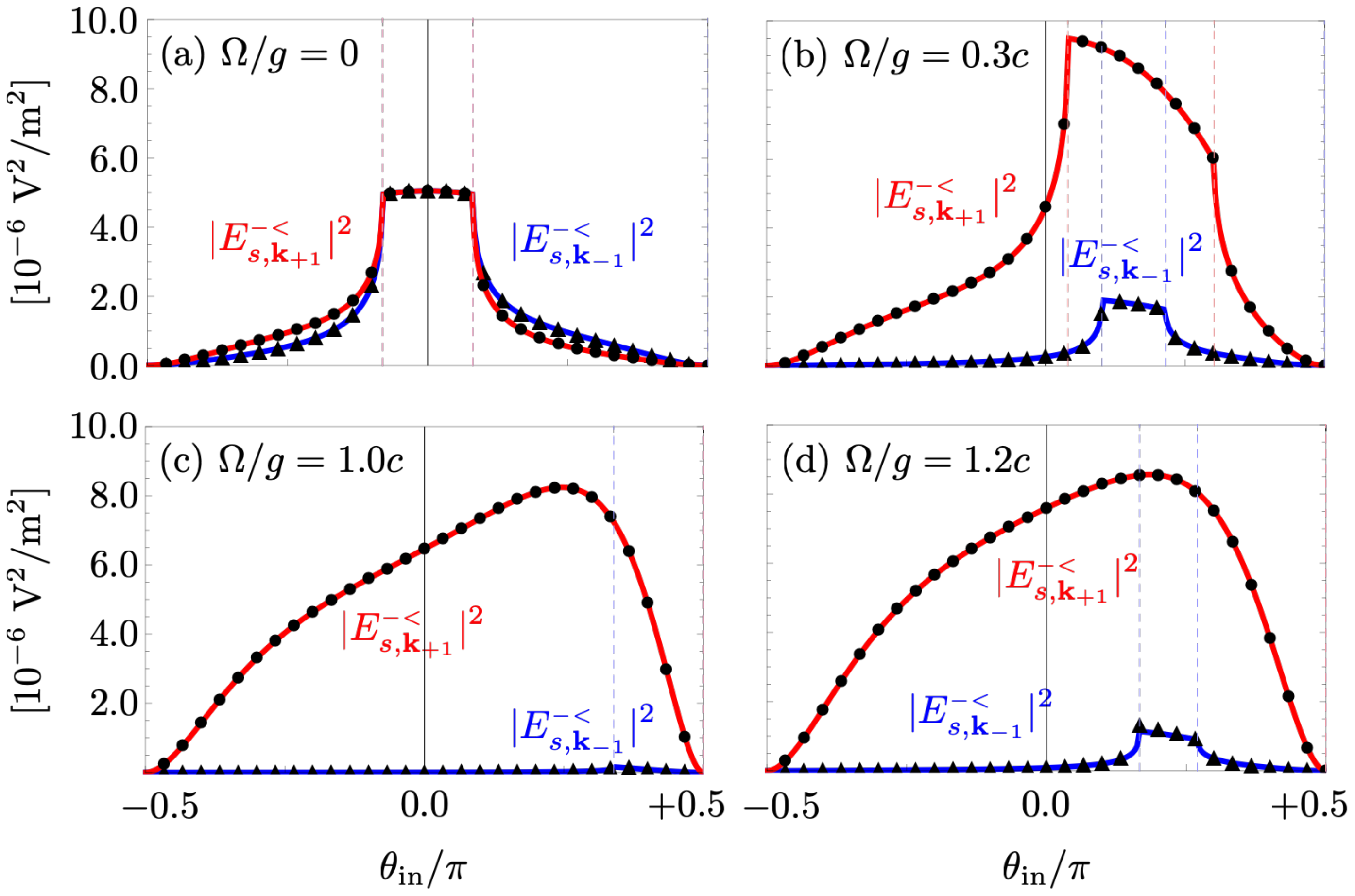}%
  \caption{
    Incident angle dependencies on the first-order diffraction intensities.
    The horizontal axes are the incident angle defined by Eq.~\eqref{eq:theta_in}.
    The blue (red) curve corresponds to the negative (positive) diffraction $|E_{s,\mathbf{k}_{-1}}^{\ssm\ssl}|^2$ ($|E_{s,\mathbf{k}_{+1}}^{\ssm\ssl}|^2$) inside the dielectric medium calculated by the dynamical differential formalism.
    The black circles and triangles are generated by the effective medium description.
    In these figures,
    we substitute the following parameters:
    the permittivity of the upper medium $\epsilon^{\ssg} = 1.0$,
    the permittivity of the lower medium $\epsilon^{\ssl} = 2.25$,
    the corrugation wavenumber $g = 2\pi\ \mathrm{[\mu m^{-1}]}$,
    the corrugation depth $A = 1\ \mathrm{[nm]}$,
    the input polarisation $\lambda = s$,
    the input electric field $E_{s,\mathbf{k}_0}^{\Inc} = 1.0\ \mathrm{[V\cdot m^{-1}]}$,
    and the input frequency $\omega_\mathrm{in} = 0.8 gc$.
    We focus on the in-plane condition $k_y = 0$.
    The cutoff number is $m_c = 3$.
  }
  \label{fig:angular}
\end{figure}

\begin{figure}[htbp]
  \centering
  \includegraphics[width=.8\linewidth]
  {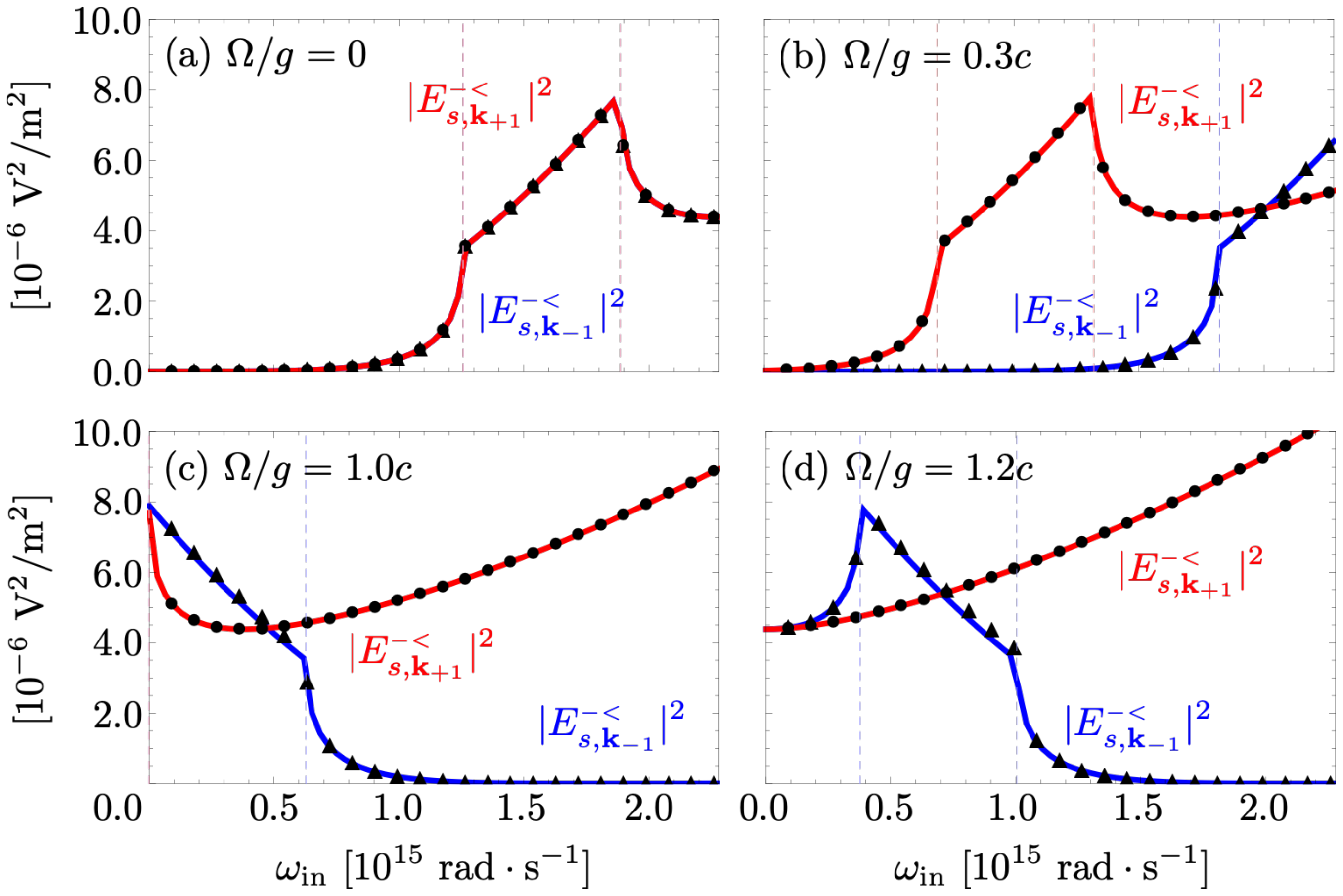}
  \caption{
    Frequency spectra of the first-order diffraction amplitudes.
    The blue (red) curve corresponds to the negative (positive) diffraction amplitude $|E_{s,\mathbf{k}_{-1}}^{\ssm\ssl}|^2$ ($|E_{s,\mathbf{k}_{+1}}^{\ssm\ssl}|^2$) calculated by the dynamical differential formalism.
    Note that red and blue curves are completely overlapping one another in (a).
    In these plots,
    we consider the normal incidence $\theta_\mathrm{in} = 0$,
    and the other parameters are the same as in \figref{fig:angular}.
  }
  \label{fig:frequency}
\end{figure}

\begin{figure}[htbp]
  \centering
  \includegraphics[width=.8\linewidth]
  {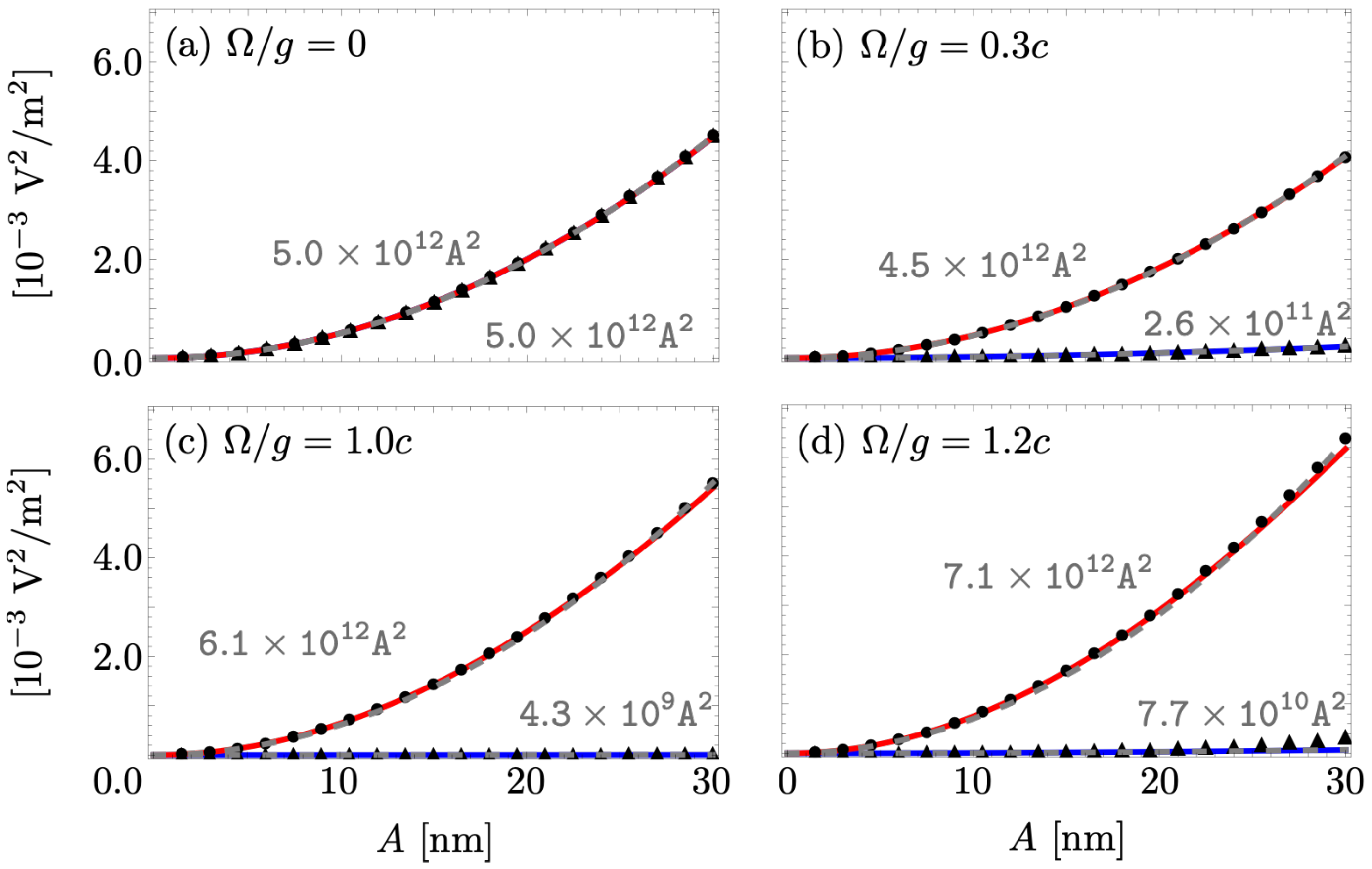}%
  \caption{
    Modulation strength dependence of the first-order diffraction intensities.
    The blue (red) curve corresponds to the negative (positive) diffraction $|E_{s,\mathbf{k}_{-1}}^{\ssm\ssl}|^2$ ($|E_{s,\mathbf{k}_{+1}}^{\ssm\ssl}|^2$) inside the dielectric medium calculated by the dynamical differential formalism.
    The black circles and triangles are produced by the effective medium approach.
    Both positive and negative intensities quadratically depend on the modulation strength as the data is fitted by parabolic curves (grey dashed curves).
    The fitting equations are shown in each figure.
    Note that positive and negative diffraction intensities are completely overlapping one another in (a).
    In these plots,
    we consider the normal incidence $\theta_\mathrm{in} = 0$,
    and the other parameters are the same as in \figref{fig:angular}.
  }
  \label{fig:amplitude}
\end{figure}

Let us compute the reflection and transmission matrices and investigate the dependences on the modulation and incident field parameters.
The results are shown in Figure \ref{fig:angular}--\ref{fig:amplitude} and checked by an effective surface description,
where the modulated surface is averaged in the $z$ direction, and an effective thin grating is considered instead of the curved boundary (\figref{fig:homogenisation}).
In the effective description,
the thin wall part is analogous to spatiotemporally modulated media 
\cite{%
  sounas2017non,%
  caloz2018electromagnetic,%
  shaltout2019spatiotemporal,%
  galiffi2019broadband,%
  huidobro2019fresnel%
}.
See Appendix \ref{app:effective} for more details of the effective surface description.

In \figref{fig:angular}, the angular dependences of the transmission amplitudes are shown for various modulation velocities $\Omega/g$.
The incident angle is defined by the radiation wavenumber,
\begin{align}
  \cos \theta_\mathrm{in}
  &= \frac{\sigma K_{\mathbf{k}}^{\ssg}}{|k_{0}|},
  \quad
  \sin \theta_\mathrm{in}
  = \frac{k_{\parallel}}{|k_{0}|}.
  \label{eq:theta_in}
\end{align}
The solid red and blue lines are the positive and negative first-order diffraction amplitudes in the dielectric medium calculated by the dynamical differential formalism.
The circle and triangle markers are produced by the effective surface description (see Appendix \ref{app:effective}).
We can recognise the results produced by the two approaches agree well.
At dashed lines corresponding to the emergence of diffraction modes
($K_{\mathbf{k}_{\pm 1}}^\tau=0$),
the angular spectra are singular.
That is a Wood grating anomaly \cite{wood1902xlii}.
From \figref{fig:angular} (a),
we can find that the spectra are symmetric with respect to the normal incidence $\theta_\mathrm{in} = 0$ if the boundary is static one ($\Omega/g = 0$).
This spectral feature recovers the fact that sinusoidal gratings show symmetric diffraction patterns.
If we turn on the temporal modulation, $\Omega/g > 0$ [\figref{fig:angular} (b--d)],
the angular spectra are pulled towards the positive $\theta_\mathrm{in}$ direction.
Besides, the positive diffraction has a larger amplitude than it has in the static case.
This is similar to the Fresnel drag and the amplification of the field travelling in the same direction as the spatiotemporal modulation in bulky materials \cite{huidobro2019fresnel,pendry2020new,pendry2021gain}.
We can also confirm that the forward scattering (the positive diffraction) at the luminal condition, $\Omega/g = c$, is dominant as it is in the bulky media \cite{galiffi2021photon}.
This is a consequence of the spatiotemporal modulation.

In \figref{fig:frequency},
the incidence frequency dependences of the transmission amplitudes are shown.
Here, we can confirm the result agree well with the effective surface description again.
For the static boundary, $\Omega/g = 0$ [\figref{fig:frequency} (a)], the positive and negative diffraction has the same input frequency dependence.
This is consistent with the ordinary diffraction by sinusoidal gratings.
As the temporal modulation frequency increases [\figref{fig:frequency} (b--d)], the positive (negative) diffraction spectrum shifts left (right) due to the Doppler frequency shift.
We can also find that the diffraction amplitudes are finite even at zero input frequency $\omega_\mathrm{in}=0$ if we turn on the temporal modulation, $\Omega/g > 0$.
In other words, if we temporally modulate the boundary with DC voltage applied, the boundary could emit radiation.
We will discuss it in detail in the next section.
This is another consequence of spatiotemporal modulation.

Let us finally check the modulation depth dependence of the diffraction amplitudes.
\figref{fig:amplitude} (a--d) are the diffraction amplitudes as functions of the modulation depth for various modulation velocity.
We can see the diffraction amplitudes are quadratically dependent on the modulation depth.
This is because our configuration is invariant under $A \mapsto -A$, and the odd order response vanishes similarly to the fact that the Casimir-induced instability is the second-order effect [recall the discussion around Eq.~\eqref{eq:DU_expanded}].
The positive and negative diffraction amplitudes have the same dependence on the modulation amplitude if the boundary is a static one [\figref{fig:amplitude} (a)].
In contrast, the positive diffraction is dominant over the negative diffraction once the modulation is turned on [\figref{fig:amplitude} (b--d)] as we have seen in the angular spectra.

\section{\v{C}erenkov radiation in vacuum from a superluminal grating}
\label{sec:cerenkov-sg}
Even in case of the zero-frequency input, $\omega_\mathrm{in} = 0$, 
which corresponds to applying an electrostatic voltage,
diffraction emerges unlike the static boundary case as we found in the previous section.
If the boundary is static one or in the subluminal regime ($v_\mathrm{ph} \equiv \Omega/g < c/\sqrt{\epsilon^\tau}$),
the wavenumber in the $z$ direction is imaginary for each diffraction order $m$, 
\begin{align}
  K_{\mathbf{k}_m}^\tau 
  = i|m|g
  \sqrt{
    1 - \frac{{v_\mathrm{ph}}^2}{(c/\sqrt{\epsilon^\tau})^2}
  },
\end{align}
where $v_\mathrm{ph} \equiv \Omega/g$ is the phase velocity with which the boundary pattern shifts.
Therefore, all the diffraction modes are evanescent ones, and no far-field radiation is observed.
In contrast, 
all the diffraction modes in each medium become propagating ones in the superluminal regime ($v_\mathrm{ph} > c/\sqrt{\epsilon^\tau}$), where the wavenumber in the $z$ direction is real for each diffraction order $m$.
The propagation direction is given by the diffraction angle,
\begin{align}
  \theta_m^\tau
  \equiv 
  \tan^{-1} \frac{K_{\mathbf{k}_m}^\tau}{k_{x,m}}
  = \tan^{-1}
  \sqrt{
    \frac{{v_\mathrm{ph}}^2}{(c/\sqrt{\epsilon^\tau})^2} - 1
  }
  \equiv 
  \theta_\mathrm{\check{C}R}^\tau.
  \label{eq:theta_m}
\end{align}
We can find that the far right-hand side is equivalent to the \v{C}erenkov angle \eqref{eq:theta_CR} that is independent of the diffraction order $m$.
This means that all the diffraction modes propagate in the same direction.
Since different diffraction has a different frequency,
all diffraction modes are superposed and result in a series of pulses.
This is analogous to the conventional \v{C}erenkov radiation.

\begin{figure}[htbp]
  \centering
  \includegraphics[width=.55\linewidth]
  {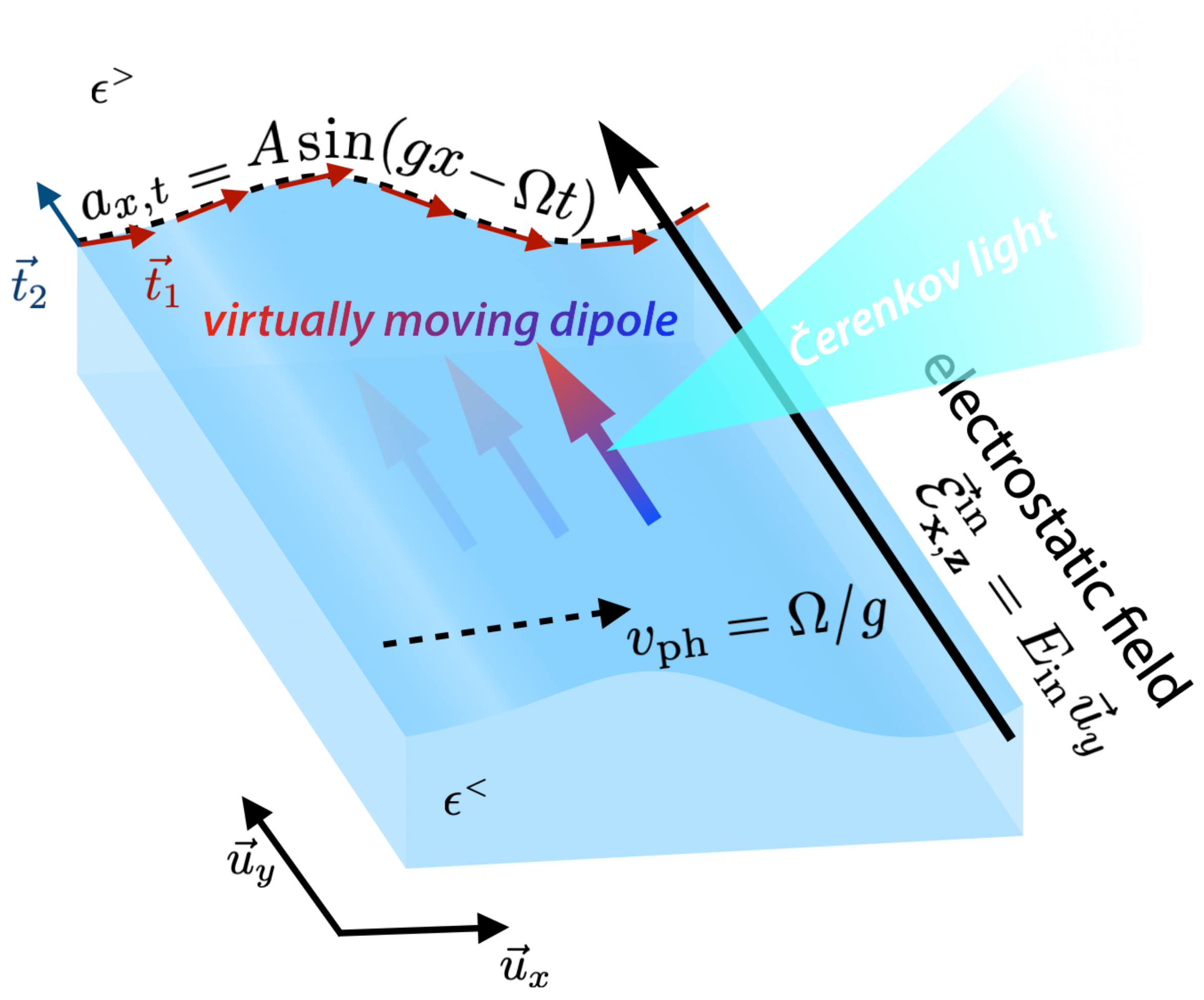}
  \caption{
    Single interface system composed of a dielectric and vacuum.
    The interface is under the spatiotemporal modulation of travelling wave type \eqref{eq:boundary}.
    The orthonormal tangential vectors, $\vec{t}_{1,2}$ can be calculated from the interface profile.
    The permittivities above and below the interface are denoted as $\epsilon^\ssg$ and $\epsilon^\ssl$,
    respectively.
    We apply a uniform electrostatic field 
    $\vec{\mathcal{E}}_{\mathbf{x},z}^\mathrm{in}$
    on the modulated interface so that there are induced dipoles that travel on the interface due to its profile of the travelling wave type.
    The velocity of the grating $v_\mathrm{ph} = \Omega/g$ can be tuned by two independent parameters so that it can exceed the speed of light.
    When the speed is faster than light,
    the induced dipoles emit \v{C}erenkov radiation.
  }
  \label{fig:fig1}
\end{figure}
In our setup, the radiation is emitted from dipoles at the boundary induced by the DC voltage (\figref{fig:fig1}).
The induced dipoles travel at the modulation phase velocity $v_\mathrm{ph}$ and mimic superluminal particles that travel faster than light.
Since the dipoles are time-varying, they are regarded as a surface electric current \eqref{eq:jsou} similarly to swift electrons in the conventional \v{C}erenkov setup (\figref{fig:Frank-Tamm}).
\v{C}erenkov emission indicates that linear momenta taken away from the surface by the radiation and hence stress on the surface as the back action.
This is closely related to the quantum friction \cite{pendry1997shearing,pendry1998can,pendry2010quantum,manjavacas2010vacuum,maghrebi2013quantum,milton2016reality,farias2019motion}.
In that setup,
dipoles on a moving surface are produced by quantum fluctuation instead of the external DC voltage,
which emit electromagnetic radiation and result in friction as the back action.

\begin{figure}[htbp]
  \centering
  \includegraphics[width=\linewidth]
  {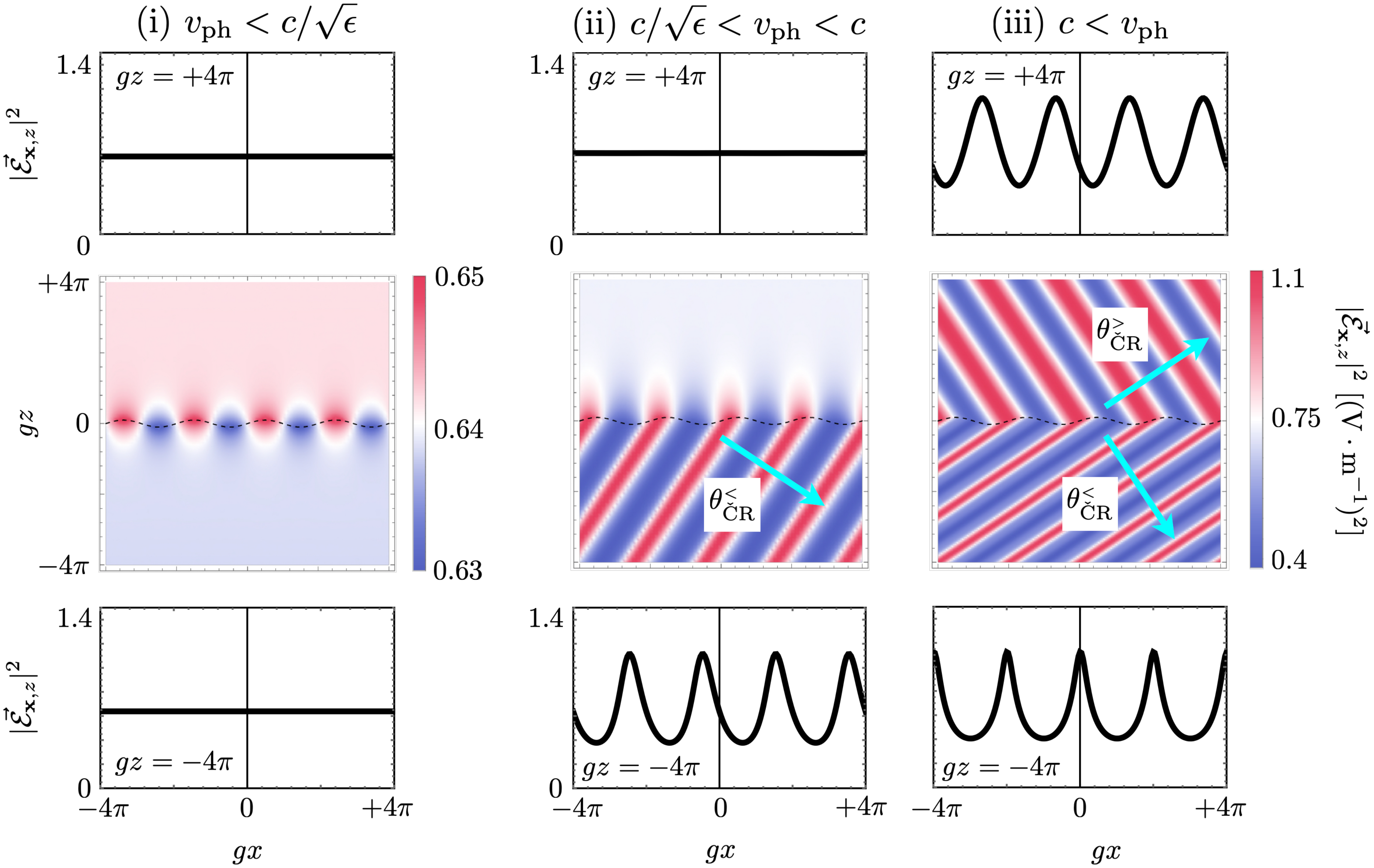}
  \caption{
    Snapshots (middle row) and their cross-sections (upper and lower rows) of the field distributions in the subluminal and superluminal regimes.
    The intensity of the total electric field $|\vec{\mathcal{E}}_{\mathbf{x},z}|^2$ is plotted.
    (i) 
    Subluminal regime in the dielectric and vacuum sides,
    $v_\mathrm{ph} = \Omega/g = 0.2c < c/\sqrt{\epsilon}$.
    (ii)
    Superluminal in the dielectric side and subluminal in the vacuum side,
    $c/\sqrt{\epsilon} < v_\mathrm{ph} = 0.8c < c$.
    (iii) 
    Superluminal regime in both sides,
    $c < v_\mathrm{ph} = 1.2c$.
    The modulation parameters are given as following:
    the spatial frequency $g = 2\pi\ [\mathrm{\mu m^{-1}}]$,
    the modulation depth $2A = 100\ [\mathrm{nm}]$.
    The input amplitude is $E_\mathrm{in} = 1\ \mathrm{[V\cdot m^{-1}]}$.
    The permittivities are $\epsilon^\ssg = 1$ and $\epsilon^\ssl = \epsilon = 2.25$.
    The cutoff is $m_c = 10$ so that we take $2m_c+1 = 21$ diffracted waves into account. 
    The color bars for (ii) and (iii) are common and shown on the right of (iii).
    Note that the scale of the color bar of (i) is different from those of (ii) and (iii),
    and the horizontal axes and the vertical axes of the color plots are normalised by the spatial period $g$ of the modulation.
  }
  \label{fig:field_cross}
\end{figure}

We can visually confirm \v{C}erenkov radiation from the modulated boundary by reconstructing the field distribution as shown in \figref{fig:field_cross}.
Once we compute the diffraction amplitudes $E_{\lambda,\mathbf{k}}^{\sigma \tau}$ as we did in the previous section, we can recover the field distribution in the real space according to the Fourier expansion,
\begin{align}
  \begin{cases}{}
    \vec{\mathcal{E}}{}_{\mathbf{x},z}^\mathrm{inc} 
    = \displaystyle{\int_{\mathbf{k}}}
    e^{i\mathbf{k}\cdot \mathbf{x}}
    \vec{E}{}_{\mathbf{k},z}^{\Inc},
    \quad
    \vec{\mathcal{H}}{}_{\mathbf{x},z}^\mathrm{inc} 
    = \displaystyle{\int_{\mathbf{k}}}
    e^{i\mathbf{k}\cdot \mathbf{x}}
    \vec{H}{}_{\mathbf{k},z}^{\Inc}, 
    & 
    (z \geq a_\mathbf{x}),
    \vspace{.5em}
    \\
    \vec{\mathcal{E}}{}_{\mathbf{x},z}^\mathrm{ref} 
    = \displaystyle{\int_{\mathbf{k}}}
    e^{i\mathbf{k}\cdot \mathbf{x}}
    \vec{E}{}_{\mathbf{k},z}^{\Refl},
    \quad
    \vec{\mathcal{H}}{}_{\mathbf{x},z}^\mathrm{ref} 
    = \displaystyle{\int_{\mathbf{k}}}
    e^{i\mathbf{k}\cdot \mathbf{x}}
    \vec{H}{}_{\mathbf{k},z}^{\Refl}
    &
    (z \geq a_\mathbf{x}),
    \vspace{.5em}
    \\
    \vec{\mathcal{E}}{}_{\mathbf{x},z}^\mathrm{tra} 
    = \displaystyle{\int_{\mathbf{k}}}
    e^{i\mathbf{k}\cdot \mathbf{x}}
    \vec{E}{}_{\mathbf{k},z}^{\Tra},
    \quad
    \vec{\mathcal{H}}{}_{\mathbf{x},z}^\mathrm{tra} 
    = \displaystyle{\int_{\mathbf{k}}}
    e^{i\mathbf{k}\cdot \mathbf{x}}
    \vec{H}{}_{\mathbf{k},z}^{\Tra}
    &
    (z \leq a_\mathbf{x}).
    \vspace{.5em}
  \end{cases}
  \tag{\ref{eq:E,H_lnc,ref,tra}}
\end{align}
In \figref{fig:field_cross}, we can confirm no far field radiation in the subluminal regime ($v_\mathrm{ph} < c/\sqrt{\epsilon^\ssl}$), radiation on both vacuum and dielectric sides in the superluminal regime ($v_\mathrm{ph} > c/\sqrt{\epsilon^\ssg}$) and an only on the dielectric side between the two regimes ($c/\sqrt{\epsilon^\ssl} < v_\mathrm{ph} < c/\sqrt{\epsilon^\ssg}$).

As for the experimental verification of the proposed effect, the spatiotemporal modulation of boundaries is the key.
One can use acoustic techniques to modulate the surfaces of materials mechanically,
where ultrafast modulation up to $1\ \mathrm{[THz]}$ is achievable by acoustic-optical modulators
\cite{%
chenu1994giant,%
mante2010generation%
}.
Using soft materials such as gels and liquids is one way to have large surface displacement up to micrometre scale
\cite{%
issenmann2006bistability,%
issenmann2008deformation,%
rambach2016visualization%
}.

One may also electrostatically modulate the permittivity of atomically thin material,
which is effectively behave as an infinitely thin sheet with a conductivity
\cite{gonccalves2016introduction}
Recently, 
in the plasmonics community,
they showed that graphene could be electrically modulated at a very high frequency around terahertz to have multi functionalities such as a broadband absorber in the terahertz range, a high-efficiency plasmonic coupler and an atomically thin lens with a tunable focal length
\cite{%
galiffi2019broadband,%
galiffi2020wood,%
kashef2020multifunctional%
}.

\chapter{Conclusion of Part I\hspace{-.1em}I}
\label{ch:conclusion_cerenkov}
In conclusion, we found asymmetric diffraction and \v{C}erenkov radiation from a spatiotemporally modulated surface.
We started by constructing a differential formalism to calculate light scattering at spatiotemporally modulated surfaces.
Our method is based on the Chandezon technique of coordinate translation generated by the surface profile.
The global coordinate translation enables us to work on a coordinate system where our modulated surface is flat.
In the translated coordinate system, we can unambiguously impose boundary conditions on the surface.
We derived the boundary conditions by integrating the Maxwell--Heaviside equations along contours that enclose the surface.
Performing the Fourier expansion of electromagnetic fields and combining with the boundary conditions,
we can write simultaneous equations that associate the scattered (diffracted) fields amplitudes with the incident ones in the reciprocal space.

We found the simultaneous equations contain a source term induced by temporal modulation.
We investigated the effects of the source term on the diffraction in a simple setup by numerically solving the simultaneous equations after truncating the Fourier expansion.
In this study, we focused on the surface modulation profile of a travelling wave type.
The spatial and temporal modulation frequencies can be independently controlled.
In the absence of temporal modulation,
the angular diffraction spectrum of a surface is symmetric with respect to the normal incidence condition.
In contrast, the diffraction symmetry is broken in the presence of the temporal modulation.
The radiation originating from the induced source term makes the diffraction spectrum asymmetric.
This is a consequence of the modulation-induced source term.

The source term can be controlled by the spatial and temporal modulation frequencies, $g$ and $\Omega$.
By choosing modulation frequencies such that $\Omega/g > c$,
our source term mimics moving dipoles along the surface whose travelling velocity is greater than the speed of light in a vacuum.
Accordingly, \v{C}erenkov radiation is emitted from the modulated surface.
We visually confirmed this radiation of \v{C}erenkov type by reconstructing the field pattern in the real space from the diffraction amplitudes in the reciprocal space.
This is another interesting consequence of the modulation-induced source term.

In this calculation, we focused on a dielectric surface where we can neglect dispersion and magnetic response.
I would leave the dispersion correction and effects of magnetic response for future works.


\chapter{Overall conclusion and future outlook}
\label{ch:conclusions}

Overall, in this thesis, we have analysed electromagnetic properties of two kinds of dynamical metasurfaces, a deformable metallic surface and a spatiotemporally modulated dielectric surface.

In the first part of this thesis,
we have discussed the Casimir effect contribution to the structural stability of a metallic film.
We found that introducing corrugation to the film surface decrease the zero-point energy of the system.
The zero-point contribution overcomes the surface tension contribution if the film is thinner than a critical thickness,
leading to the structural instabilities of a mercury film in the nanoscale.
Our findings will provide a fundamental thickness limit of nanoscale materials.

In the second part of this thesis,
we have investigated the asymmetric diffraction and \v{C}erenkov radiation from a dielectric surface modulated in space and time.
We developed a dynamical differential formalism where we apply global coordinate translation to map the dynamically corrugated surface to a flat one, and we can unambiguously calculate the reflection and transmission at the surface.
We found that the dynamical modulation induces a source term (surface electric current).
The source term brings about the diffraction symmetry breaking and \v{C}erenkov type radiation.
These findings will pave the way to light generation and control by dynamic modulations. 

There are several possible future works related to the second part, as listed below.
\begin{itemize}
  \item magnetic response ($\mu > 1$):\\ 
    When the input field frequency is in the microwave region,
    we cannot neglect the magnetic response of materials sometimes.
    In this case,
    temporally modulating the surface of a medium induces a surface magnetic current as we had the surface electric current.
  \item out-of-plane incidence:\\
    In order to tackle Casimir type problems,
    it is important to consider all possible modes.
    Although we focused on the in-plane incidence ($k_y = 0$) in the second part,
    the field can be incident on the surface with $k_y \neq 0$.
    In such a case,
    the grating mixes TE and TM modes,
    and we need to solve the two modes simultaneously.
  \item deep grating:\\
    Handling deep gratings calls for local coordinate distortion instead of global translation.
    The dynamical coordinate distortion could be done with the help of the general theory of relativity.
  \item dispersion correction:\\
    In order to take the dispersion into consideration,
    we have to consider the dispersive response at each point of the medium and track the motion of each point.
    As a consequence, the permittivity is modified as following:
    \begin{align}
      \epsilon_{\vec{r}} 
      \rightarrow 
      \epsilon_{\vec{r}}(t-t'),
    \end{align}
    where $\vec{r}$ specifies a point in a medium,
    and the argument $t-t'$ represents the dispersive response.
    Since temporal modulation puts each point of the medium in motion,
    $\vec{r}$ is also a time-dependent quantity [i.e.~$\vec{r} = \vec{r}(t)$].
\end{itemize}

\appendix
\chapter{Krein's formula}
\label{app:Krein}
Here, I provide a breif derivation of Krein's formula,
The mode density can be defined by the imaginary part of the retarded Green's function,
\begin{align}
  \rho 
  \equiv 
  \frac{-1}{\pi}
  \operatorname{Im} \operatorname{Tr} G^\mathrm{R}.
\end{align}
Substitute the Dyson equation $G^\mathrm{R} = G_0^\mathrm{R} + G_0^\mathrm{R} V G^\mathrm{R}$,
where $G_0^\mathrm{R}=(E+i0-H_0)^{-1}$ is Green's function for free space,
we can obtain
\begin{align}
  \operatorname{Im} \operatorname{Tr} G^\mathrm{R} 
  &=
  \operatorname{Im} \operatorname{Tr} G_0^\mathrm{R} +
  \operatorname{Im} \operatorname{Tr} G_0^\mathrm{R} V G^\mathrm{R},
  \\
  -\pi(\rho - \rho_0)
  &=
  \operatorname{Im} \operatorname{Tr} G_0^\mathrm{R} V G^\mathrm{R},
  \\ 
  \Delta \rho
  \equiv
  \rho - \rho_0
  &=
  \frac{-1}{\pi}
  \operatorname{Im} \operatorname{Tr} 
  G_0^\mathrm{R} V (1 + G_0^\mathrm{R} V + (G_0^\mathrm{R} V)^2 + \cdots) G_0^\mathrm{R} 
  \\
  &=
  \frac{-1}{\pi}
  \operatorname{Im} \operatorname{Tr} 
  G_0^\mathrm{R} V \sum_{n=0}^{\infty} (G_0^\mathrm{R} V)^n G_0^\mathrm{R}
  \\
  &=
  \frac{-1}{\pi}
  \operatorname{Im} \operatorname{Tr}
  G_0^\mathrm{R} G_0^\mathrm{R} V 
  \sum_{n=0}^{\infty} (G_0^\mathrm{R} V)^n. 
\end{align}
Remind that the trace is invariant under the cyclic permutation of the arguments $\operatorname{Tr}(ABC) = \operatorname{Tr}(CAB)$,
Since The derivative of the retarded Green's function in the free space is given by 
\begin{align}
  \frac{\mathrm{d}G_0^\mathrm{R}}{\mathrm{d}E}
  &= -(E+i0-H_0)^{-2} = -G_0^\mathrm{R} G_0^\mathrm{R},
\end{align}
we can write 
\begin{align}
  \Delta \rho
  &=
  \frac{-1}{\pi}
  \operatorname{Im} \operatorname{Tr}
  \left(
    -\frac{\mathrm{d}G_0^\mathrm{R}}{\mathrm{d}E}
  \right)V 
  \sum_{n=0}^{\infty} (G_0^\mathrm{R} V)^n,
\end{align}
Applying the chain rule,
we can rewrite
\begin{align}
  \Delta \rho
  &=
  \frac{-1}{\pi}
  \operatorname{Im} \operatorname{Tr}
  \sum_{n=1}^{\infty} \frac{-1}{n} \frac{\mathrm{d}}{\mathrm{d}E}(G_0^\mathrm{R} V)^n 
  =
  \frac{-1}{\pi}
  \operatorname{Im} 
  \frac{\mathrm{d}}{\mathrm{d}E}
  \operatorname{Tr}
  \sum_{n=1}^{\infty} \frac{-1}{n} (G_0^\mathrm{R} V)^n.
\end{align}
Recalling the Taylor expansion of the natural logarithmic function,
\begin{align}
  \ln (1-X) = \sum_{n=1}^\infty \frac{-1}{n}X^n,
\end{align}
we can obtain
\begin{align}
  \Delta \rho
  &=
  \frac{-1}{\pi}
  \operatorname{Im}
  \frac{\mathrm{d}}{\mathrm{d}E}
  \operatorname{Tr}
  \ln (1 - G_0^\mathrm{R} V),
  \\
  &=
  \frac{-1}{2\pi i}
  \frac{\mathrm{d}}{\mathrm{d}E}
  \operatorname{Tr}
  \left[
  \ln (1 - G_0^\mathrm{R} V)
  - \ln (1 - G_0^\mathrm{A} V)
  \right],
  \\
  &=
  \frac{1}{2\pi i}
  \frac{\mathrm{d}}{\mathrm{d}E}
  \operatorname{Tr}
    \ln 
    (1 - G_0^\mathrm{A} V)
    (1 - G_0^\mathrm{R} V)^{-1},
  \\
  &= 
  \frac{1}{2\pi i}
  \frac{\mathrm{d}}{\mathrm{d}E}
  \operatorname{Tr}
    \ln 
  \left[
    \sum_{n=0}^\infty (G_0^\mathrm{R}V)^n
    - G_0^\mathrm{A} V \sum_{n=0}^\infty (G_0^\mathrm{R}V)^n
  \right],
  \\
  &=
  \frac{1}{2\pi i}
  \frac{\mathrm{d}}{\mathrm{d}E}
  \operatorname{Tr}
    \ln 
  \left[
    1 + G_0^\mathrm{R}T
    - G_0^\mathrm{A} V V^{-1} T
  \right]
  \\
  &=
  \frac{1}{2\pi i}
  \frac{\mathrm{d}}{\mathrm{d}E}
  \operatorname{Tr}
    \ln 
  \left[
    1 -
    (G_0^\mathrm{A}
    -
    G_0^\mathrm{R})T
  \right]
\end{align}
where the advanced Green's function is given by 
$G_0^\mathrm{A} = (E-i0-H_0)^{-1}$,
and the transfer matrix is defined by
$T:=V\displaystyle{\sum_{n=0}^\infty }(G_0^\mathrm{R}V)$.
Note that we have used the following property:
\begin{align}
  V^{-1}T = 1 + G_0^\mathrm{R}T = \displaystyle{\sum_{n=0}^\infty }(G_0^\mathrm{R}V)^n.
\end{align}
Remind that we can write the Dirac delta function in terms of complex functions,
\begin{align}
  \frac{1}{x-i0} - \frac{1}{x+i0}
  \equiv
  2\pi i\delta(x). 
\end{align}
We can write 
\begin{align}
  \Delta \rho
  &=
  \frac{1}{2\pi i}
  \frac{\mathrm{d}}{\mathrm{d}E}
  \operatorname{Tr}
    \ln 
  \left[
    1 - \delta(E-H_0)T
  \right]
  =
  \frac{1}{2\pi i}
  \frac{\mathrm{d}}{\mathrm{d}E}
  \ln \operatorname{Det} S(E),
\end{align}
where we have defined the scattering matrix on an energy shell,
$S(E) = (1 - \delta(E-H_0)T)^{-1}$
and used a matrix identity $\operatorname{Tr} \ln S = \ln \operatorname{Det} S$.

Using the property of the scattering matrix,
\begin{align}
  S^\dagger S = 1,
\end{align}
we can write
\begin{align}
  \frac{\mathrm{d}}{\mathrm{d}E}
  (S^\dagger S) 
  &= 0,
  \\
  \left(
    \frac{\mathrm{d}}{\mathrm{d}E}
    S^\dagger
  \right) S
  + 
  S^\dagger
  \left(
    \frac{\mathrm{d}}{\mathrm{d}E}S 
  \right)
  &= 0,
  \\
  \left(
    S^\dagger
    \frac{\mathrm{d}}{\mathrm{d}E}
    S
  \right)^\dagger
  + 
  S^\dagger
  \frac{\mathrm{d}}{\mathrm{d}E}S 
  &= 0,
  \\
  \left(
    \frac{\mathrm{d}}{\mathrm{d}E}
    \ln S
  \right)^\dagger
  + 
  \frac{\mathrm{d}}{\mathrm{d}E}
  \ln S
  &= 0,
  \\
  \left(
  \operatorname{Tr}
    \frac{\mathrm{d}}{\mathrm{d}E}
    \ln S
  \right)^*
  + 
  \operatorname{Tr}
  \frac{\mathrm{d}}{\mathrm{d}E}
  \ln S
  &= 0,
  \\
  \left(
    \frac{\mathrm{d}}{\mathrm{d}E}
    \ln \operatorname{Det} S
  \right)^*
  + 
  \frac{\mathrm{d}}{\mathrm{d}E}
    \ln \operatorname{Det} S
  &= 0.
\end{align}
This implies that $\ln \operatorname{Det} S$ is a pure imaginary number,
\begin{align}
  \left(
    \ln \operatorname{Det} S
  \right)^*
  &= 
  \ln \operatorname{Det} S.
  \label{eq:lnDetS_imag}
\end{align}

Since we can obtain the dispersion relation 
($\mathfrak{F}=0$)
from the determinant of the scattering matrix,
we can write,
\begin{align}
  \frac{1}{\operatorname{Det}S}
  &= \# \mathfrak{F}
  \\
  \ln \frac{1}{\operatorname{Det}S}
  &= \ln (\# \mathfrak{F})
  \\
  -\ln \operatorname{Det}S
  &= \ln \# + \ln \mathfrak{F}.
\end{align}
Since the left-hand side is a pure imaginary number [Eq.~\eqref{eq:lnDetS_imag}],
the right-hand side should be written as following:
\begin{align}
  -\ln \operatorname{Det}S
  &= - \ln \mathfrak{F}^* + \ln \mathfrak{F},
\end{align}
and we have
\begin{align}
  \ln \operatorname{Det}S
  &= \ln \frac{\mathfrak{F}^*}{\mathfrak{F}}.
  \\
  \Delta \rho
  = 
  \frac{1}{2\pi i}
  \frac{\mathrm{d}}{\mathrm{d}E}
  \ln \operatorname{Det}S
  &=
  \frac{1}{2\pi i}
  \frac{\mathrm{d}}{\mathrm{d}E}
  \ln \frac{\mathfrak{F}^*}{\mathfrak{F}}.
\end{align}

\chapter{Quasistatic approximation}
\label{app:quasistatic}
The quasistatic approximation 
corresponds to the nonrelativistic limit 
($c \rightarrow \infty$),
\begin{align}
  \K{k}{}{\tau}
  &=
  \sgn(\omega)
  \operatorname{Re}
  \sqrt{
    \frac{\omega^2}{c^2}\epstau - {\kpara}^2
  }
  +
  i\operatorname{Im}
  \sqrt{
    \frac{\omega^2}{c^2}\epstau - {\kpara}^2
  }
  \simeq
  i\kpara,
  \\
  \xi_{\mathbf{k}}^\mathrm{mv}
  &= 
  -\frac{\K{k}{}{\mathrm{m}}/\epsm}{\K{k}{}{\mathrm{v}}/\epsv} 
  \simeq 
  -\frac{i\kpara/\epsm}{i\kpara/\epsv}
\end{align}
Applying this approximation 
to the implicit dispersion relation of the plasmon modes
at a metallic film standing in free space \eqref{eq:xi_vmv},
we can evaluate the explicit dispersion relation,
\begin{align}
  \cfrac{1 - \xi_{\mathbf{k}}^\mathrm{mv}}{1 + \xi_{\mathbf{k}}^\mathrm{mv}} 
  &= \pm e^{-\operatorname{Im} \K{k}{}{\mathrm{m}} d},
  \tag{\ref{eq:xi_vmv}}
  \\
  \cfrac{1 + \epsv/\epsm}{1 - \epsv/\epsm} 
  &\simeq \pm e^{-\kpara d},
  \\
  \omega 
  &\simeq \omega_\mathrm{sp}\sqrt{1\pm e^{-\kpara d}}.
  \tag{\ref{eq:omega_quasistatic}}
\end{align}

\chapter{Differential formalism: TM mode (p polarisation)}
\label{app:p-pol}
In the $p$ polarisation case,
we should deal with the following boundary matching equations:
\begin{align}
  \begin{cases}{}
    \eta
    \vec{t}_1 \cdot 
    (
    \vec{\mathcal{E}}{}_{\mathbf{x},a_\mathbf{x}}^\mathrm{inc} 
    + 
    \vec{\mathcal{E}}{}_{\mathbf{x},a_\mathbf{x}}^\mathrm{ref} 
    -
    \vec{\mathcal{E}}{}_{\mathbf{x},a_\mathbf{x}}^\mathrm{tra}
    ) = 0,
    \vspace{.5em}
    \\
    \vec{t}_2 \cdot 
    Z_0 (
    \vec{\mathcal{H}}{}_{\mathbf{x},a_\mathbf{x}}^\mathrm{inc} 
    +
    \vec{\mathcal{H}}{}_{\mathbf{x},a_\mathbf{x}}^\mathrm{ref} 
    - 
    \vec{\mathcal{H}}{}_{\mathbf{x},a_\mathbf{x}}^\mathrm{tra}
    )
    =
    -\eta\vec{t}_1
    \cdot
    Z_0 \vec{j}_\mathbf{x}^\mathrm{sou}.
  \end{cases}
  \tag{\ref{eq:boundary_conditions_p_pol}}
\end{align}
Let us begin with the electric field.
The tangential component of the electric field is evaluated as following:
\begin{align}
  \eta
  \vec{t}_1
  \cdot
  \vec{\mathcal{E}}_{\mathbf{x},a_\mathbf{x}}^{\Lambda}
  &=
  \int_\mathbf{k}
  e^{i\mathbf{k}\cdot\mathbf{x}}
  e^{i\sigma K_\mathbf{k}^\tau a_\mathbf{x}}
  Z_{p,\mathbf{k}}^\tau
  H_{p,\mathbf{k}}^{\sigma \tau}
  \eta \vec{t}_1
  \cdot
  \vec{e}_{p,\mathbf{k}}^{\hspace{.2em}\sigma\tau}
  \\
  &=
  \int_\mathbf{k}
  e^{i\mathbf{k}\cdot\mathbf{x}}
  \frac{Z_0}{\epsilon^\tau}
  e^{i\sigma K_\mathbf{k}^\tau a_\mathbf{x}}
  \frac{\sigma K_\mathbf{k}^\tau k_x - a_\mathbf{x}' {k_\parallel}^2}
  {k_\parallel |k_0|}
  H_{p,\mathbf{k}}^{\sigma \tau},
\end{align}
where we have used the fact that the electric and magnetic amplitude can be associated with each other,
\begin{align}
  \begin{cases}{}
    E_{s,\mathbf{k}}^{\sigma \tau}
    =
    Z_{s,\mathbf{k}}^\tau
    H_{s,\mathbf{k}}^{\sigma \tau},
    \vspace{.5em}
    \\
    E_{p,\mathbf{k}}^{\sigma \tau}
    =
    Z_{p,\mathbf{k}}^\tau
    H_{p,\mathbf{k}}^{\sigma \tau},
  \end{cases}
\end{align}
via the characteristic impedance,
\begin{align}
  Z_{\lambda,\mathbf{k}}^\tau
  =
  Z_0
  \begin{cases}{}
    \displaystyle{
      \frac{1}{\kappa^\tau}
      \sqrt{\frac{{k_0}^2}{|K_{\mathbf{k}}^\tau|^2 + {k_\parallel}^2}}
    }
    &
    (\lambda = s),
    \vspace{0.5em}
    \\
    \displaystyle{
      \frac{1}{\epsilon^\tau}
      \sqrt{\frac{|K_{\mathbf{k}}^\tau|^2 + {k_\parallel}^2}{{k_0}^2}}
    }
    &
    (\lambda = p),
  \end{cases}
\end{align}
where $Z_0=\sqrt{\mu_0/\epsilon_0}$ is the impedance of free space.
Note also that we substitute $\kappa^\tau = (\mu^\tau)^{-1} = 1$ in the current case.

We use the Jacobi-Anger identity and perform the Fourier transform,
\begin{align}
  \mathscr{F}
  \left[
    \eta
    \vec{t}_1
    \cdot
    \vec{\mathcal{E}}_{\mathbf{x},z}^{\Lambda}
  \right]_{\mathbf{k}_l}
  &=
  Z_0
  \left[
    \widetilde{\mathsf{N}}_{\mathbf{k}}^{\sigma\tau}
    \mathbb{H}_{p,\mathbf{k}}^{\sigma\tau}
  \right]_l,
\end{align}
where we have defined 
$\widetilde{\mathsf{N}}_{\mathbf{k}}^{\sigma\tau} 
= 
\mathsf{N}_{\mathbf{k}}^{\sigma\tau}/\epsilon^\tau$,
and introduced the magnetic modal amplitude vector,
\begin{align}
  \mathbb{H}_{\lambda,\mathbf{k}}^{\sigma\tau}
   &=
   \begin{pmatrix}
     \vdots
     \\
     H_{\lambda,\mathbf{k}_{-1}}^{\sigma\tau}
     \\
     H_{\lambda,\mathbf{k}_{0}}^{\sigma\tau}
     \\
     H_{\lambda,\mathbf{k}_{+1}}^{\sigma\tau}
     \\
     \vdots
   \end{pmatrix}.
   \label{eq:modal_amp_vec_H}
\end{align}

Applying the Fourier transform to Eq.~\eqref{eq:t_1_E} gives a matrix equation,
\begin{align}
  \widetilde{\mathsf{N}}_{\mathbf{k}}^{\Inc}
  \mathbb{H}_{p,\mathbf{k}}^{\Inc}
  +
  \widetilde{\mathsf{N}}_{\mathbf{k}}^{\Refl}
  \mathbb{H}_{p,\mathbf{k}}^{\Refl}
  -
  \widetilde{\mathsf{N}}_{\mathbf{k}}^{\Tra}
  \mathbb{H}_{p,\mathbf{k}}^{\Tra}
  &= 0.
  \label{eq:t_1_E_p_FT}
\end{align}

We derive the other matrix equation that is simultaneously solved with \eqref{eq:t_1_E_p_FT} from the boundary condition for the magnetic field.
We start from evaluating the tangential component of the magnetic field,
\begin{align}
  \vec{t}_2 
  \cdot 
  \vec{\mathcal{H}}_{\mathbf{x},a_\mathbf{x}}^{\Lambda}
  &=
  -\int_\mathbf{k}
  e^{i\mathbf{k}\cdot\mathbf{x}}
  \operatorname{sgn}(\omega)
  \frac{k_x}{k_\parallel}
  e^{i\phi_{\mathbf{k}}^{\sigma\tau}\sin \mathbf{q}\cdot\mathbf{x}}
  H_{p,\mathbf{k}}^{\sigma\tau}.
\end{align}
Using Jacobi-Anger expansion and applying the Fourier transform,
we can obtain
\begin{align}
  \mathscr{F}
  \left[
    \vec{t}_2 
    \cdot 
    \vec{\mathcal{H}}{}_{\mathbf{x},a_\mathbf{x}}^{\Lambda} 
  \right]_{\mathbf{k}_l}
  &=
  \left[
    \mathsf{M}_{\mathbf{k}}^{\sigma\tau}
    \mathbb{H}_{p,\mathbf{k}}^{\sigma\tau}
  \right]_l,
\end{align}
where the $\mathsf{M}$ matrix is given in Eq.~\eqref{eq:M}.

The surface current contribution in Eq.~\eqref{eq:boundary_conditions_p_pol} is evaluated as
\begin{align}
  \mathscr{F}
  \left[
    -\displaystyle{
      \frac{
        \dot{a}_\mathbf{x}  
      }{c}
    }
    \alpha
    \eta\vec{t}_1
    \cdot
    \frac{
      \vec{\mathcal{E}}{}_{\mathbf{x},a_\mathbf{x}}^\mathrm{tra}
    }{Z_0}
  \right]_{\mathbf{k}_l}
  &=
  \left[
    \widetilde{\mathsf{L}}_{\mathbf{k}}
    \mathbb{H}_{p,\mathbf{k}}^{\Tra}
  \right]_l,
\end{align}
where we have introduced 
\begin{align}
  \widetilde{\mathsf{L}}_\mathbf{k}
  &=
  \frac{A\Omega}{c}\alpha
  \left\{
    \left[
      \mathsf{N}_{\mathbf{k}}^{\Tra}
    \right]_{l-1,m}
    +
    \left[
      \mathsf{N}_{\mathbf{k}}^{\Tra}
    \right]_{l+1,m}
  \right\}.
\end{align}

Substituting those result into the boundary matching equation \eqref{eq:boundary_conditions_p_pol},
we can reach the following matrix equation,
\begin{align}
  \begin{pmatrix}
    \mathsf{M}_{\mathbf{k}}^{\Refl} 
    &
    -(\mathsf{M}_{\mathbf{k}}^{\Tra}
    +\widetilde{\mathsf{L}}_{\mathbf{k}})
    \\
    \widetilde{\mathsf{N}}_{\mathbf{k}}^{\Refl} 
    &
    -\widetilde{\mathsf{N}}_{\mathbf{k}}^{\Tra} 
  \end{pmatrix}
  \begin{pmatrix}
    \mathbb{H}_{p,\mathbf{k}}^{\Refl}
    \\
    \mathbb{H}_{p,\mathbf{k}}^{\Tra}
  \end{pmatrix}
  = 
  \begin{pmatrix}
    -\mathsf{M}_{\mathbf{k}}^{\Inc} 
    \mathbb{H}_{p,\mathbf{k}}^{\Inc}
    \\
    -\widetilde{\mathsf{N}}_{\mathbf{k}}^{\Inc} 
    \mathbb{H}_{p,\mathbf{k}}^{\Inc}
  \end{pmatrix},
  \label{eq:p_pol}
\end{align}

Inverting the coefficient matrix,
we can obtain the reflection and transmission matrices,
\begin{align}
  \begin{pmatrix}
    \mathsf{R}_{p,\mathbf{k}}
    \\
    \mathsf{T}_{p,\mathbf{k}}
  \end{pmatrix}
  &=
  \begin{pmatrix}
    \mathsf{M}_{\mathbf{k}}^{\Refl} 
    &
    -(\mathsf{M}_{\mathbf{k}}^{\Tra}
    +\widetilde{\mathsf{L}}_{\mathbf{k}})
    \\
    \widetilde{\mathsf{N}}_{\mathbf{k}}^{\Refl} 
    &
    -\widetilde{\mathsf{N}}_{\mathbf{k}}^{\Tra} 
  \end{pmatrix}^{-1}
  \begin{pmatrix}
    -\mathsf{M}_{\mathbf{k}}^{\Inc}
    \\
    -\mathsf{N}_{\mathbf{k}}^{\Inc}
  \end{pmatrix}.
  \label{eq:RTp}
\end{align}

We can use the impedance matrix,
\begin{align}
  \mathsf{Z}_{p,\mathbf{k}}^\tau 
  &=
  \operatorname{diag}
  (
  \cdots,
  Z_{p,\mathbf{k}_{-1}}^\tau,
  Z_{p,\mathbf{k}_{0}}^\tau,
  Z_{p,\mathbf{k}_{+1}}^\tau,
  \cdots
  ),
\end{align}
in order to give the reflection matrix in terms of the electric field amplitude instead of the magnetic one,
\begin{align}
  \mathbb{H}_{p,\mathbf{k}}^{\Refl}
  &=
  \mathsf{R}_{p,\mathbf{k}}
  \mathbb{H}_{p,\mathbf{k}}^{\Inc},
  \\
  \mathbb{E}_{p,\mathbf{k}}^{\Refl}
  &=
  \mathsf{Z}_{p,\mathbf{k}}^\ssg
  \mathsf{R}_{p,\mathbf{k}}
  \mathsf{Z}_{p,\mathbf{k}}^{\ssg\inv}
  \mathbb{E}_{p,\mathbf{k}}^{\Inc},
\end{align}
and the transmission matrix,
\begin{align}
  \mathbb{H}_{p,\mathbf{k}}^{\Tra}
  &=
  \mathsf{T}_{p,\mathbf{k}}
  \mathbb{H}_{p,\mathbf{k}}^{\Inc},
  \\
  \mathbb{E}_{p,\mathbf{k}}^{\Tra}
  &=
  \mathsf{Z}_{p,\mathbf{k}}^\ssl
  \mathsf{T}_{p,\mathbf{k}}
  \mathsf{Z}_{p,\mathbf{k}}^{\ssg\inv}
  \mathbb{E}_{p,\mathbf{k}}^{\Inc}.
\end{align}

\chapter{Effective surface description}
\label{app:effective}
\begin{figure}[htbp]
  \centering
  \includegraphics[width=.6\linewidth]
  {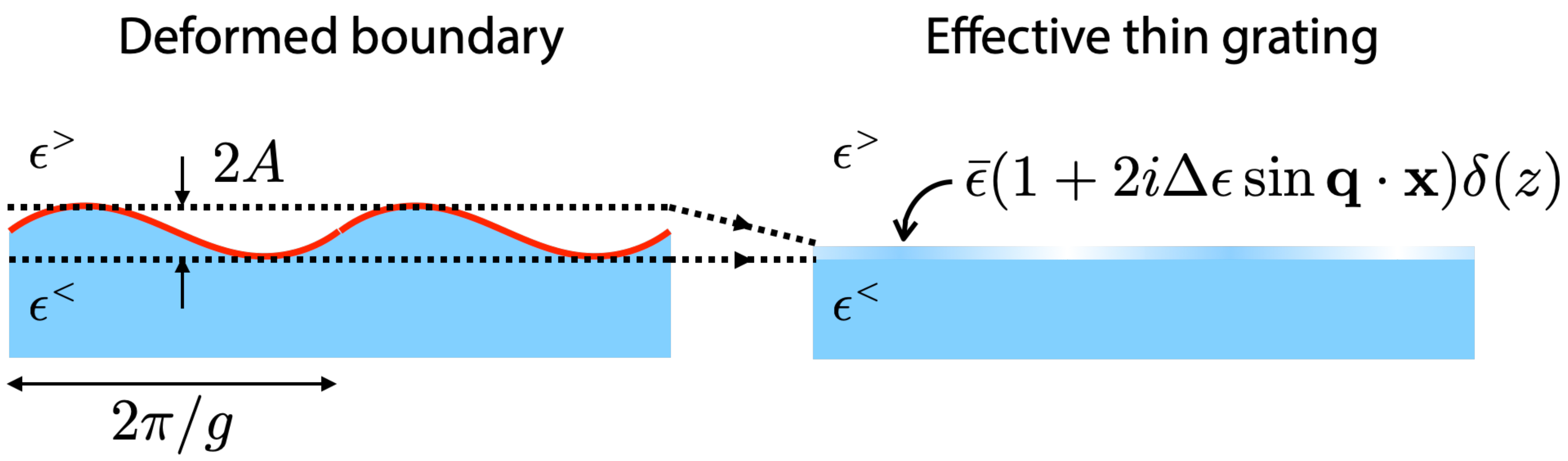}
  \caption{
    Homogenisation.
    Averaging the permittivity in the $z$ direction within the modulated region ($-A \leq z \leq +A$) generates a very thin grating model.
  }
  \label{fig:homogenisation_app}
\end{figure}

Here, we describe effective surface approximation in order to check whether the method developed in Sec.~\ref{ch:df} is consistent with the homogenisation theory at the shallow corrugation limit.
In the case of shallow corrugation,
the structure is homogenised by the electromagnetic fields in the $z$ direction and hence can be replaced by an effective flat surface with inhomogeneous permittivity
\cite{%
  aspnes1982local,%
  meade1993accurate,%
  johnson2001block,%
  kidwai2012effective%
}
(see \figref{fig:homogenisation_app}),
\begin{align}
  \epsilon_{\mathbf{x}}^\mathrm{sf}
  &= \int_{-A}^{+A} 
  \epsilon_{\mathbf{x},z}
  \hspace{.1em}
  \mathrm{d}z
  =
  \bar{\epsilon} (1 + 2i\Delta\epsilon \sin \mathbf{q} \cdot \mathbf{x}),
  \label{eq:homogenisation}
\end{align}
where we have defined the averaged and varying parts as following:
\begin{align}
  \bar{\epsilon} 
  &= (\epsilon^\ssl + \epsilon^\ssg)A,
  \quad
  \Delta\epsilon 
  = \frac{1}{2i}\frac{\epsilon^\ssl - \epsilon^\ssg}{\epsilon^\ssl + \epsilon^\ssg}.
\end{align}

\section{s polarisation}
Thanks to the homogenisation,
the surface is no longer corrugated so that we can use the conventional boundary conditions.
One is the continuity of the electric field at the surface,
\begin{align}
  \vec{u}_y
  \cdot
  \left(
    \vec{\mathcal{E}}_{\mathbf{x},z}^\mathrm{inc}
    +
    \vec{\mathcal{E}}_{\mathbf{x},z}^\mathrm{ref}
    -
    \vec{\mathcal{E}}_{\mathbf{x},z}^\mathrm{tra}
  \right)
  = 0,
  \label{eq:uy_E}
\end{align}
and the other is the discontinuity of the magnetic field,
\begin{align}
  \vec{u}_x
  \cdot
  \left(
    \vec{\mathcal{H}}_{\mathbf{x},0}^\mathrm{inc}
    +
    \vec{\mathcal{H}}_{\mathbf{x},0}^\mathrm{ref}
    -
    \vec{\mathcal{H}}_{\mathbf{x},0}^\mathrm{tra}
  \right)
  = 
  \vec{u}_y
  \cdot
  \vec{j}_\mathbf{x}^\mathrm{sou}.
  \label{eq:ux_H}
\end{align}
Note that the effective flat surface has finite permittivity,
which is space and time-dependent,
and results in the effective surface electric current on the right-hand side of Eq.~\eqref{eq:ux_H},
\begin{align}
  \vec{j}_\mathbf{x}^\mathrm{sou}
  &=
  \frac{1}{cZ_0}
  \frac{\partial}{\partial t}
  \epsilon_{\mathbf{x}}^\mathrm{sf}
  \vec{\mathcal{E}}_{\mathbf{x},0}^\mathrm{tra}.
  \label{eq:j^sou_effective}
\end{align}

We can apply the Fourier transform as following:
\begin{align}
  \mathscr{F}
  \left[
    \vec{u}_y
    \cdot
    \vec{\mathcal{E}}_{\mathbf{x},0}^\mathrm{\Lambda}
  \right]_{\mathbf{k}_l}
  &= -\sum_m 
  \operatorname{sgn}(\omega_m)
  \frac{k_{x,m}}{k_{\parallel,m}}
  \delta_{l,m}
  E_{s,\mathbf{k}_m}^{\sigma\tau}
  =
  -\left[
    \mathsf{M}_\mathbf{k}^{\star \sigma \tau}
    \mathbb{E}_{s,\mathbf{k}}
  \right]_l,
  \\
  \mathscr{F}
  \left[
    \vec{u}_x
    \cdot
    \vec{\mathcal{H}}_{\mathbf{x},0}^\mathrm{\Lambda}
  \right]_{\mathbf{k}_l}
  &=
  \sum_m
  \frac{k_{x,m}}{k_{\parallel,m}}
  \frac{\sigma K_{\mathbf{k}_m}^\tau}{|k_{0,m}|}
  \delta_{l,m}
  \frac{E_{s,\mathbf{k}_m}^{\sigma\tau}}{Z_0}
  =
  Z_0^{-1}
  \left[
    \mathsf{N}_{\mathbf{k}}^{\star\sigma\tau} 
    \mathbb{E}_{s,\mathbf{k}}^{\sigma\tau}
  \right]_l,
  \\
  \mathscr{F}
  \left[
    \vec{u}_y
    \cdot
    \frac{\partial}{\partial t}
    \epsilon_{\mathbf{x}}^\mathrm{sf}
    \frac{
      \vec{\mathcal{E}}_{\mathbf{x},0}^\mathrm{tra}
    }{Z_0}
  \right]_{\mathbf{k}_l}
  &= 
  \sum_m
  \bar{\epsilon}
  \left(
    k_{0,m} \delta_{l,m}
    +
    \Delta\epsilon 
    k_{0,m-1} \delta_{l,m-1}
    +
    \Delta\epsilon 
    k_{0,m+1} \delta_{l,m+1}
  \right)
  \operatorname{sgn}(\omega_m)
  \frac{k_{x,m}}{k_{\parallel,m}}
  \frac{E_{s,\mathbf{k}_m}^{\Tra}}{Z_0}
  \notag \\
  &= 
  Z_0^{-1}
  \left[
    \mathsf{L}_{\mathbf{k}}^{\star}
    \mathbb{E}_{s,\mathbf{k}}^{\Tra}
  \right]_l.
\end{align}
Here, we introduced a matrix representation,
where the matrix elements lead
\begin{align}
  \left[
    \mathsf{M}_\mathbf{k}^{\star \sigma \tau}
  \right]_{lm}
  &=
  \operatorname{sgn}(\omega_m)
  \frac{k_{x,m}}{k_{\parallel,m}}
  \delta_{l,m},
  \\
  \left[
    \mathsf{N}_\mathbf{k}^{\star \sigma \tau}
  \right]_{lm}
  &=
  \frac{k_{x,m}}{k_{\parallel,m}}
  \frac{\sigma K_{\mathbf{k}_m}^\tau}{|k_{0,m}|}
  \delta_{l,m},
  \\
  \left[
    \mathsf{L}_{\mathbf{k}}^{\star}
  \right]_{lm}
  &=
  \bar{\epsilon}
  \left(
    k_{0,m} \delta_{l,m}
    +
    \Delta\epsilon 
    k_{0,m-1} \delta_{l,m-1}
    +
    \Delta\epsilon 
    k_{0,m+1} \delta_{l,m+1}
  \right)
  \operatorname{sgn}(\omega_m)
  \frac{k_{x,m}}{k_{\parallel,m}},
\end{align}
and the modal electric amplitude vector,
\begin{align}
  \mathbb{E}_{\lambda,\mathbf{k}}^{\sigma\tau}
   &=
   \begin{pmatrix}
     \vdots
     \\
     E_{\lambda,\mathbf{k}_{-1}}^{\sigma\tau}
     \\
     E_{\lambda,\mathbf{k}_{0}}^{\sigma\tau}
     \\
     E_{\lambda,\mathbf{k}_{+1}}^{\sigma\tau}
     \\
     \vdots
   \end{pmatrix}.
   \tag{\ref{eq:modal_amp_vec}}
\end{align}

Therefore, we can transform Eqs.~(\ref{eq:uy_E}, \ref{eq:ux_H}) to write the scattering equation in the reciprocal space,
\begin{align}
  \begin{cases}{}
    \mathsf{M}_\mathbf{k}^{\star \Refl}
    \mathbb{E}_{s,\mathbf{k}}^{\Refl}
    -
    \mathsf{M}_\mathbf{k}^{\star \Tra}
    \mathbb{E}_{s,\mathbf{k}}^{\Tra}
    = 
    -\mathsf{M}_\mathbf{k}^{\star \Inc}
    \mathbb{E}_{s,\mathbf{k}}^{\Inc},
    \\
    \mathsf{N}_{\mathbf{k}}^{\star\Refl}
    \mathbb{E}_{s,\mathbf{k}}^{\Refl}
    -
    \left(
      \mathsf{N}_{\mathbf{k}}^{\star\Tra}
      + 
      \mathsf{L}_{\mathbf{k}}^{\star}
    \right)
    \mathbb{E}_{s,\mathbf{k}}^{\Tra}
    =
    -\mathsf{N}_{\mathbf{k}}^{\star\Inc}
    \mathbb{E}_{s,\mathbf{k}}^{\Inc}.
  \end{cases}
\end{align}
Arranging this in the matrix form,
\begin{align}
  \begin{pmatrix}
    \mathsf{M}_{\mathbf{k}}^{\star\Refl}
    &
    -\mathsf{M}_{\mathbf{k}}^{\star\Tra}
    \\
    \mathsf{N}_{\mathbf{k}}^{\star\Refl}
    &
    -(\mathsf{N}_{\mathbf{k}}^{\star\Tra}
    + 
    \mathsf{L}_{\mathbf{k}}^{\star})
  \end{pmatrix}
  \begin{pmatrix}
    \mathbb{E}_{s,\mathbf{k}}^{\Refl}
    \\
    \mathbb{E}_{s,\mathbf{k}}^{\Tra}
  \end{pmatrix}
  &=
  \begin{pmatrix}
    -\mathsf{M}_{\mathbf{k}}^{\star\Inc}
    \mathbb{E}_{s,\mathbf{k}}^{\Inc}
    \\
    -\mathsf{N}_{\mathbf{k}}^{\star\Inc}
    \mathbb{E}_{s,\mathbf{k}}^{\Inc}
  \end{pmatrix}
  \label{eq:s_pol_effective}
\end{align}
and inverting Eq.~\eqref{eq:s_pol_effective} gives reflection and transmission matrices,
\begin{align}
  \begin{pmatrix}
    \mathsf{R}_{s,\mathbf{k}}^\star
    \\
    \mathsf{T}_{s,\mathbf{k}}^\star
  \end{pmatrix}
  &=
  \begin{pmatrix}
    \mathsf{M}_{\mathbf{k}}^{\star\Refl}
    &
    -\mathsf{M}_{\mathbf{k}}^{\star\Tra}
    \\
    \mathsf{N}_{\mathbf{k}}^{\star\Refl}
    &
    -(\mathsf{N}_{\mathbf{k}}^{\star\Tra}
    + 
    \mathsf{L}_{\mathbf{k}}^{\star})
  \end{pmatrix}^{-1}
  \begin{pmatrix}
    -\mathsf{M}_{\mathbf{k}}^{\star\Inc}
    \\
    -\mathsf{N}_{\mathbf{k}}^{\star\Inc}
  \end{pmatrix}.
  \label{eq:RTs_effective}
\end{align}

\section{p polarisation}
Following the same procedure as the $s$ polarised case,
we can derive the reflection and transmission matrices.
Recalling that the surface is homogenised by the field \eqref{eq:homogenisation},
we impose the conventional boundary conditions:
the continuity of the electric field,
\begin{align}
  \vec{u}_x
  \cdot
  \left(
    \vec{\mathcal{E}}_{\mathbf{x},z}^\mathrm{inc}
    +
    \vec{\mathcal{E}}_{\mathbf{x},z}^\mathrm{ref}
    -
    \vec{\mathcal{E}}_{\mathbf{x},z}^\mathrm{tra}
  \right)
  = 0,
  \label{eq:ux_E}
\end{align}
and the magnetic field discontinuity,
\begin{align}
  \vec{u}_y
  \cdot
  \left(
    \vec{\mathcal{H}}_{\mathbf{x},z}^\mathrm{inc}
    +
    \vec{\mathcal{H}}_{\mathbf{x},z}^\mathrm{ref}
    -
    \vec{\mathcal{H}}_{\mathbf{x},z}^\mathrm{tra}
  \right)
  =
  (-\vec{u}_x)
  \cdot
  \vec{j}_\mathbf{x}^\mathrm{sou}.
  \label{eq:uy_H}
\end{align}
Note that we have the modulation induced source term on the right-hand side,
which is given by Eq.~\eqref{eq:j^sou_effective}.

We can apply the Fourier transform to have
\begin{align}
  \mathscr{F}
  \left[
    \vec{u}_y
    \cdot
    \vec{\mathcal{H}}_{\mathbf{x},0}^\Lambda
  \right]_{\mathbf{k}_l}
  &=
  -\sum_m
  \operatorname{sgn}(\omega_m)
  \frac{k_{x,m}}{k_{\parallel,m}}
  H_{p,\mathbf{k}_m}^{\sigma \tau}
  =
  -\left[
    \mathsf{M}_{\mathbf{k}}^{\star \sigma\tau}
    \mathbb{H}_{p,\mathbf{k}}^{\sigma\tau}
  \right]_l,
  \\
  \mathscr{F}
  \left[
    \vec{u}_x
    \cdot
    \vec{\mathcal{E}}_{\mathbf{x},0}^\mathrm{\Lambda}
  \right]_{\mathbf{k}_l}
  &=
  \frac{Z_0}{\epsilon^\tau}
  \sum_m
  \frac{k_{x,m}}{k_{\parallel,m}}
  \frac{\sigma K_{\mathbf{k}_m}^\tau}{|k_{0,m}|}
  \delta_{l,m}
  H_{s,\mathbf{k}_m}^{\sigma\tau}
  =
  Z_0
  \left[
    \widetilde{\mathsf{N}}_{\mathbf{k}}^{\star\sigma\tau} 
    \mathbb{H}_{s,\mathbf{k}}^{\sigma\tau}
  \right]_l,
  \\
  \mathscr{F}
  \left[
    -\vec{u}_x
    \cdot
    \frac{\partial}{\partial t}
    \epsilon_{\mathbf{x}}^\mathrm{sf}
    \frac{
      \vec{\mathcal{E}}_{\mathbf{x},0}^\mathrm{tra}
    }{Z_0}
  \right]_{\mathbf{k}_l}
  &= 
  \frac{1}{\epsilon^\ssl}
  \sum_m
  \bar{\epsilon}
  \left(
    k_{0,m} \delta_{l,m}
    +
    \Delta\epsilon 
    k_{0,m-1} \delta_{l,m-1}
    +
    \Delta\epsilon 
    k_{0,m+1} \delta_{l,m+1}
  \right)
  \frac{k_{x,m}}{k_{\parallel,m}}
  \frac{\sigma K_{\mathbf{k}_m}^\tau}{|k_{0,m}|}
  H_{s,\mathbf{k}_m}^{\Tra}
  \notag \\
  &= 
  \left[
    \widetilde{\mathsf{L}}_{\mathbf{k}}^{\star}
    \mathbb{H}_{p,\mathbf{k}}^{\Tra}
  \right]_l,
\end{align}
where we have introduced the matrix representation and defined 
\begin{align}
  \left[
    \widetilde{\mathsf{N}}_{\mathbf{k}}^{\star\sigma\tau}
  \right]_{lm}
  &=
  \frac{1}{\epsilon^{\tau}}
  \left[
    \mathsf{N}_{\mathbf{k}}^{\star\sigma\tau}
  \right]_{lm},
  \quad
  \left[
  \widetilde{\mathsf{L}}_{\mathbf{k}}^{\star\sigma\tau}
  \right]_{lm}
  =
  \frac{1}{\epsilon^{\tau}}
  \frac{\sigma K_{\mathbf{k}_m}^\tau}{|k_{0,m}|}
  \left[
    \mathsf{L}_{\mathbf{k}}^{\star\sigma\tau}
  \right]_{lm},
\end{align}
and the magnetic modal amplitude vector,
\begin{align}
  \mathbb{H}_{\lambda,\mathbf{k}}^{\sigma\tau}
   &=
   \begin{pmatrix}
     \vdots
     \\
     H_{\lambda,\mathbf{k}_{-1}}^{\sigma\tau}
     \\
     H_{\lambda,\mathbf{k}_{0}}^{\sigma\tau}
     \\
     H_{\lambda,\mathbf{k}_{+1}}^{\sigma\tau}
     \\
     \vdots
   \end{pmatrix}.
   \tag{\ref{eq:modal_amp_vec_H}}
\end{align}

We can derive matrix equations by transforming Eqs.~(\ref{eq:ux_E},\ref{eq:uy_H}),
\begin{align}
  \begin{cases}{}
    \mathsf{M}_\mathbf{k}^{\star \Refl}
    \mathbb{H}_{p,\mathbf{k}}^{\Refl}
    -(
      \mathsf{M}_\mathbf{k}^{\star \Tra}
      +
      \widetilde{\mathsf{L}}_{\mathbf{k}}^{\star}
    )
    \mathbb{H}_{p,\mathbf{k}}^{\Tra}
    = 
    -\mathsf{M}_\mathbf{k}^{\star \Inc}
    \mathbb{H}_{p,\mathbf{k}}^{\Inc},
    \\
    \widetilde{\mathsf{N}}_{\mathbf{k}}^{\star\Refl}
    \mathbb{H}_{p,\mathbf{k}}^{\Refl}
    -
    \widetilde{\mathsf{N}}_{\mathbf{k}}^{\star\Tra}
    \mathbb{H}_{p,\mathbf{k}}^{\Tra}
    =
    -\widetilde{\mathsf{N}}_{\mathbf{k}}^{\star\Inc}
    \mathbb{H}_{p,\mathbf{k}}^{\Inc}.
  \end{cases}
\end{align}

Rearranging these equations in the matrix form,
\begin{align}
  \begin{pmatrix}
    \mathsf{M}_{\mathbf{k}}^{\star\Refl}
    &
    -(\mathsf{M}_{\mathbf{k}}^{\star\Tra}
    +
    \widetilde{\mathsf{L}}_{\mathbf{k}}^{\star})
    \\
    \widetilde{\mathsf{N}}_{\mathbf{k}}^{\star\Refl}
    &
    -\widetilde{\mathsf{N}}_{\mathbf{k}}^{\star\Tra}
  \end{pmatrix}
  \begin{pmatrix}
    \mathbb{H}_{p,\mathbf{k}}^{\Refl}
    \\
    \mathbb{H}_{p,\mathbf{k}}^{\Tra}
  \end{pmatrix}
  &=
  \begin{pmatrix}
    -\mathsf{M}_{\mathbf{k}}^{\star\Inc}
    \mathbb{H}_{p,\mathbf{k}}^{\Inc}
    \\
    -\widetilde{\mathsf{N}}_{\mathbf{k}}^{\star\Inc}
    \mathbb{H}_{p,\mathbf{k}}^{\Inc}
  \end{pmatrix},
\end{align}
and inverting provides the reflection and transmission matrices,
\begin{align}
  \begin{pmatrix}
    \mathsf{R}_{p,\mathbf{k}}^{\star}
    \\
    \mathsf{T}_{p,\mathbf{k}}^{\star}
  \end{pmatrix}
  &=
  \begin{pmatrix}
    \mathsf{M}_{\mathbf{k}}^{\star\Refl}
    &
    -(\mathsf{M}_{\mathbf{k}}^{\star\Tra}
    +
    \widetilde{\mathsf{L}}_{\mathbf{k}}^{\star})
    \\
    \widetilde{\mathsf{N}}_{\mathbf{k}}^{\star\Refl}
    &
    -\widetilde{\mathsf{N}}_{\mathbf{k}}^{\star\Tra}
  \end{pmatrix}^{-1}
  \begin{pmatrix}
    -\mathsf{M}_{\mathbf{k}}^{\star\Inc}
    \\
    -\widetilde{\mathsf{N}}_{\mathbf{k}}^{\star\Inc}
  \end{pmatrix}.
\end{align}

\addcontentsline{toc}{chapter}{Bibliography}
\bibliographystyle{unsrt}
\bibliography{%
  paper/all,%
  textbook/textbook_all%
}

\end{document}